\documentclass{amsart}
\usepackage{amsmath}
\usepackage{amsxtra}
\usepackage{amscd}
\usepackage{amsthm}
\usepackage{amsfonts}
\usepackage{amssymb}
\usepackage{eucal}

\newcommand{\U}{U}
\newcommand{\Ur}{\U_A}
\newcommand{\Ure}{\U_{\epsilon}}
\newcommand{\Uri}{\U_{\scriptscriptstyle\sqrt{-1}}}
\newcommand{\Urip}{\U^{+}_{\scriptscriptstyle\sqrt{-1}}}

\newcommand{\Xc}{\mathcal{X}}
\newcommand{\Skew}{\mathop{\rm Skew}}
\newcommand{\Sym}{\mathop{\rm Sym}}
\newcommand{\Vq}{V_{K}}

\newcommand{\Jb}{\overline{J}}
\newcommand{\sN}{{\scriptscriptstyle N}}
\newcommand{\aff}{{\rm aff}}

\newcommand{\cnorm}{c^{\rm norm}}

\newcommand{\slt}{\mathfrak{sl}_2}
\newcommand{\slth}{\widehat{\mathfrak{sl}}_2}
\newcommand{\vpi}{\varpi_N}

\newcommand{\chih}{\widehat{\chi}}
\newcommand{\nn}{\nonumber}
\newcommand{\bea}{\begin{eqnarray}}
\newcommand{\ena}{\end{eqnarray}}
\newcommand{\be}{\begin{eqnarray*}}
\newcommand{\en}{\end{eqnarray*}}

\newcommand{\sgn}{{\mathop{\rm sgn}}}

\newcommand{\ch}{{\rm ch}}
\newcommand{\End}{\mathop{{\rm End}}}

\newcommand{\gr}{\mathop{{\rm gr}}}

\newcommand{\qi}[1]{[\![ #1 ]\!]}    
\newcommand{\qbin}[2]{{\left[
\begin{matrix}{\displaystyle #1}\\
{\displaystyle #2}\end{matrix}
\right]
}}

\newcommand{\vep}{\varepsilon}

\newcommand{\C}{{\mathbb C}}
\newcommand{\Z}{{\mathbb Z}} 
\newcommand{\Zb}{\overline{Z}} 

\newcommand{\R}{{\mathbb R}}
 
\newcommand{\F}{\mathcal D}

\newcommand{\A}{\mathcal A}
\newcommand{\f}{\mathbf{f}}
\newcommand{\Cc}{{\mathcal C}_N}
\newcommand{\Rc}{{\mathcal F}_N}
\newcommand{\Ic}{{\mathcal I}'}
\newcommand{\Ii}{{\mathcal I}}
\newcommand{\ds}[1]{\displaystyle #1}
\newcommand{\BW}[5]{W\left[
\begin{matrix}#1&#2\\
#3&#4\\\end{matrix}\right](#5)}
\def\sh{{\rm sh}\,}
\numberwithin{equation}{section}
\newtheorem{thm}{Theorem}[section]
\newtheorem{prop}[thm]{Proposition}
\newtheorem{lem}[thm]{Lemma}
\newtheorem{cor}[thm]{Corollary}
\newtheorem{rem}[thm]{Remark}
\newtheorem{definition}[thm]{Definition}
\newtheorem{dfn}[thm]{Definition}

\def\theenumi{\roman{enumi}}

\newcommand{\G}{\mathcal{G}}
\newcommand{\Zc}{\mathcal{Z}}
\newcommand{\Pc}{\mathcal{P}}
\newcommand{\Ker}{\mathop{\rm Ker}}
\renewcommand{\Im}{\mathop{\rm Im}}
\newcommand{\wt}{{\rm wt}\,}
\begin{document} 

\title[]
{Counting minimal form factors of the 
restricted sine-Gordon model}
\author{M. Jimbo, T. Miwa and Y. Takeyama}
\address{MJ: Graduate School of Mathematical Sciences, The 
University of Tokyo, Tokyo 153-8914, Japan}\email{jimbomic@ms.u-tokyo.ac.jp}
\address{TM: Department of Mathematics, Graduate School of Science, 
Kyoto University, Kyoto 606-8502
Japan}\email{tetsuji@kusm.kyoto-u.ac.jp}
\address{YT: Department of Mathematics, Graduate School of Science, 
Kyoto University, Kyoto 606-8502
Japan}\email{takeyama@kusm.kyoto-u.ac.jp}

\dedicatory{Dedicated to Boris Feigin on the occasion of his fiftieth birthday}

\date{\today}

\begin{abstract}
We revisit the issue of counting all local 
fields of the restricted sine-Gordon model, 
in the case corresponding to a perturbation of minimal 
unitary conformal field theory. 
The problem amounts to the study 
of a quotient of certain space of polynomials  
which enter the integral representation for form factors. 
This space may be viewed as a $q$-analog of the space of 
conformal coinvariants associated with 
$U_q(\slth)$ with $q=\sqrt{-1}$. 
We prove that its character is given by the 
restricted Kostka polynomial multiplied by a simple factor.   
As a result, we obtain a formula for 
the truncated character of the total
space of local fields in terms of the Virasoro characters.
\end{abstract}
\maketitle


\renewcommand\O{{\mathcal O}}
\newcommand\E{{\mathcal E}}

\setcounter{section}{0}
\setcounter{equation}{0}

\section{Introduction}\label{sec:Intro}

Integrable perturbation of conformal field theory initiated in \cite{Zam} 
has been a subject of intensive study over the last 15 years, 
and many rich structures have been revealed. 
{}From physics point of view, it is 
natural to expect that the space of local fields 
in a perturbed theory is `isomorphic' to its conformal limit. 
Here, by `isomorphic' we mean that their 
characters with respect to natural gradings coincide 
\footnote{Actually, in massive theory,  
the character of the whole space does not literally make sense, 
and one needs to consider a certain truncation.
See eq.\eqref{eq:negative-power} below and the discussion there.}.
The form factor bootstrap \cite{Sbk}
offers an appropriate framework 
to examine the validity of this picture.  
Favorable results have been obtained for simple models 
where the $S$ matrix is a scalar \cite{CM,Kou}.  
For models with internal degrees of freedom, 
the problem becomes far more complicated. 
Important progress in this direction has been made 
for the sine-Gordon (SG) model by Smirnov and 
Babelon-Bernard-Smirnov
\cite{Sm,BBS}. 
Nevertheless, we think that the issue of determining 
the character of the space of local fields 
has not been settled in these works.  
Recently Nakayashiki \cite{N} 
solved this problem (under certain assumptions)
for the $SU(2)$-invariant Thirring model, 
which is a rational degeneration of the SG model.    
The aim of this paper is to perform a similar 
analysis for the SG model at a generic coupling, 
and the restricted sine-Gordon (RSG) model  
corresponding to the perturbation of minimal unitary series. 

Let us describe the problem in more detail. 
Recall that in the bootstrap
approach, a local operator $\mathcal{O}$ in the 
theory is specified by its form factor. 
A form factor is a tower $f=(f_n)_{n=0}^\infty$ 
of meromorphic functions $f_n=f_n(\beta_1,\cdots,\beta_n)$, satisfying certain axioms. 
In physical terms, these functions for real values of $\beta_j$'s 
are the matrix elements of $\O$ between the vacuum and the $n$-particle 
asymptotic states with rapidities $\beta_1,\cdots,\beta_n$. 
We will refer to $f_n$ as an $n$-particle form factor. 
General matrix elements between 
$m$- and $n$-particle states 
are obtained by analytic continuation from $f_{m+n}$ \cite{Sbk}. 
When the operator $\O$ has Lorentz spin $s$, the corresponding 
form factor has the homogeneity property
\bea
&&f_n(\beta_1+\Lambda,\cdots,\beta_n+\Lambda)
=e^{s\Lambda}f_n(\beta_1,\cdots,\beta_n)
\label{Lspin}
\\
&&\hbox{ for any }\Lambda\in\R
\hbox{ and }n\in\Z_{\geq0}.
\nn
\ena
We say that $f$ has degree $s$ if \eqref{Lspin} holds. 

We consider the SG model with the coupling parameter $\xi>1$,
so that breathers do not appear in the space of physical states.
The $n$-particle form factors $f_n$ then 
take values in $(\C^2)^{\otimes n}$.
We regard $(\C^2)^{\otimes n}$ as a representation space of the quantum loop 
algebra $U_q(\widetilde{\mathfrak{sl}}_2)$, wherein 
$e^{-\beta_1/\xi},\cdots,e^{-\beta_n/\xi}$ play the role of spectral parameters. 
The parameter $q$ of the algebra and the parameter $\xi$ of the SG model are related by
\bea\label{QXI}
q=e^{-\frac{\pi i}\xi}.
\ena
In this paper, we consider the SG model in a restricted sector:
We impose the constraints that $f_n(\beta_1,\cdots,\beta_n)\in
(\C^2)^{\otimes n}$ satisfies
\bea
&&h_1f_n(\beta_1,\cdots,\beta_n)=mf_n(\beta_1,\cdots,\beta_n),\label{WEIGHT}\\
&&e_1f_n(\beta_1,\cdots,\beta_n)=0,\label{KerE}
\ena
for some $m\in \Z_{\ge 0}$, 
where $e_i,f_i,t_i=q^{h_i}$ are the Chevalley generators of 
$U_q(\widetilde{\mathfrak{sl}}_2)$. Note that we have chosen one of the two
$U_q(\slt)$ symmetries of the model in choosing the restricted
sector. In Section \ref{sec:FormFactorsofSG},
we formulate Smirnov's axioms for form factors 
in the SG model along with these constraints. 
A large class of functions satisfying these axioms
is afforded by the theory of hypergeometric integrals \cite{TV,NPT,T}. 
In counting form factors, we make the basic Ansatz that 
`all solutions' to the form factor axioms 
can be found within this category of functions. 
Henceforth we will consider only hypergeometric integrals. 
As we explain below, the problem 
reduces to the study of certain quotient spaces of polynomials
(under further assumptions to be mentioned).

Let $n,l$ be non-negative integers with $n\ge 2l$. Let $C_{n,l}$
denote the space of polynomials $P(X_1,\ldots,X_l;z_1,\ldots,z_n)$
which are skew-symmetric in
$X_1,\ldots,X_l$, symmetric in $z_1,\ldots,z_n$, and have degree less than
$n$ in each variable $X_p$. 
For $P_i\in C_{n,l_i}$ ($i=1,2$)
we will use the wedge product notation 
\bea\label{WEDGE}
&&(P_1\wedge P_2)(X_1,\cdots,X_{l_1+l_2})
=\Skew\bigl(P_1(X_1,\cdots,X_{l_1})
P_2(X_{l_1+1},\cdots,X_{l_1+l_2})\bigr),
\ena
where $\Skew$ stands for the skew symmetrization. 
The hypergeometric integral is a linear map which associates 
to each $P\in C_{n,l}$ a meromorphic function
$\psi_P(\beta_1,\cdots,\beta_n)$ with values in $(\C^2)^{\otimes n}$. 
It has the form 
\be
&&\psi_P(\beta_1,\cdots,\beta_n)=\int_{C^l}\prod_{p=1}^ld\alpha_p
\prod_{p=1}^l\phi(\alpha_p;\beta_1,\cdots,\beta_n)
\\
&&\times
v(\alpha_1,\cdots,\alpha_l;\beta_1,\cdots,\beta_n)
P(X_1,\cdots,X_l;z_1,\cdots,z_n), 
\en
where $\phi$ is a certain special function, 
$v$ is a fixed vector-valued function, 
\be
X_p=e^{-\alpha_p},\quad z_j=e^{\beta_j}, 
\en
and the integral is over certain contour $C$. 
We give the details in Section \ref{sec:FormFactorsofSG}. 
We call an element $P$ of $C_{n,l}$ a deformed cycle,
since in a certain limit the integral reduces to
a hyperelliptic integral whose cycle of integration 
is determined by $P$ \cite{Abelian}. 
The constraints \eqref{WEIGHT},\eqref{KerE} are also satisfied
by $\psi_P$, where 
the $\slt$-weight $m$ in (\ref{WEIGHT}) is related to $n$ and $l$ by 
\bea
m=n-2l.
\label{mnl}
\ena

To construct form factors, we choose $P_{n,l}\in C_{n,l}$ for each
$n, l$ satisfying \eqref{mnl}, and define
$f_n$ from $\psi_{P_{n,l}}$ by multiplying a common scalar function. 
Up to a shift depending on $n,l,\xi$, the degree $s$ in \eqref{Lspin} 
is equal to the degree of the polynomial $P_{n,l}$ given by
\bea
\deg X_p=-1,\quad \deg z_j=1.
\label{def:degXz}
\ena
Among the three main axioms for form factors, the first two 
give conditions on each $f_n$ separately. 
The $f_n$ constructed above 
automatically satisfies them for any choice of $P_{n,l}\in C_{n,l}$. 
The third axiom relates 
the residue of $f_n$ at $\beta_n=\beta_{n-1}+\pi i$ to $f_{n-2}$.
In order to satisfy this condition,  
we must choose a tower of polynomials $\{P_{n,l}\}$ properly.
In this paper we do not address this question. 
Instead, we follow the approach of minimal 
form factors \cite{Sm,N}.  

We say a form factor $f$ is $N$-minimal if
\be
f_n=0\hbox{ for all }n<N.
\en
The residue axiom then implies that 
\bea\label{NULLRES}
{\rm Res}_{\beta_N=\beta_{N-1}+\pi i}f_N=0.
\ena
We assume that, conversely,  
any function $f_N$ satisfying \eqref{NULLRES} and the rest of the axioms 
can always be extended to a form factor $f=(f_n)_{n=0}^\infty$ 
containing $f_N$ as a member.
Then the counting of local fields, or equivalently of form factors, 
is reduced to that of a single function $f_N$. 
In the context of the $SU(2)$-invariant Thirring model,  
Nakayashiki \cite{N} pointed out that 
\bea\label{RESPSI}
{\rm Res}_{\beta_N=\beta_{N-1}+\pi i}\psi_{P_{N,l}}=0
\ena
follows from the condition 
\bea\label{NULLPSI}
P_{N,l}|_{X_1^{-1}=z_1=-z_2}=0,
\ena
and proposed the assumption that it is also necessary for \eqref{RESPSI}. 
We make the same assumption in the SG model.

Let $W_{N,l}$ be the subspace of $C_{N,l}$ consisting of polynomials 
satisfying \eqref{NULLPSI}.
Under the assumptions made above, 
the space of $N$-minimal form factors can be 
identified with the quotient space $M_{N,l}=W_{N,l}/\Ker\psi\cap W_{N,l}$. 
We note that the kernel of the hypergeometric map 
$\psi:P\mapsto\psi_{P}$ is known explicitly
\cite{Sm,T}. 
It is generated by two homogeneous cycles 
$\Sigma_1(X)$ and $\Sigma_2(X_1,X_2)$
in the sense of the wedge product. 
Both $W_{N,l}$ and $M_{N,l}$ are graded 
by the degree assignment \eqref{def:degXz}.
The counting of local fields in the SG model 
is reduced to the problem of determining the character of $M_{N,l}$. 
Here and in what follows, by a character 
of a graded vector space $V=\oplus_{d}V_d$ 
we mean the generating series
\bea
\ch_q V=\sum_{d}q^d \dim V_d. 
\label{def:characterofV}
\ena
The following results are proved in \cite{N}:
\bea
\ch_q W_{N,l}&=&\frac{1}{(q)_N}\qbin{N}{l},\label{CHIW}\\
\ch_q M_{N,l}&=&\frac{1}{(q)_N}\left(\qbin{N}{l}-\qbin{N}{l-1}\right),
\label{CHIM}
\ena
where
\be
\left[N\atop l\right]=\frac{(q)_N}{(q)_l(q)_{N-l}},
\quad(q)_l=\prod_{i=1}^l(1-q^i).
\en
In this paper we 
give an alternative proof of Nakayashiki's results \eqref{CHIM} 
by clarifying the algebraic structure of $W_{N,l}$, $M_{N,l}$. 

We mentioned the quantum algebra $U_q(\widetilde{\mathfrak{sl}}_2)$ 
where $q=e^{-\pi i/\xi}$. 
This algebra describes the  
$(\C^2)^{\otimes n}$ structure of form factors. 
It controls the $(\C^2)^{\otimes n}$-valued $l$-form 
(`deformed cocycles') in the hypergeometric integral.
There appears another quantum algebra
$U_{\sqrt{-1}}(\widetilde{\mathfrak{sl}}_2)$, which is the specialization
of $U_q(\widetilde{\mathfrak{sl}}_2)$ at $q=\sqrt{-1}$ in the sense of Lustzig.
It acts on the space of deformed cycles \cite{TV,T}. 
Set 
\be
R_N=\C[z_1,\cdots,z_N]^{\mathfrak{S}_N},
\en
and denote by $x^{\pm }_n$, $a_n$ the Drinfeld generators of 
$U_q(\widetilde{\mathfrak{sl}}_2)$.  
Let $\Rc$ be the $R_N$-subalgebra
of $U_{\sqrt{-1}}(\widetilde{\mathfrak{sl}}_2)\otimes R_N$ generated by the
currents which are obtained by the specialization at
$q=\sqrt{-1}$ from
\begin{equation}
\Xc(z)=\sum_{k=1}^\infty x^-_k (q^{-1}z)^k,
\quad
\Xc(z)^{(2)}=\frac{\Xc(z)^2}{q+q^{-1}}, 
\end{equation}
along with $x_0^-$ and $(x_0^-)^{(2)}=(x^-_0)^2/(q+q^{-1})$.
At $q=\sqrt{-1}$, the generators $x^-_k$ anti-commute
by the definition of $U_{\sqrt{-1}}(\widetilde{\mathfrak{sl}}_2)$.

Set
\[
W_N=\oplus_{l=0}^NW_{N,l}.
\]
This is an $R_N$-algebra by the product given by (\ref{WEDGE}).

We will show that there exists an $R_N$-algebra homomorphism
\bea
\rho_N:\Rc\rightarrow W_N,
\ena
and determine the kernel $\Ii_N$ of the mapping $\rho_N$ explicitly
by applying a super-symmetric analog of the argument in \cite{FF}.
As a byproduct, we obtain the first identity \eqref{CHIW}.
The second identity \eqref{CHIM} is obtained as follows.
The generators $\Sigma_1,\Sigma_2$ of $\Ker\psi$ are given by 
$\rho_N(x_0^-),\rho_N((x_0^-)^{(2)})$ \cite{BBS,T}. To be precise, in \cite{T}
it is proved that $\Sigma_1$ and $\Sigma_2$ generate the kernel in the
case of the $SU(2)$-invariant Thirring model.
We assume that the same statement is valid in the sine-Gordon case.

By applying a super-symmetric version of 
the standard argument of filtration in the dual functional spaces \cite{FS}
we obtain an estimate for $\ch_qM_{N,l}$ from above. 
By counting dimensions, which follows from Tarasov's result \cite{T}, 
we obtain the estimate from below.
These are the main results for the SG model.

When the parameter $\xi$ is rational, 
a reduction takes place in the SG model. 
In this paper we consider the RSG model taking $\xi$ to be an integer $r\geq3$,
which corresponds to the $\phi_{1,3}$-perturbation of minimal unitary series. 
We call a sequence $J=(j_1,\ldots,j_n)$ of non-negative half integers 
$r$-restricted path if for all $i$ we have $j_{i+1}=j_i\pm 1$ and $2j_i\le r-2$.
As explained in \cite{RS}, asymptotic states of the RSG model are parametrized 
by $r$-restricted paths. 
In the SG model considered so far, $n$-particle form factors take values
in the space $\Omega_{n,l}$ of $(\C^2)^{\otimes n}$ 
consisting of highest weight vectors of $U_q(\slt)$ with weight $m=n-2l$.
For convenience, we make a gauge transformaton
$\tilde f_n=e^{(1/2r)\sum_j\beta_j\sigma^z_j}f_n$, 
so that the action of $U_q(\slt)$ on $\tilde{f}_n$ 
becomes independent of $\beta_1,\cdots,\beta_n$. 
The parameter $q$ is now a root of unity 
\be
\epsilon=e^{-\frac{\pi i}r}.
\en
We have a decomposition 
\bea
(\C^2)^{\otimes n}={\mathcal G}^{(r)}_n\oplus{\mathcal B}^{(r)}_n
\label{gb}
\ena
into `good' subspace ${\mathcal G}^{(r)}_n$ and `bad' subspace ${\mathcal B}^{(r)}_n$,  
the latter being a direct sum of modules with $0$ quantum dimension \cite{RT}. 
Set 
\be
\Omega^{(r)}_{n,l}=\Omega_{n,l}/\Omega_{n,l}\cap \mathcal{B}^{(r)}_n,
\en
and denote by $\mathcal{P}:\Omega_{n,l}\rightarrow \Omega^{(r)}_{n,l}$
the projection along \eqref{gb}. 
By definition, $n$-particle form factors of the RSG model are 
the projection $\mathcal{P}\tilde{f}_n$ of the one $f_n$ in the SG model. 

Under similar assumptions as in the SG case,
the space of the minimal form factors $f_N$ is identified with the 
quotient space of $M_{N,l}$ modulo those $P_{N,l}$ which satisfy
\bea\label{CUT}
\mathcal{P}\tilde{\psi}_{P_{N,l}}=0.
\ena
Set
\be
\mu=r-1-(N-2l).
\en
Setting $\Gamma_1=\rho_N(x^-_1)$ and $\Gamma_2=\rho_N((x^-_1)^{(2)})$, 
we show that \eqref{CUT} holds if $P_{N,l}$ belongs to the subspace
\be
&&\Gamma_1\wedge(\wedge^\nu\Gamma_2)\wedge M_{N,l-2\nu-1}
+(\wedge^{\nu+1}\Gamma_2)\wedge M_{N,l-2\nu-2}
\quad \hbox{ if }\mu=2\nu+1,\\
\hbox{or}&&\\
&&(\wedge^\nu\Gamma_2)\wedge M_{N,l-2\nu}\quad 
\hbox{ if }\mu=2\nu.
\en
We denote by $M_{N,l}^{(r)}$ the quotient space of $M_{N,l}$ by the above
subspace. Again, we assume that the kernel of the mapping 
$\mathcal{P}\tilde{\psi}$ is equal to this subspace.

>From the definition, $M_{N,l}^{(r)}$ is a $q$-analog of the space 
of conformal coinvariants in the level $r-2$ $SU(2)$ WZW conformal field theory,
with the deformation parameter $q$ being $\sqrt{-1}$.   
We prove that
\bea\label{RESKOS}
\ch_q M^{(r)}_{N,l}= \frac1{(q)_N}K^{(r-2)}_{N-2l,(1^N)}(q),
\ena
where $K^{(r-2)}_{N-2l,(1^N)}(q)$ is the restricted Kostka polynomial
corresponding to the tensor product $(\C^2)^{\otimes N}$. 
We note that the right hand side of (\ref{CHIM}) is equal to the Kostka
polynomial $K_{N-2l,(1^N)}(q)$, which is the limit of
$K^{(r-2)}_{N-2l,(1^N)}(q)$ when $r\rightarrow\infty$.
Apart from the factor $1/(q)_N$, 
formula \eqref{RESKOS} is a direct analog 
of the corresponding result for the conformal case 
obtained in \cite{FJKLM}. 

Our proof of the equality (\ref{RESKOS}) consists of two parts;
the estimate from above and the estimate from below. The argument for the
former is a super-symmetric version of the conformal case \cite{FJKLM}. 
To show the latter in the conformal case, we used in \cite{FJKLM}
the fusion rule proved by Tsuchiya-Ueno-Yamada \cite{TUY}.
In the present case, we employ results from 
Kashiwara's theory of global basis for level $0$-modules \cite{Kas}.

If we extend the minimal form factors to those obtained 
from the deformed cycles in the extended space
$C_{n,l}\times(z_1\cdots z_n)^{-\frac{L}{2}}$ $(L=0,1,2,\ldots)$,
we obtain an increasing sequence of the space of minimal form factors
\bea
F^{(r)}_m[0]\subset F^{(r)}_m[1]\subset F^{(r)}_m[2]\subset
\cdots.
\label{eq:negative-power} 
\ena
Under the assumptions made so far, 
the total space of form factors in the RSG model 
should be represented by the union of them.
In the above, we have considered the case $L=0$. 
{}From \eqref{RESKOS}, it follows that 
\bea
\ch_q F^{(r)}_m[0]=q^{c/24}\chi^{(r,r+1)}_{m+1,1}(q),
\label{vir0}
\ena
where $\chi^{(r,r+1)}_{b,a}(q)$ 
denotes the irreducible character of
the Virasoro minimal unitary series
with central charge and highest weight
\be
c=1-\frac{6}{r(r+1)},
\quad
h_{b a}=\frac{((r+1)b-ra)^2-1}{4r(r+1)}.
\en
For general $L$, formula \eqref{vir0} generalizes to 
\bea
\ch_q F^{(r)}_m[L]=\sum_{1\le a\le r\atop a\equiv L-1~\bmod 2}
\chi^{(r,r+1)}_{m+1,a}(q)
\chi^{(r,r+1)}_{1,a}(q^{-1};L),
\label{eq:character-formula-F} 
\ena
where 
$\chi^{(r,r+1)}_{b,a}(q;L)$ is a polynomial 
first found in \cite{ABF} as a 
finitization of $\chi^{(r,r+1)}_{b,a}(q)$.  
Formulas of this sort have been 
observed earlier in the RSG model corresponding
to the $(2,p)$ minimal series, where the $S$ matrices 
are scalar \cite{Kou}.
The appearance of $q$ and $q^{-1}$ is interpreted 
there as mixing of
the two chiralities in massive field theory. 
Here we have confirmed the validity of this picture 
in a general setting.

Our counting is based on the assumption that any $N$-particle minimal form
factor $f_N$ can be lifted to a tower $f=(f_n)_{n=0}^\infty$.
We plan to address this problem in our next paper.

The plan of the paper is as follows.
In Section \ref{sec:FormFactorsofSG}, 
we review the integral formula for form factors
of the SG model. 
After a brief review of the bootstrap approach, 
we give the hypergeometric pairing, 
describe null cycles and discuss the minimality condition. 
The materials in this section follow basically 
\cite{NPT, NT} with minor modifications. 
In Section \ref{sec:SpaceWnl}, we study the space
$W_N$. We give the action of the algebra 
$\Rc$ mentioned above, and determine the complete set of 
relations for the currents $\Xc(z),\Xc(z)^{(2)}$. 
The main results are stated in Theorems \ref{thm:WN} 
and \ref{thm:chWN}. 
In Section \ref{sec:Gordon}, we determine the structure of $M_{N,l}$. 
The formula \ref{CHIM} is proved in Theorem \ref{thm:Nakayashiki}.
The RSG model is discussed in Section \ref{sec:RSGdef}.
The formula \eqref{RESKOS} is proved in Theorem \ref{thm:4.1}. 
Section \ref{sec:Virch} is devoted to the derivation of the 
character formula \eqref{eq:character-formula-F}.

In the appendices we collect some facts concerning
the quantum loop algebra and its representations. 
In Appendix \ref{app:aff} we give our convention 
concerning $U_q(\widetilde{\mathfrak{sl}}_2)$. 
In Appendix \ref{app:JW} we discuss realization of
the currents $\Xc(z),\Xc(z)^{(2)}$ using the creation part of the
Jordan-Wigner fermions. Appendix \ref{app:TrigHGS} is an exposition of 
the action of $U_q(\widetilde{\mathfrak{sl}}_2)$
on the trigonometric hypergeometric space 
of Tarasov-Varchenko \cite{TV}. 
Appendix \ref{app:goodbad} is concerned with representations
of $U_\epsilon(\slt)$ at $\epsilon^r=-1$.



\section{Form factors of the sine-Gordon model}\label{sec:FormFactorsofSG}

\subsection{The bootstrap axiom}\label{subsec:bootstrap}
In this subsection, we review briefly
the general setting for form factors of the SG model.
Our aim is to present the main `axioms'
and motivate the subsequent discussions,
thereby introducing our notation.
For more details and the physical background,
the reader is referred to \cite{Sbk}.

Fix a real parameter $\xi>1$ throughout.
Consider the quantum loop algebra
$U=U_q(\widetilde{\mathfrak{sl}}_2)$ with $q=e^{-\pi i/\xi}$.
We will use the convention concerning $U$
in Appendix \ref{app:aff}.
Let
\be
V=\C v_+\oplus \C v_-\simeq \C^2.
\en
For $\zeta\in\C^{\times}$, let
$\pi_\zeta:U\rightarrow \End(V)$
be the representation given by
\be
e_0\mapsto \zeta \sigma^-,~~
f_0\mapsto \zeta^{-1} \sigma^+,~~
e_1\mapsto \zeta \sigma^+,~~
f_1\mapsto \zeta^{-1} \sigma^-,~~
t_0\mapsto q^{-\sigma^z},~~
t_1\mapsto q^{\sigma^z},
\en
where $\sigma^\pm$, $\sigma^z$ are the Pauli matrices.
We regard the tensor product
\bea\label{PARTICLE}
\pi_{\zeta_1}\otimes\cdots\otimes\pi_{\zeta_n}
\ena
as a $U$-module via the coproduct
\bea
&&\Delta'(e_i)=e_i\otimes t_i+1\otimes e_i,\quad
\Delta'(f_i)=f_i\otimes 1+t_i^{-1}\otimes f_i,
\quad\Delta'(t_i)=t_i\otimes t_i.
\label{def:OPP}
\ena
This coproduct is opposite to the one given in \eqref{copro}.
We will use the symmetric bilinear form
$(~,~)$ such that the vectors $v_{\vep_1}\otimes\cdots
\otimes v_{\vep_n}$ are orthonormal.
We have
\be
(xu,v)=(u,\alpha(x)v)
\qquad
(\forall u\in
\pi_{\zeta_1^{-1}}\otimes\cdots\otimes\pi_{\zeta_n^{-1}},
~~\forall v\in \pi_{\zeta_1}\otimes\cdots\otimes\pi_{\zeta_n}),
\en
where $\alpha:U\rightarrow U$ denotes
the anti-involution given by
\be
\alpha(e_i)=qt_if_i,
\quad\alpha(f_i)=q^{-1}e_it_i^{-1},
\quad\alpha(t_i)=t_i.
\en

At a heuristic level, the space of physical states
of the SG model is a `direct sum' of
$\pi_{\zeta_1}\otimes\cdots\otimes \pi_{\zeta_n}$
where $n\in\Z_{\ge 0}$, $\zeta_j=e^{\beta_j/\xi}$,
and $\beta_1,\cdots,\beta_n$ are real parameters
called rapidities.
We use the symbol
$|\beta_1,\ldots,\beta_n\rangle_{\varepsilon_1,\ldots,\varepsilon_n}$
to represent the vector
$v_{\varepsilon_1}\otimes\cdots\otimes v_{\varepsilon_n}$
in the module
$\pi_{\zeta_1}\otimes\cdots\otimes\pi_{\zeta_n}$.
There is an exchange relation
which identifies the vectors when $\beta_j$
and $\beta_{j+1}$ are interchanged.
This relation is given by the $S$-matrix.

The $S$-matrix of the SG model is
\be
S(\beta)=S_{0}(\beta)\widehat{S}(\beta).
\en
Here $\widehat{S}(\beta)$ is the linear operator defined by
\be
&&
\widehat{S}(\beta)(v_{\epsilon_{1}}\otimes v_{\epsilon_{2}})
=\sum_{\epsilon_{1}', \epsilon_{2}'}
\widehat{S}_{\epsilon_{1}, \epsilon_{2}}^{\epsilon_{1}',
\epsilon_{2}'}(\beta)
(v_{\epsilon_{1}'} \otimes v_{\epsilon_{2}'}), \\
&&
\widehat{S}_{\pm, \pm}^{\pm, \pm}(\beta)=1,\quad
\widehat{S}_{\pm, \mp}^{\pm, \mp}(\beta)=
\frac{{\rm sh}\frac{\beta}{\xi}}
     {{\rm sh}\frac{1}{\xi}(\beta-\pi i)}, \quad
\widehat{S}_{\pm, \mp}^{\mp, \pm}(\beta)=
\frac{{\rm sh}\frac{-\pi i}{\xi}}
     {{\rm sh}\frac{1}{\xi}(\beta-\pi i)}, \\
&&
\widehat{S}_{\epsilon_{1}, \epsilon_{2}}^{\epsilon_{1}',
\epsilon_{2}'}(\beta)=0,
\quad {\rm otherwise},
\en
and $S_{0}(\beta)$ is the normalization factor
\bea
S_{0}(\beta)=
\frac{S_{2}(-i\beta)S_{2}(\pi +i\beta)}{S_{2}(\pi-i\beta)S_{2}(i\beta)},
\label{def:normalization-factor}
\ena
where $S_{2}(\beta)=S_{2}(\beta | 2\pi , \xi\pi)$ is
the double sine function.
(See \cite{JM} for properties of the double sine function.)
The exchange relation is given by
\be
&&
|\cdots,\beta_j,\beta_{j+1},\cdots
\rangle_{\cdots,\epsilon_j,\epsilon_{j+1},\cdots}
\\
&&
\quad 
=\sum_{\epsilon'_j,\epsilon'_{j+1}}
|\cdots,\beta_{j+1},\beta_{j},\cdots
\rangle_{\cdots,\epsilon'_{j+1},\epsilon'_{j},\cdots}
\times 
S^{\epsilon'_j,\epsilon'_{j+1}}_{\epsilon_j,\epsilon_{j+1}}(\beta_{j+1}-\beta_j).
\en
The map 
\be
PS(\beta_2-\beta_1): v_{\epsilon_1}\otimes v_{\epsilon_2}\mapsto 
\sum_{\epsilon'_1,\epsilon'_2}v_{\epsilon'_2}\otimes v_{\epsilon'_1}
\times S^{\epsilon'_1,\epsilon'_2}_{\epsilon_1,\epsilon_2}(\beta_2-\beta_1)
\en
is an intertwiner from 
$\pi_{\zeta_1}\otimes\pi_{\zeta_2}$
to 
$\pi_{\zeta_2}\otimes\pi_{\zeta_1}$,
where $P(v_1\otimes v_2)=v_2\otimes v_1$. 
Hence the above
relation is compatible with the action of $U$.

We denote the vacuum vector by $|\rm vac\rangle$.
This is a generator of the space of $0$ particle state,
i.e., $n=0$ in (\ref{PARTICLE}).
We denote its dual vector by $\langle\rm vac|$.
A local operator $\O$ is an operator
acting on the space of physical states.
It is uniquely specified by the matrix element \cite{Sbk}
\[
f_n(\beta_1,\ldots,\beta_n)_{\varepsilon_1,\ldots,\varepsilon_n}
=\langle{\rm vac}|\O
|\beta_1,\ldots,\beta_n\rangle_{\varepsilon_1,\ldots,\varepsilon_n}.
\]
They are encapsulated into a dual vector
\be
f_n(\beta_1,\ldots,\beta_n)
=\sum_{\varepsilon_1,\ldots,\varepsilon_n}
f_n(\beta_1,\ldots,\beta_n)_{\varepsilon_1,\ldots,\varepsilon_n}
v_{\varepsilon_1}\otimes\cdots\otimes v_{\varepsilon_n},
\en
on which $U$ acts by
$\pi_{\zeta_1^{-1}}\otimes\cdots\otimes\pi_{\zeta_n^{-1}}$.
We call the tower of vector-valued functions
$f=\bigl(f_n(\beta_1,\ldots,\beta_n)\bigr)_{n\ge 0}$
the form factor of the local oparator $\O$.

In this paper we consider only form factors
satisfying the conditions
\be
({\rm A0}) &&\!\!\!h_1f_n(\beta_{1}, \ldots , \beta_{n})=
mf_n(\beta_{1}, \ldots , \beta_{n}), \\
&&\!\!\!
e_1f_n(\beta_{1}, \ldots , \beta_{n})=0
\en
for some $m\in\Z_{\geq0}$.
In other words, the vector $f_n(\beta_{1}, \ldots , \beta_{n})$
is the highest weight vector of the
$m+1$ dimensional sub-representation of
$V^{\otimes n}$ as $U_q(\frak{sl}_2)$-module,
where $U_q(\frak{sl}_2)$ signifies the subalgebra
of $U$ generated by $e_1,f_1,t_1$.
As it is explained in \cite{RS},
this amounts to considering a certain
subsector of local operators.
This class is further subdivided into two subsectors
labeled by the index $\varepsilon\in\{0,1\}$, 
which enters the axioms (A2), (A3) below. 

Physical consideratons for local operators
lead to several conditions on form factors.
They are required to have the following
analyticity and asymptotic properties.
\begin{enumerate}
{\renewcommand{\theenumi}{\alph{enumi}}
\item $f_{n}(\beta_1,\ldots, \beta_n)$ extend
to meromorphic functions in $\beta_1, \cdots,\beta_n$
on the whole complex plane,
\item When $\beta_j\in\R$ ($j\neq n$),
they are holomorphic in the domain $0<{\rm Im}\beta_n<2\pi$
except for possible simple poles at $\beta_n=\beta_j+\pi i$,
\item When $\beta_j\to \pm\infty$ we have
$|f_{n}(\beta_1,\ldots, \beta_n)|=O(e^{K|\beta_j|})$
for some $K>0$.
}
\end{enumerate}
In addition, the following main `axioms' are imposed
\cite{Sbk}:
\be
({\rm A1}) &&\!\!\!
f_{n}(\ldots , \beta_{j+1}, \beta_{j}, \ldots)=
P_{j, j+1}S_{j, j+1}(\beta_{j}-\beta_{j+1})
f_{n}(\ldots , \beta_{j}, \beta_{j+1}, \ldots), \\
({\rm A2}) &&\!\!\!
f_{n}(\beta_{1}, \ldots , \beta_{n-1}, \beta_{n}+2\pi i) \\
&& {}=
e^{(\frac{n}{2}+\varepsilon)\pi i}
e^{\frac{m\pi i}{2\xi}\sigma_{n}^{z}}
P_{n, n-1} \cdots P_{2, 1}
f_{n}(\beta_{n}, \beta_{1}, \ldots , \beta_{n-1}), \\
({\rm A3}) &&\!\!\!
{\rm res}_{\beta_{n}=\beta_{n-1}+\pi i}f_{n}(\beta_{1}, \ldots , \beta_
{n})
\\
&& {}=
(I+e^{-(\frac{n}{2}+\varepsilon)\pi i}
S_{n-1, n-2}(\beta_{n-1}-\beta_{n-2})
\cdots S_{n-1, 1}(\beta_{n-1}-\beta_{1})e^{-\frac{m\pi i}{2\xi}\sigma_{n
-1}^{z}}) \\
&&\quad\times f_{n-2}(\beta_1,\ldots,\beta_{n-2})
\otimes(v_+\otimes v_--v_-\otimes v_+).
\en
In the above, the subscripts of the operators
refer to the components in the tensor product $V^{\otimes n}$
on which the operators act.

A large family of functions with these properties
can be constructed in terms of the hypergeometric integrals \cite{Sbk,TV}
(see \eqref{def:formfactor-P} below).
We expect that `all' form factors can be obtained in this way.
Motivated by these considerations, henceforth we restrict ourselves
to form factors obtained by hypergeometric integrals.

In the next section, we define the hypergeometric
integral following \cite{Sbk,TV}, and using it
we construct $N$-minimal form factors of the $\mathfrak{sl}_2$ weight $N-2l$.
We fix $N$ and $l$ until we start the discussion on Virasoro characters
in Section \ref{sec:Virch} 
(except in the proof of Proposition \ref{FeiginFeigin:upper}).

The hypergeometric integral consists of three ingredients;
the phase function
$\phi(\alpha;\beta_1,\ldots,\beta_N)$, a deformed cocycle
$w(a_1,\ldots,a_l;\beta_1,\ldots,\beta_N)$ and
a deformed cycle $P(X_1,\ldots,X_l;z_1,\ldots,z_N)$.
It gives a pairing
between deformed cocycles and deformed cycles.
Here and in what follows, we set
\begin{equation}
X_p=e^{-\alpha_p},z_j=e^{\beta_j},
a_p=e^{\frac{2\alpha_p}\xi},b_j=e^{\frac{2\beta_j}\xi},
\omega=e^{\frac{2\pi i}\xi}.
\end{equation}
The variables $X_p$ and $z_j$ are $2\pi i$ periodic
with respect to $\alpha_p$ and $\beta_j$, respectively,
while the variables $a_p$ and $b_j$ are $\xi\pi i$ periodic.

\subsection{Phase function}
\par\noindent
We use the function
\[
\varphi(\alpha)=\frac{1}{S_2(\frac\pi2-i\alpha)S_2(\frac\pi2+i\alpha)}.
\]
Define the phase function
\begin{equation}
\phi(\alpha;\beta_1,\ldots,\beta_N)=\prod_{j=1}^N
\Bigl(e^{\frac{\xi-1}{2\xi}(\alpha-\beta_j)}\varphi(\alpha-\beta_j+
{\scriptstyle\frac{3\pi i}2})\Bigr).
\label{def:phase-function}
\end{equation}
We have
\begin{eqnarray}
&&\frac{\phi(\alpha;\beta_1,\ldots,\beta_N+2\pi i)}
{\phi(\alpha;\beta_1,\ldots,\beta_N)}=
\frac{\omega a-b_N}{a-b_N},\label{QP1}\\
&&\frac{\phi(\alpha-2\pi i;\beta_1,\ldots,\beta_N)}
{\phi(\alpha;\beta_1,\ldots,\beta_N)}=\prod_{j=1}^N
\frac{\omega a-b_j}{a-b_j},\label{QP2}\\
&&\frac{\phi(\alpha-\xi\pi i;\beta_1,\ldots,\beta_N)}
{\phi(\alpha;\beta_1,\ldots,\beta_N)}=\prod_{j=1}^N
\frac{e^{-(\alpha-\beta_j)}+1}{e^{-(\alpha-\beta_j)+\xi\pi i}-1},\label{QP3}
\end{eqnarray}
where $a=e^{2\alpha/\xi}$.
We have the estimates,
\begin{eqnarray}
|\phi(\alpha)|=
\begin{cases}
O(e^{-\frac{N\alpha}\xi})&\hbox{when }\alpha\rightarrow\infty;\\
O(e^{N\alpha})&\hbox{when }\alpha\rightarrow-\infty.
\end{cases}
\end{eqnarray}

For each integer $l$ such that $0\leq 2l\leq N$, we define
the space of deformed cocycles and that of deformed cycles.

\subsection{Deformed cocycles}
\par\noindent
A deformed cocycle $w$ is a function of the variables
$a_1,\ldots,a_l$ and $\beta_1,\ldots,\beta_N$ such that
\begin{eqnarray}
&&w(a_1,\ldots,a_l;\beta_1,\ldots,\beta_N)\label{WQ}\\
&&=\frac{Q(a_1,\ldots,a_l;\beta_1,\ldots,\beta_N)}
{\prod_{p=1}^l\prod_{j=1}^N(a_p-b_j)},\nonumber
\end{eqnarray}
where $Q$ is a polynomial in $a_1,\ldots,a_l$ satisfying the conditions,
\begin{eqnarray}
&&Q\hbox{ is skew-symmetric in }a_1,\ldots,a_l,\label{SKEW}\\
&&{\rm deg}_{a_p}Q\leq N+l-1,\label{DEG}\\
&&Q|_{a_p=0}=0\hbox{ for any }p,\label{NULL}\\
&&Q|_{a_p=\omega a_{p'}=b_j}=0\hbox{ for any }p,p',j.\label{SERRE}
\end{eqnarray}
We define special cocycles $w_M$ indexed by a subset
$M\subset\{1,\ldots,N\}$ such that $M=\{m_1<\cdots<m_l\}$, and we mainly
use them:
\begin{equation}
w_M={\rm Skew}_{a_1,\ldots,a_l}g_M. 
\end{equation}
Here the function $g_{M}$ is defined by 
\begin{eqnarray}
&&
g_M(a_1,\ldots,a_l;\beta_1,\ldots,\beta_N)\label{G}=
e^{\frac{N-4l}{4\xi}\sum_{j=1}^N\beta_j
+\frac{1}{\xi}\sum_{p=1}^{l}\beta_{m_{p}}} \label{def:function-g}\\ 
&& \qquad {}\times 
\prod_{p=1}^l
\frac{a_p\prod_{j<m_p}(\omega a_p-b_j)\prod_{j>m_p}(a_p-b_j)}
{\prod_{j=1}^N(a_p-b_j)}
\prod_{p<p'}(\omega a_p-a_{p'}), \nonumber
\end{eqnarray}
and ${\rm Skew}_{a_{1}, \ldots , a_{l}}$ is the skew-symmetrization 
with respect to $a_{1}, \ldots , a_{l}$:
\be 
{\rm Skew}_{a_{1}, \ldots , a_{l}}g_{M}=\sum_{\sigma\in \frak{S}_l}(\sgn\,\sigma)\,
g_{M}(a_{\sigma(1)},\cdots,a_{\sigma(l)}),
\en
where $\frak{S}_l$ stands for the symmetric group on $l$ letters. 

In the construction of the form factors, we use the special vector $v_{N,l}$
in $\left(V^{\otimes N}\right)_l=\{v\in V^{\otimes N};h_1v=(N-2l)v\}$
given by
\begin{equation}
v_{N,l}=\sum_{\# M=l}q^{\nu(M)}w_Mv_M
\end{equation}
where $N,l$ are fixed in the right hand side, and
\begin{eqnarray}
&&\nu(M)=\sum_{p=1}^l(m_p-1),\label{NU}\\
&&v_M=v_{\varepsilon_1}\otimes\cdots\otimes v_{\varepsilon_N}, \label{vM}
\end{eqnarray}
where
\begin{equation}
M=\{j|1\le j \le N, \varepsilon_j=-\}.\label{VAREP}
\end{equation}

When we specialize $\beta_1,\ldots,\beta_N$ to generic values,
the vector space of the deformed cocycles is $\binom Nl$-dimensional,
and it is spanned by $w_M$. This is known in \cite{TV}.

\subsection{Deformed cycles}
\par\noindent
A deformed cycle $P$ is a polynomial of $X_1,\ldots,X_l$ and
$z_1,\ldots,z_N$ satisfying the conditions
\begin{eqnarray}
&&P\hbox{ is skew-symmetric in }X_1,\ldots,X_l,\\
&&P\hbox{ is symmetric in }z_1,\ldots,z_N,\\
&&{\rm deg}_{X_p}P\leq N-1.\label{DEGP}
\end{eqnarray}
Denote by $C_{N, l}$ the space of deformed cycles with fixed $N,l$.
\subsection{Hypergeometric pairing}
The hypergeometric pairing of a deformed cocycle $w$ and a deformed cycle $P$
is  given by
\begin{eqnarray}
&&I(w,P)=\int_{C^l}\prod_{p=1}^ld\alpha_p\prod_{p=1}^l
\phi(\alpha_p;\beta_1,\ldots,\beta_N) \label{def:hypergeometric-pairing}\\
&&\times w(a_1,\ldots,a_l;\beta_1,\ldots,\beta_N)
P(X_1,\ldots,X_l;z_1,\ldots,z_N),\nonumber
\end{eqnarray}
where the integration contour $C$  goes along real axis except that
the simple poles of the integrand at
\begin{equation}
\alpha_p=\beta_j-2\pi i \Z_{\geq0}-\xi\pi i \Z_{\geq0}
\end{equation}
are located below $C$, and those at
\begin{equation}
\alpha_p=\beta_j-\pi i+2\pi i \Z_{\geq0}+\xi\pi i \Z_{\geq0}
\end{equation}
above $C$. These are the only poles of the integrand.

Recall that we have the restriction $0\leq2l\leq N$ from the axiom (A5).
The convergence of the integral follows from the following estimates when
$\alpha_p\rightarrow\pm\infty$.
\begin{eqnarray}
&&|w|=
\begin{cases}
O(e^{\frac{2(l-1)\alpha_p}\xi})&
\hbox{when }\alpha_p\rightarrow\infty;\\
O(e^{\frac{2\alpha_p}\xi})&
\hbox{when }\alpha_p\rightarrow-\infty,
\end{cases}\\
&&|P|=
\begin{cases}
O(1)&\hbox{when }\alpha_p\rightarrow\infty;\\
O(e^{-(N-1)\alpha_p})&\hbox{when }\alpha_p\rightarrow-\infty.
\end{cases}
\end{eqnarray}

\begin{rem}\label{WEAK}
{\rm 
The convergence of the integral $I(w,P)$ is valid even if
we weaken the condition $(\ref{DEGP})$ to
\begin{equation}
{\rm deg}_{X_p}P\leq N.\label{DEGPP}
\end{equation}
This is because of the condition $(\ref{NULL})$ for $w$.
If we drop this condition for $w$, the convergence is still valid
under the assumption $(\ref{DEGP})$ for $P$.
} 
\end{rem}

\begin{thm}\label{QKZ}
For each deformed cycle $P$, the hypergeometric integral
\begin{equation}
\psi_P(\beta_1,\ldots,\beta_N)=I(v_{N,l},P)
\end{equation}
satisfies the following: 
\be 
&& 
h_{1} \psi_{P}(\beta_{1}, \ldots , \beta_{N})=
(N-2l)\psi_{P}(\beta_{1}, \ldots , \beta_{N}), 
\label{eq:axiom0} \\ 
&& 
\psi_{P}(\ldots , \beta_{j+1}, \beta_{j}, \ldots)= 
P_{j, j+1}\widehat{S}_{j, j+1}(\beta_{j}-\beta_{j+1})
\psi_{P}(\ldots , \beta_{j}, \beta_{j+1}, \ldots), 
\label{eq:axiom1} \\ 
&&
\psi_{P}(\beta_{1}, \ldots , \beta_{N-1}, \beta_{N}+2\pi i)=
e^{\frac{(N-2l)\pi i}{2\xi}\sigma_{N}^{z}} 
P_{N, N-1} \cdots P_{2, 1} 
\psi_{P}(\beta_{N}, \beta_{1}, \ldots , \beta_{N-1}).
\label{eq:axiom2}
\en 
\end{thm}
The proof is similar to that of Theorem 6.3 in \cite{NPT}. 
We omit the proof here.

When we specialize $\beta_1,\ldots,\beta_N$, the space of the deformed cycles
is nothing but the space of skew-symmetric polynomials in $X_1,\ldots,X_l$
satisfying the degree restriction (\ref{DEGP}). This is an
$\binom Nl$-dimensional vector space. If the values of $\beta_1,\ldots,\beta_N$
are generic, the space spanned by $w_M$ is also $\binom Nl$-dimensional.
The integral $I(w,P)$ defines a pairing between these two spaces.
This pairing is degenerate. There exist cocycles $w_0$ such that $I(w_0,P)=0$
for all $P$, and vice versa. We call them null (co)cycles.

We define the twisted difference operator $\nabla_{\alpha,c}$
acting on a function $f(\alpha)$ by
\begin{equation}
\nabla_{\alpha,c}(f)=f(\alpha)-f(\alpha+c)\times
\frac{\phi(\alpha+c)}{\phi(\alpha)}.
\end{equation}
Similarly, for two variables $\alpha_1,\alpha_2$, we define
\begin{equation}
\nabla_{\alpha_1,\alpha_2,c}(f)
=f(\alpha_1,\alpha_2)-f(\alpha_1+c,\alpha_2+c)\times
\frac{\phi(\alpha_1+c)\phi(\alpha_2+c)}{\phi(\alpha_1)\phi(\alpha_2)}.
\end{equation}

If the integral $\int_Cd\alpha\phi f$ is convergent and the integrand
has no pole between the contours $C$ and $C+c$, then
\[
\int_Cd\alpha\phi\nabla_{\alpha,c}(f)=(\int_C-\int_{C+c})d\alpha\phi f=0.
\]
We use $c=-2\pi i$ for the construction of null cocycles,
and $c=-\xi\pi i$ for null cycles. The ratio $\phi(\alpha+c)/\phi(\alpha)$
for these $c$ is given by $(\ref{QP2})$ and $(\ref{QP3})$.
We call this type of argument for the vanishing of the integral
`the twisted difference method'.
\subsection{Null cocycles}\label{NCC}
We can extend the definition of the pairing $I(w,P)$ to a wider class
of functions $w$ by dropping the skew-symmetry of $Q$ and the conditions
(\ref{NULL}) and (\ref{SERRE}). In this case, we must restrict $P$ by
(\ref{DEGP}) for convergence. Since $P$ is skew-symmetric, the integral
$I(w,P)$ is multiplied by $l!$ if we replace $w$ by its skew-symmetrization
with respect to $a_1,\ldots,a_l$.
\begin{lem}
Let $w$ be of the form $(\ref{WQ})$ where $Q$ is a polynomial in
$a_1,\ldots,a_l$ satisfying $(\ref{DEG})$. We assume that $Q$
has zero at $a_1=b_j$ for all $1\leq j\leq N$.
We have
\[
\nabla_{\alpha_1,-2\pi i}(w)
=w(\alpha_1)-w(\alpha_1-2\pi i)\times
\prod_{j=1}^N\frac{\omega a_1-b_j}{a_1-b_j}.
\]
and
\[
I(\nabla_{\alpha_1,-2\pi i}(w),P)=0
\]
for any deformed cycle $P$.
\end{lem}
\begin{proof}
We apply the twisted difference method.
The integrand of $I(w,P)$ has no pole in $\alpha_1$ between the contours $C$
and $C-2\pi i$. The deformed cycle $P$ is $2\pi i$ periodic, and
the phase function $\phi$ satisfies $(\ref{QP2})$. The assertion follows from
these properties.
\end{proof}
\begin{lem}
Fix $N$ and $l$, and let
\[
\tilde g_M=(\omega-1)e^{-\frac{N-4l}{4\xi}
\sum_{j=1}^N\beta_j-\frac{1}{\xi}\sum_{p=1}^l\beta_{m_p}}g_{M}
\]
where $g_M$ is given by $(\ref{G})$.
For each $k\not\in M$, we similarly define $\tilde g_{M\cup\{k\}}$
with $l$ replaced by $l+1$. We define $\varepsilon_j$ by $(\ref{VAREP})$
for $M$, and $\nu({M\cup\{k\}})$ by $(\ref{NU})$ for $M\cup\{k\}$.
Set
\[
w(\alpha,\alpha_1,\ldots,\alpha_l)=-\prod_{p=1}^l(\omega a-a_p)\times
\tilde g_M(\alpha_1,\ldots,\alpha_l;\beta_1,\ldots,\beta_N).
\]
Then, we have
\[
\nabla_{\alpha,-2\pi i}(w)\equiv q^{-\nu(M)-1}\sum_{k\not\in M}
q^{\sum_{j\leq k}\varepsilon_j+\nu(M\cup\{k\})}\tilde g_{M\cup\{k\}},
\]
where the equivalence relation $A\equiv B$ means
${\rm Skew}_{a,a_1,\ldots,a_l}(A-B)=0$.
\end{lem}
\begin{proof}
This is a straightforward generalization of (3.5) in Lemma 3.5 of \cite{NPT}.
\end{proof}
By these lemmas we conclude that the deformed cocycles
\begin{equation}\label{NULLCO}
\sum_{k\not\in M}q^{\sum_{j\leq k}\varepsilon_j+\nu(M\cup\{k\})}
e^{-\frac{1}{\xi}\beta_{k}}w_{M\cup\{k\}}
\end{equation}
with $\sharp(M)=l-1$ are null cocycles.

Consider an action of $U_q(\mathfrak{sl}_2)$ with the canonical generators
$e_{1}, f_{1}, t_{1}$ and $q=e^{-\frac{\pi i}\xi}$,
on $V^{\otimes N}$ through the opposite coproduct
$\Delta'$ given by (\ref{def:OPP}). The function $\psi_{P}$ takes values
in the representation space $V^{\otimes N}$ 
of $\pi_{\zeta_{1}^{-1}} \otimes \cdots \otimes \pi_{\zeta_{N}^{-1}}$ 
with $\zeta_{j}=e^{\beta_{j}/\xi}$. 
\begin{cor}\label{INV}
The vector $\psi_{P}=I(v_{N,l},P)$ is a highest weight vector for any deformed
cycle $P:$
\begin{equation}
e_{1} \psi_{P}(\beta_{1}, \ldots , \beta_{N})=0. 
\end{equation}
\end{cor}

Now we introduce the function $\zeta(\beta)$ defined by 
\be 
\zeta(\beta)=
\frac{S_{3}(-i\beta+2\pi)S_{3}(i\beta)}{S_{3}(-i\beta+3\pi)S_{3}(i\beta+\pi)}, 
\quad 
S_{3}(\beta)=S_{3}(\beta | 2\pi , 2\pi , \xi\pi). 
\en 
Here $S_{3}(\beta)$ is the triple sine function 
(see \cite{JM} for the definition and properties). 

For a deformed cycle $P \in C_{N, l}$ we set 
\bea 
&& 
\label{def:formfactor-P} \\
&&
f_{P}(\beta_{1}, \ldots , \beta_{N})= 
e^{(\frac{N}{4}-\frac{\varepsilon}{2})\sum_{j=1}^{N}\beta_{j}}
\prod_{1 \le j<j' \le N}\!\!\!\zeta(\beta_{j}-\beta_{j'}) \cdot
\psi_{P}(\beta_{1}, \ldots , \beta_{N}).  \nn
\ena 
Here we write down the formula for the function $\psi_{P}$: 
\be 
&& 
\psi_{P}(\beta_{1}, \ldots , \beta_{N})=\sum_{\# M=l}v_{M} 
\int_{C^{l}}\prod_{p=1}^{l}d\alpha_{p} 
\prod_{p=1}^{l}\phi(\alpha_{p}; \beta_{1}, \ldots , \beta_{N}) \\ 
&& {} \quad {}\times 
\left( 
{\rm Skew}_{\alpha_{1}, \ldots , \alpha_{l}} 
g'_{M}(\alpha_{1}, \ldots , \alpha_{l}; \beta_{1}, \ldots , \beta_{N}) \right) 
P(X_{1}, \ldots , X_{l} ; z_{1}, \ldots , z_{N}). 
\en 
The phase function $\phi(\alpha ; \beta_{1}, \ldots , \beta_{N})$ is defined by 
\eqref{def:phase-function}. 
See \eqref{def:hypergeometric-pairing} below for the definition of the contour $C$. 
We have set $g'_{M}=q^{\nu(M)}g_M$ where $g_M$ given by
\eqref{def:function-g}. More explicitly, 
\be 
&& 
g'_{M}(\alpha_{1}, \ldots , \alpha_{l})=c_{l} 
e^{\frac{N-4l}{4\xi}\sum_{j=1}^{N}\beta_{j}
+\frac{l}{\xi}\sum_{p=1}^{l}\alpha_{p}} \\ 
&& {}\times 
\prod_{p=1}^{l}\left( 
\frac{1}{\sh{\frac{1}{\xi}(\alpha_{p}-\beta_{m_{p}})}} 
\prod_{j<m_{p}}
\frac{\sh{\frac{1}{\xi}(\alpha_{p}-\beta_{j}+\pi i)}}{\sh{\frac{1}{\xi}(\alpha_{p}-\beta_{j})}}
\right) 
\prod_{1 \le p<p' \le l}\sh{\frac{1}{\xi}(\alpha_{p}-\alpha_{p'}+\pi i)}, 
\en 
where $c_{l}:=2^{l(l-3)/2}e^{l(l-1)\pi i/2\xi}$ is a constant. 

{}From Theorem \ref{QKZ} and Corollary \ref{INV} we find 
\begin{prop} 
For any deformed cycle $P \in C_{N, l}$ 
the function $f_{P}$ satisfies the axioms {\rm (A0), (A1)} and {\rm (A2)} with $m=N-2l$. 
\end{prop} 

\subsection{Minimality condition} 
Let us consider the $N$-minimality condition: 
\be 
{\rm res}_{\beta_{N}=\beta_{N-1}+\pi i}f_{P}(\beta_{1}, \ldots , \beta_{N})=0. 
\en 
A deformed cycle $P \in C_{N, l}$ is called {\it minimal} if
\begin{equation}
P|_{z_1=-z_2=X_1^{-1}}=0.\label{MIN}
\end{equation}
We denote by $W_{N,l}$ the space of the minimal deformed cycles
with fixed $N,l$.

\begin{thm}
Let $\xi > 1$ be a generic value. For each minimal deformed cycle $P$,
the hypergeometric integral $\psi_{P}=I(v_{N,l},P)$ satisfies 
\be 
{\rm res}_{\beta_{N}=\beta_{N-1}+\pi i}\psi_{P}(\beta_{1}, \ldots , \beta_{N})=0. 
\en 
Hence the function $f_{P}$ associated with a minimal deformed cycle $P \in W_{N, l}$ 
gives an $N$-particle minimal form factor of the SG model. 
\end{thm}
\begin{proof}
It is enough to show the cancellation of the residues at
$\beta_N=\beta_{N-1}+\pi i$. 
The proof is given by repeating the argument in \cite{N}.
\end{proof}

\subsection{Null cycles}
Our aim in this section is to construct minimal deformed cycle $P$
such that
\[
I(v_{N,l},P)=0.
\]

If $P_1\in C_{N, l_{1}}$ is a null cycle, then for any $P_2\in C_{N, l_{2}}$,
the deformed cycle $P_1\wedge P_2\in C_{N, l_{1}+l_{2}}$ is a null cycle.
Following \cite{T}, we find the following minimal null cycles.

Set
\bea
&& 
\Theta(X)=\prod_{j=1}^{N}(1-z_{j}X), 
\label{def:thetaX} \\
&& 
\Theta(X_{1}, X_{2})=\Theta(X_{1})\Theta(X_{2})-\Theta(-X_{1})\Theta(-X_{2}). 
\label{def:sigmaX} 
\ena 
and 
\begin{eqnarray}
&&\Sigma_1(X)=\Theta(-X)-(-1)^{N}\Theta(X),\label{N1}\\
&&\Sigma_2(X_1,X_2)=\frac{X_1-X_2}{X_1+X_2}
\Theta(X_{1}, X_{2})+(-1)^{N}\Theta(X_{1}, -X_{2}). \label{N2} 
\end{eqnarray}
Here we use $X=e^{-\alpha}$, $z_j=e^{\beta_j}$, etc.. 
The polynomial $\Theta(X_{1}, X_{2})$ is divisible by $X_1+X_2$. The degree
$N$ terms cancel in both $\Sigma_{1}$ and $\Sigma_{2}$ .
Note also that both $\Sigma_1$ and $\Sigma_2$ are minimal cycles.

\begin{prop}\cite{BBS, T}\label{SIGMA}
The deformed cycles $\Sigma_1$ and $\Sigma_2$ are null cycles.
In fact, $I(w,\Sigma_1)=0$ for all $w(a)$ which is given by 
\eqref{WQ} $(l=1)$ with $Q(a)$ satisfying 
\eqref{DEG} and \eqref{NULL}).
We have also $I(w,\Sigma_2)=0$ if $w=w(a_1,a_2)$ is 
given by \eqref{WQ} $(l=2)$ with 
$Q(a_1,a_2)$ satisfying \eqref{DEG}, \eqref{NULL}
and \eqref{SERRE}.
\end{prop}
\begin{proof}
Set $P_1(X)=\Theta(X)$. 
Note that $\Sigma_1=\nabla_{\alpha,-\xi\pi i}(P_1)$.
By Remark \ref{WEAK}, the integral $I(w,P_1)$ 
is convergent for any $w$
as given in the statement of the lemma. 
The choice of $P_1$ is such that
the poles of the integrand is canceled between the contours
$C$ and $C-\xi\pi i$. Using \eqref{QP3}, we obtain
$I(w,\nabla_{\alpha,-\xi\pi i}(P_1))=0$.

The proof for $\Sigma_2$ is slightly more involved. Since
\[
\Theta(X_1,-X_2)=
\Sigma_1(X_1)\prod_{j=1}^N(z_jX_2+1)-
\prod_{j=1}^N(z_jX_1+1)\cdot\Sigma_1(X_2),
\]
we have $I(w,\Theta(X_{1}, -X_{2}))=0$. 
It remains to show that 
$I(w,\frac{X_1-X_2}{X_1+X_2}\Theta(X_{1}, X_{2}))=0$. 
Set
\begin{equation}
P_2(X_1,X_2)=\frac{X_1-X_2}{X_1+X_2}\Theta(X_{1})\Theta(X_{2}),
\end{equation}
and denote the integrand of $I(w,P_2)$ by   
$F(\alpha_1,\alpha_2)=
\phi(\alpha_1)\phi(\alpha_2)w(a_1,a_2)P_2(X_1,X_2)$. 
It has poles at 
$\alpha_p=\beta_j-2\pi i\Z_{\ge 0}-\xi\pi i\Z_{>0}$,
$\alpha_p=\beta_j-\pi i+2\pi i\Z_{\ge 0}+\xi\pi i\Z_{\ge0}$,
and $\alpha_1=\alpha_2+\pi i+2\pi i\Z$. 
Let $C_1$ be the union of small circles going clockwise 
around $\beta_j-\xi\pi i$ ($1\le j\le N$), 
and let 
$C_2$ be the contour going from $-\infty$ to $\infty$, 
such that $\beta_j-\xi \pi i$ is above
and $\beta_j-\pi i-\xi\pi i$ is below $C_2$. 
Since the poles at $\alpha_p=\beta_j-2\pi i\Z$ is 
cancelled by $P_2$, we can deform the contour
$C\times C$ to $(C_1+C_2)\times (C_1+C_2)$. 
If we move $C_2$ further below $\beta_j-\pi i-\xi\pi i$, 
the contour becomes $(C-\xi\pi i)\times(C-\xi\pi i)$, and we
obtain
\be
&&\int_{C_1+C_2}\!\!\int_{C_1+C_2}
F(\alpha_1,\alpha_2)d\alpha_1d\alpha_2
\\
&&
=\int_{C-\xi\pi i}\!\!\int_{C-\xi\pi i}
F(\alpha_1,\alpha_2)d\alpha_1d\alpha_2
-2\pi i(I_{12}+I_{21}),
\en
where 
$I_{jk}=\int_{C_1}d\alpha_j{\rm res}_{\alpha_k=\alpha_j-\pi i}
F(\alpha_1,\alpha_2)$.
Because of the condition \eqref{SERRE}, 
we have $I_{jk}=0$. 
Therefore, 
\be
&&\int_{C}\!\!\int_{C}
\bigl(F(\alpha_1,\alpha_2)-F(\alpha_1-\xi\pi i,\alpha_2-\xi\pi i)\bigr)
d\alpha_1d\alpha_2=0.
\en
Noting that
\[
\nabla_{\alpha_1,\alpha_2,-\xi\pi i}(P_2)=
\frac{X_1-X_2}{X_1+X_2}\Theta(X_{1}, X_{2}),
\]
we have $I(w,\frac{X_{1}-X_{2}}{X_{1}+X_{2}}\Theta(X_{1}, X_{2}))=0$. 
This completes the proof of $I(w,\Sigma_2)=0$.
\end{proof}

\subsection{The degree of minimal deformed cycles} 
Define the degree on $W_{N, l}$ by 
\bea 
\deg{X_{a}}=-1, \quad \deg{z_{j}}=1. 
\label{def:degree-onW} 
\ena 

Set 
\bea\label{MNL}
&&M_{N, l}=W_{N, l}/(\Sigma_{1}\wedge W_{N, l-1}+\Sigma_{2}\wedge W_{N, l-2}). 
\ena
We identify $M_{N, l}$ with the space of $N$-particle minimal form factors 
by the map $P \mapsto f_{P}$ \eqref{def:formfactor-P}. 
Note that $\deg{\Sigma_{1}}=0$ and $\deg{\Sigma_{2}}=0$, 
and hence $M_{N, l}$ is also graded by the degree. 

In the following sections we calculate the characters
\be 
\ch_{q}W_{N, l}=\sum_dq^d\dim_{\C}(W_{N, l})_d, \quad 
\ch_{q}M_{N, l}=\sum_dq^d\dim_{\C}(M_{N, l})_d,  
\en 
where $(W_{N, l})_{d}$ and $(M_{N, l})_{d}$ are the homogeneous 
components with the degree $d$ of $W_{N, l}$ and $M_{N, l}$, respectively.


\section{Algebraic structure of the space of deformed cycles}
\label{sec:SpaceWnl}
In this section we determine the algebraic structure of the space of 
minimal deformed cycles 
\be 
W_N=\oplus_{l=0}^{N}W_{N, l}.
\en 
This space is naturally embedded in the associative algebra
$A_N\otimes\C[z_1^{\pm1},\ldots,z_N^{\pm1}]$, where $A_N$
is the exterior algebra generated by the space $\oplus_{j=0}^{N-1}\C X^j$ 
of polynomials in $X$ of degree less than $N$. 
We denote the ring of symmetric polynomials in $N$ variables
by $R_N$:
\begin{equation}
R_N=\C[z_1,\ldots,z_N]^{{\frak S}_N}.
\end{equation}
Note that if $P_1,P_2$ are minimal deformed cycles, then $P_1\wedge P_2$
is also a minimal deformed cycle. Hence $W_N$ is an $R_N$-algebra.

First we outline the content of this section. 
We consider the quantum algebra $\Uri$, and its action $\vpi$ on
$V^{\otimes N}\otimes\C[z_1^{\pm1},\ldots,z_N^{\pm1}]$
given by the coproduct $\Delta$ \eqref{copro}
\[
\vpi:\Uri\rightarrow
\End(V^{\otimes N}\otimes\C[z_1^{\pm1},\ldots,z_N^{\pm1}]).
\]
{}Following \cite{T}, we define an embedding (see \eqref{Cc0})
\begin{equation}\label{C}
\Cc:V^{\otimes N}\otimes\C[z_1^{\pm1},\ldots,z_N^{\pm1}]
\rightarrow
A_N\otimes\C[z_1^{\pm1},\ldots,z_N^{\pm1}].
\end{equation}
In particular, we have
\be
\Cc v_+^{\otimes N}=1.
\en

Let $F^{(+)}$ be the subalgebra of $\Uri$ generated by 
$x^-_j$ $(j\in\Z_{\geq0})$, $(x^-_0)^{(2)}$ and 
the coefficients of the
divided power $\Xc(z)^{(2)}$ of the generating series
\begin{equation}
\Xc(z)=\sum_{k=1}^\infty x^-_k (q^{-1}z)^k.
\end{equation}
The algebra $F^{(+)}$ is a super-symmetric analog of the abelian subalgebra
$\C F\otimes\C[t]$ of the loop algebra $\mathfrak{sl}_2\otimes\C[t,t^{-1}]$.
The action of an element $x\in F^{(+)}$ is given by left 
multiplication by $\Cc(\vpi(x)(v_+^{N}\otimes1))$
in $A_N\otimes\C[z_1^{\pm1},\ldots,z_N^{\pm1}]$ 
and the space $W_{N}$ is invariant by this action.

Define the $R_N$-algebra $\Rc=F^{(+)}\otimes R_N$. 
There is an $R_N$-algebra homomorphism 
\bea
\rho_N:\Rc\longrightarrow W_N, 
\label{rhotil}
\ena
through which $\Rc$ acts on $W_N$ by left multiplication. 
We will define a two-sided ideal $\Ii_N$ of $\Rc$ which belongs to
the kernel of $\rho_N$ (see the end of Section \ref{subsec:4.2}). 
The main result in this section is
\begin{thm}\label{thm:WN}
The map \eqref{rhotil} is surjective, and we have the isomorphism
\begin{equation}
\Rc/\Ii_N \simeq W_N.
\end{equation}
\end{thm}
This is a super-symmetric analog of the result by Feigin-Feigin \cite{FF}.
In the process of its proof, we rederive Nakayashiki's result
on the character of $W_N$,
\begin{thm}$($\cite{N}$)$\label{thm:chWN} 
The space of the minimal deformed cycles 
$W_{N,l}$ is a free $R_N$-module. Its character is given by
\begin{equation}
\ch_q W_{N,l}=\frac{1}{(q)_{N}}{N \atopwithdelims[] l}.
\end{equation}
\end{thm}
\subsection{Action of the algebra $F^+$ on the space 
of the deformed cycles}\label{subsec:4.1}
We have the action of $\Uri\otimes R_N$ 
on $(V^{\rm aff})^{\otimes N}\simeq V^{\otimes N}
\otimes\C[z_1^{\pm1},\ldots,z_N^{\pm1}]$. 
We abuse the notation to denote it by the same letter
\begin{equation}
\vpi:\Uri\otimes R_N 
\longrightarrow \End((V^\aff)^{\otimes N}).
\end{equation}
In fact, the image $\vpi(\Uri\otimes R_N)$ is contained in
$\End(V^{\otimes N})\otimes\C[z_1^{\pm1},\ldots,z_N^{\pm1}]$
where $\C[z_1^{\pm1},\ldots,z_N^{\pm1}]$ acts as multiplication.
We rewrite the action $\vpi$ of the subalgebra $\Rc$ by using the
Jordan-Wigner transformation. The details are given in Appendix 
\ref{app:JW}.
Here we summarize the results which we use in the further discussion.

We have the Grassmann variables $\psi_1,\ldots,\psi_N$. We denote by
$\Lambda_N$ the exterior algebra generated by them over $\C$. 
It acts on $V^{\otimes N}$.
There is an inclusion $\Lambda_N\subset\End(V^{\otimes N})$ of the algebra
induced from this action, and an isomorphism $\Lambda_N\simeq V^{\otimes N}$
of vector spaces given by $\psi\mapsto\psi v_+^{\otimes N}$.
We make these identifications throughout the paper.
{}For the algebra $\Rc$ we have
\[
\vpi(\Rc)\subset\Lambda_N\otimes\C[z_1,\ldots,z_N].
\]
Explicitly, the actions of the generators are given by
\begin{prop}\label{prop:fermi}
We have
\begin{eqnarray}
&&\vpi(x^-_0)=\sum_{a=1}^N(-1)^{N-a}\psi_a,\\
&&\vpi((x^-_0)^{(2)})=-i\sum_{1\leq a<b\leq N}(-1)^{a+b}\psi_a\psi_b,\\
&&\vpi(\Xc(z))=\sum_{a=1}^Nc_a(z)\psi_a,\\
&&\vpi(\Xc(z)^{(2)})=i\sum_{1\leq a<b\leq N}
c_{a,b}(z)\psi_a\psi_b,
\end{eqnarray}
\end{prop}
where
\be 
c_a(z)=\frac{z_{a}z}{1-z_{a}z} \prod_{j=a+1}^{N}\frac{1+z_{j}z}{1-z_{j}z},  
\quad 
c_{a,b}(z)=\frac{z_{a}z}{1-z_{a}z} \prod_{j=a+1}^{b-1}\frac{1+z_{j}z}{1-z_{j}z} 
\frac{z_{b}z}{1-z_{b}z}.
\en

Let $A_{N,l}$ be the space of skew-symmetric polynomials in
the variables $X_1,\ldots,X_l$ of degree less than $N$ in each variable $X_p$. 
We identify
\[
A_N\simeq\oplus_{l=0}^NA_{N,l}.
\]
Note that the space of the deformed cycles $W_{N,l}$ is a subspace of
$A_{N,l}\otimes\C[z_1,\ldots,z_N]$.

The map (\ref{C}) is $\C[z_1^{\pm 1},\cdots,z_N^{\pm 1}]$-linear
and is given by
\bea
\Cc(\psi_{m_1}\cdots\psi_{m_l}v_+^{\otimes N})
=\Skew\Bigl(G_{m_1}(X_1)\cdots G_{m_l}(X_l)\Bigr),
\label{Cc0}
\ena
where $G_m(X)$ denotes the polynomial
\be
G_m(X)=\prod_{j=1}^{m-1}(1+z_jX)\prod_{j=m+1}^N(1-z_jX).
\en
If we write $G_m(X)=\sum_{j=0}^{N-1}G_{mj}X^j$, 
then it is easy to see that
\be
\det\bigl(G_{mj}\bigr)_{1\le m\le N,0\le j\le N-1}
=\prod_{1\le i<j\le N}(z_i+z_j).
\en
{}From this it follows that \eqref{Cc0} is an embedding.

By the mapping $\Cc$, the action of $\psi\in\Lambda_N$ on
$V^{\otimes N}\otimes\C[z_1^{\pm1},\ldots,z_N^{\pm1}]$ is intertwined with
the wedge product 
$\Cc(\psi v_+^{\otimes N})\wedge$ on
$A_N\otimes\C[z_1^{\pm1},\ldots,z_N^{\pm1}]$.

Let $\Urip$ denote the subalgebra of $\Uri$ generated by 
$t^{\pm 1}_1$ and $e_0^{(s)},e_1^{(s)},f_1^{(s)}$ $(s\ge 0)$.  The algebra
$\Rc$ is a subalgebra of $\Urip\otimes R_N$. 
In Appendix \ref{app:TrigHGS} we give a proof
of the following proposition along with its background following \cite{TV}. 
\begin{prop}\label{prop:2.2}
We have
\be
\Cc\bigl(\vpi(\Urip)(v_+^{\otimes N}\otimes1)\bigr)\subset W_N.
\en
\end{prop}
We set 
\begin{eqnarray}
\rho_N:\Rc\longrightarrow W_{N},
\quad x\mapsto\Cc\bigl(\vpi(x)(v_+^{\otimes N}\otimes1)\bigr).
\label{rtil}
\end{eqnarray}

Combining Proposition \ref{prop:fermi} and the formula (\ref{Cc0}), we obtain
\begin{prop}\label{prop:current-cycle-correspondence}  
The image of the generators of $F^{(+)}\simeq F^{(+)}\otimes1\subset\Rc$
by \eqref{rtil} is given as follows.
\bea
x_0^{-}&\mapsto&\frac12\Sigma_1(X),\\
(x_0^-)^{(2)}&\mapsto&\frac i4\Sigma_2(X_1,X_2),\\
\Xc(z)&\mapsto&\frac{1}{\Theta(z)}\frac{z}{2(X-z)}\Theta(z, -X), \\ 
(\Xc(z))^{(2)}&\mapsto&
\frac i4\Bigl\{\frac{X_1-X_2}{X_1+X_2}\frac{z}{X_1+z}\frac{z}{X_2+z}
\Theta(X_{1}, X_{2}) \\ 
&&+{\rm Skew}_{X_1,X_2}\frac{\Theta(-X_2)}{\Theta(z)}
\frac{z}{X_2+z}\frac{z}{X_1-z}\Theta(z, -X_{1})\Bigr\}.\nonumber
\ena
where $\Sigma_{1}(X),\Sigma_{2}(X_1,X_2),\Theta(X_{1}, X_{2})$ are given 
by \eqref{N1},\eqref{N2},\eqref{def:sigmaX}, respectively.
\end{prop} 

\subsection{Relations of the super-symmetric currents}
\label{subsec:4.2}
Our goal is to determine the relations satisfied by $\Rc$
when it acts on 
$V^{\otimes N}\otimes\C[z_1^{\pm1},\ldots,z_N^{\pm1}]$ through
$\vpi$. For this purpose we introduce formal symbols
which represent the generators of $F^{(+)}$.

Consider the Grassmann variables $\xi_n$ $(n\geq0)$ and
denote by $\Lambda(\xi_0,\xi_1,\xi_2,\ldots)$ the
exterior algebra generated by them over $\C$. 
We also consider the commuting variables $\eta_n$ $(n\geq0)$ and set
\begin{equation}\label{LC}
Z=\Lambda(\xi_0,\xi_1,\xi_2,\ldots)\otimes\C[\eta_0,\eta_1,\eta_2,\ldots].
\end{equation}
We define the degree and weight of the generators by
\begin{equation}\label{DEGWEI}
\deg\xi_n=n,\quad\deg\eta_n=n,\quad\wt\xi_n=1,\quad\wt\eta_n=2.
\end{equation}
We introduce the currents in $Z$:
\begin{equation}
\xi(z)=\sum_{n=1}^\infty \xi_n z^n,
\quad\eta(z)=\sum_{n=1}^\infty\eta_nz^n.
\end{equation}

Recall Proposition \ref{prop:fermi} which describes the action
of $\Rc$ on $\Lambda_N\otimes\C[z_1,\ldots,z_N]$.
Let us consider the algebra homomorphism over $R_N$
\begin{equation}
\rho'_N:Z\otimes R_N\rightarrow\Lambda_N\otimes\C[z_1,\ldots,z_N]
\end{equation}
given by
\begin{eqnarray}
&&\rho'_N(\xi_0)=\sum_{a=1}^N(-1)^{N-a}\psi_a,\\
&&\rho'_N(\eta_0)=\sum_{1\leq a<b\leq N}(-1)^{a+b}\psi_a\psi_b,\\
&&\rho'_N(\xi(z))=\sum_{a=1}^Nc_a(z)\psi_a,\\
&&\rho'_N(\eta(z))=\sum_{1\leq a<b\leq N}c_{a,b}(z)\psi_a\psi_b.
\end{eqnarray}
If we count the degree and weight in $\Lambda_N$ and $\C[z_1,\ldots,z_N]$ by
\begin{equation}
\deg\psi_j=0,\quad\wt\psi_j=1,\quad
\deg z_j=1,\quad\wt z_j=0,
\end{equation}
then the mapping $\rho'_N$ respects them.

Let us consider the kernel of $\rho'_N$.
\begin{prop}\label{prop:current-rel1}
We have the relations
\bea 
\xi(z)\xi(-z)+\eta(z)-\eta(-z)\in{\rm Ker}\,\rho'_N. 
\ena 
\end{prop}
The proof is straightforward.
These relations 
stem from an identity between the currents $\Xc(z)^{(s)}$ 
proved in \cite{CP}, Proposition 4.1.

Note, in particular, that
\[
\eta_1\in{\rm Ker}\,\rho'_N. 
\]

{}For $A=\{a_{1}, \ldots , a_{\mu}\}$ ($a_{1}< \cdots <a_{\mu}$), we set 
\be 
\psi_{A}=\psi_{a_{1}} \cdots \psi_{a_{\mu}} 
\en 
and  
\be 
c_{A}(z)=
\begin{cases}
c_{a_{1}, a_{2}}(z) \cdots c_{a_{2\nu-3}, a_{2\nu-2}}(z)c_{a_{2\nu-1}}(z)
&\hbox{if }\mu=2\nu-1; \\ 
c_{a_{1}, a_{2}}(z)\cdots c_{a_{2\nu-3}, a_{2\nu-2}}(z)c_{a_{2\nu-1}, a_{2\nu}}(z)
&\hbox{if }\mu=2\nu. 
\end{cases}
\en
The following is also straightforward.
\begin{lem}\label{lem:polynomial-rep}
We have
\[
\frac{1}{\nu!}\rho'_N(\xi(z)\eta(z)^\nu)= 
\sum_{\# A=2\nu+1}\psi_{A}c_{A}(z), \quad 
\frac{1}{\nu!}\rho'_N(\eta(z)^{\nu})=
\sum_{\# A=2\nu}\psi_{A}c_{A}(z). 
\]
In particular, we see that
\[
\prod_{j=1}^{N}(1-z_{j}z)\cdot \rho'_N\hskip-2pt\left(\xi(z)\eta(z)^\nu\right)
\quad{\it and}\quad
\prod_{j=1}^{N}(1-z_{j}z)\cdot \rho'_N\hskip-2pt\left(\eta(z)^{\nu}\right)
\]
are polynomials in $z$. 
\end{lem}

{}For a current $x(z)=\sum_{n=0}^{\infty} x_{n}z^{n}$, we set 
\be 
[x(z)]_{\ge \mu}=\sum_{n=\mu}^{\infty}x_nz^n. 
\en 
Introduce the currents $I_\mu^{(N)}(z)$ ($\mu=1,2,\ldots$) given by 
\be 
I_{2\nu+1}^{(N)}(z)&=& 
\prod_{j=1}^N(1-z_jz)\cdot(\xi_0+\xi(z))(\eta_0+\eta(z))^\nu,\\
I_{2\nu}^{(N)}(z)&=&\prod_{j=1}^N(1-z_jz)\cdot
\left\{(\eta_0+\eta(z))^\nu+\nu\xi_0\xi(z)(\eta_0+\eta(z))^{\nu-1}\right\}. 
\en 
{}From Lemma \ref{lem:polynomial-rep}, these are mapped by $\rho'_N$
to polynomials in $z$. In fact, we have the following. 
\begin{prop}\label{prop:current-rel2}
The currents $I_\mu^{(N)}(z)$ $(\mu=1,2,\ldots)$ satisfy that 
\be 
[I_\mu^{(N)}(z)]_{\ge N-\mu+1}\in{\rm Ker}\,\rho'_N \otimes \mathbb{C}[z]. 
\en  
\end{prop}
\begin{proof} 
We prove the assertion by induction on $N$. Set
\be 
&&a_0^{(N)}(z)=\prod_{j=1}^N(1-z_jz),\\ 
&&a_{2\nu+1}^{(N)}(z)
=\prod_{j=1}^N(1-z_jz)\cdot
\frac1{\nu!}\rho'_N(\xi(z)\eta(z)^\nu),\\
&&a_{2\nu}^{(N)}(z)=\prod_{j=1}^N(1-z_jz)\cdot
\frac1{\nu!}\rho'_N(\eta(z)^\nu).
\en 
Consider the case $N=1$. We have 
\be 
&& 
a_{0}^{(1)}(z)=1-z_{1}z, \quad a_{1}^{(1)}(z)=z_{1}z \psi_{1}, \quad 
a_{\mu}^{(1)}(z)=0\quad (\mu>1), \\ 
&& 
\rho_{1}'(\xi_0)=\psi_1,\quad\rho_{1}'(\eta_0)=0. 
\en 
{}From these formulas 
it is easy to check that $\rho_{1}'(I_1^{(1)}(z))=\psi_1$ and 
$\rho_{1}'(I_\mu^{(1)}(z))=0$ ($\mu\geq2$).  

Next consider the case of $N>1$. We make a natural identification of
$\Lambda_{N-1}\otimes\C[z_1,\ldots,z_{N-1}]$ with the subspace of
$\Lambda_N\otimes\C[z_1,\ldots,z_N]$. From Lemma \ref{lem:polynomial-rep}
we see that 
\be
&&a_{2\nu+1}^{(N)}(z)=(1+z_Nz)a_{2\nu+1}^{(N-1)}(z)
+a_{2\nu}^{(N-1)}(z)z_Nz\psi_N, \\ 
&&a_{2\nu}^{(N)}(z)=
(1-z_{N}z)a_{2\nu}^{(N-1)}(z)+a_{2\nu-1}^{(N-1)}(z) z_{N}z\psi_{N}, \\ 
&&\rho'_N(\xi_0)=-\rho_{N-1}'(\xi_0)+\psi_N, \quad 
\rho'_N(\eta_0)=\rho_{N-1}'(\eta_0)-\rho_{N-1}'(\xi_0)\psi_N. 
\en  
{}From these recursions we find
\be 
&&\rho_{\scriptscriptstyle N}'(I_{2\nu+1}^{(\scriptscriptstyle N)}(z))
=(1+z_{\scriptscriptstyle N}z)
\rho_{\scriptscriptstyle N-1}'(I_{2\nu+1}^{(\scriptscriptstyle N-1)}(z))+
(\psi_{\scriptscriptstyle N}-2\pi_{\scriptscriptstyle N-1}(\xi_0))
\rho_{\scriptscriptstyle N-1}'(I_{2\nu}^{(\scriptscriptstyle N-1)}(z)), \\ 
&&\rho_{\scriptscriptstyle N}'(I_{2\nu}^{(\scriptscriptstyle N)}(z))
=(1-z_{\scriptscriptstyle N}z)
\rho_{\scriptscriptstyle N-1}'(I_{2\nu}^{(\scriptscriptstyle N-1)}(z))+
\nu(\psi_{\scriptscriptstyle N}-2\pi_{\scriptscriptstyle N-1}(\xi_0))
\rho_{\scriptscriptstyle N-1}'(I_{2\nu-1}^{(\scriptscriptstyle N-1)}(z)). 
\en 
{}From these relations and the assumption of induction that 
$\rho_{N-1}'([I_{\mu}^{(N-1)}(z)]_{\geq N-\mu})=0$, 
we get $\rho'_N([I_{\mu}^{(N)}(z)]_{\geq N-\mu+1})=0$. 
\end{proof}

\begin{definition}\label{def:JN}  
Let $\Ic_N\subset Z\otimes R_N$ be the two-sided ideal generated by
the coefficients of
\bea
&&\xi(z)\xi(-z)+\eta(z)-\eta(-z)
\ena
and
\bea\label{BZ}
[I_{\mu}^{(N)}(z)]_{\ge N-\mu+1}\quad(\mu=1,2,\ldots). 
\ena
We define the $R_N$-algebra $\Zc_N$ by  
\be 
\Zc_N=(Z\otimes R_N)/\Ic_N. 
\en 
\end{definition}

Let $\Ii_N$ be the two-sided ideal of $\Rc$ generated by the
coefficients of (\ref{BZ}) where $\xi_0,\xi(z),\eta_0,\eta(z)$
are replaced by $x^-_0,-i\Xc(z),(x^-_0)^{(2)},-i\Xc(z)^{(2)}$, respectively.
By the definition and the remark after Proposition 3.6, 
we have surjections
\begin{eqnarray}
\Zc_N\longrightarrow \Rc/\Ii_N\longrightarrow 
\vpi(\Rc).
\label{ZRc}
\end{eqnarray}
In Section \ref{subsec:4.4}, we will
prove that these are isomorphisms
(Corollary \ref{cor:iso-rho-N} and \eqref{rho}). 

\subsection{Isomorphism of $R_N$ algebras}
\label{subsec:FeiginFeigin} 
The algebra $Z\otimes R_N$ is bi-graded by degree and weight.
The ideal $\Ic_N$ is also bi-graded, and, therefore, 
the algebra $\Zc_N$ is bi-graded. 
Notice that the variables $z_j$ do not enter 
the definition of the algebra $Z$.
We denote the degree $s$ component of $Z$ by $Z_s$, and set
$Z_{\leq s}=\oplus_{t=0}^s Z_t$.
Then we define the $R_N$-submodule $F_s(\Zc_N)$ of $\Zc_N$ by 
\bea 
F_s(\Zc_N)=(Z_{\leq s}\otimes R_N)/((Z_{\leq s}\otimes R_N)\cap\Ic_N). 
\label{def:filtration} 
\ena 
These submodules satisfy 
\be 
\Zc_N=\cup_{s=0}^{\infty}F_s(\Zc_N),\quad 
0=F_{-1}(\Zc_N)
\subset F_0(\Zc_N)\subset F_1(\Zc_N)\subset F_2(\Zc_N)\subset\cdots. 
\en 
Hence $\Zc_N$ is a filtered $R_N$-module. 
We consider the associated graded module 
\be 
\gr\Zc_N=\oplus_{s=0}^\infty F_s(\Zc_N)/F_{s-1}(\Zc_N).
\en 
\begin{dfn}\label{def:J_N}
We denote by $J_N$ the ideal of $Z$ generated by the coefficients of
\[
\xi(z)\xi(-z)+\eta(z)-\eta(-z)\quad\hbox{and}\quad
[J_{\mu}(z)]_{\ge N-\mu+1}\quad(\mu=1,2,\ldots),
\]
where 
\be 
&& 
J_{2\nu-1}(z)=(\xi_{0}+\xi(z))(\eta_{0}+\eta(z))^{\nu-1}, \\ 
&& 
J_{2\nu}(z)=
(\eta_{0}+\eta(z))^{\nu}+\nu\xi_{0}\xi(z)(\eta_{0}+\eta(z))^{\nu-1}. 
\en 
\end{dfn}
{}For $s\geq N-\mu+1$ the generator $[I^{(N)}_{\mu}(z)]_s$ of the ideal $\Ic_N$ 
belongs to $F_s(\Zc_N)$ and satisfies 
$[I^{(N)}_{\mu}(z)]_s=[J_{\mu}(z)]_s$ in $F_{s}(\Zc_N)/F_{s-1}(\Zc_N)$.

We set $\bar Z_N=Z/J_N$. This is a bi-graded $\C$-algebra. 
We denote by $(\bar Z_N)_{s,l}$ the component of degree $s$ and weight $l$.

\begin{prop}
There exists a surjective $R_N$-module homomorphism 
\begin{equation} \label{SURJ}
\bar Z_N\otimes R_N\longrightarrow
\gr\Zc_N\longrightarrow0. 
\end{equation}
\end{prop} 

\begin{proof} 
We have the exact sequence of $R_N$-modules: 
\be 
0 \longrightarrow J_N\otimes R_N
\longrightarrow Z\otimes R_N
\longrightarrow(Z/J_N)\otimes R_N
\longrightarrow 0. 
\en 

Note that the associated graded module ${\rm gr}\,\Zc_N$ 
is canonically isomorphic to the quotient 
\be 
\gr\Zc_N\simeq(Z\otimes R_N)/\Ic_N{}^{{\rm top}}, 
\en 
where 
\be 
\Ic_N{}^{{\rm top}}=
{\rm span}_{R_N}\{b_s|{}^{\exists}b=b_0+b_1+\cdots+b_s\in\Ic_N\hbox{ where }
b_t\in Z_t\otimes R_N\}. 
\en 

We have $J_N\otimes R_N\subset\Ic_N{}^{{\rm top}}$. 
Hence we get 
\be 
(Z/J_N) \otimes R_N\simeq(Z\otimes R_N)/(J_N\otimes R_N)
\twoheadrightarrow(Z\otimes R_N)/\Ic_N{}^{{\rm top}}\simeq
\gr\Zc_N. 
\en 
\end{proof}
We will prove that the above mapping is an isomorphism 
(see Proposition \ref{prop:isom}).
In other words, we have $J_N\otimes R_N=\Ic_N{}^{{\rm top}}$. 

Let us calculate the character of $\bar{Z}_N$, 
\bea 
\chi_N(q,z)
=
\sum_{s,l\ge0}q^sz^l\dim_\C(\bar Z_N)_{s,l}.
\label{chiN}
\ena
Our strategy is to follow
the idea in \cite{FF}. 
{}First we find the upper bound of the character. 
To this end we prove the following two lemmas. 

\begin{lem}\label{FeiginFeigin1}
Let $\iota: Z \longrightarrow Z$ be the $\C$-algebra homomorphism defined by 
\bea
&& 
\iota(\xi_{0}+\xi(z))=\frac{\xi(z)}{z},\label{IOTA1} \\ 
&& 
\iota(\eta_{0}+\eta(z))= 
\frac{1}{z^2}(\eta(-z)+\eta_{1}z+\xi(z)\xi_{1}z).\label{IOTA2}
\ena
This map induces the $\C$-algebra homomorphism 
$\iota_N:\bar{Z}_{N-1}\longrightarrow\bar{Z}_N$. 
\end{lem} 
\begin{proof} 
It is enough to prove that 
\bea 
&& 
\iota(\xi(z)\xi(-z)+\eta(z)-\eta(-z))=0 \quad {\rm and} 
\label{pf:FeiginFeigin-to-prove0} \\ 
&& 
\iota ([J_{\mu}(z)]_{\ge N-\mu})=0 \quad {\rm in} \,\, \bar{Z}_{N}. 
\label{pf:FeiginFeigin-to-prove1} 
\ena 
We can check (\ref{pf:FeiginFeigin-to-prove0}) easily. 
Let us prove (\ref{pf:FeiginFeigin-to-prove1}). 
The image of the current $J_{\mu}(z)$ is given as follows: 
\be
\iota(J_{2\nu-1}(z))&=&z^{-(2\nu-1)}\xi(z)(\eta(-z)+\eta_{1}z)^{\nu-1}, \\ 
\iota(J_{2\nu}(z))&=&z^{-2\nu}(\eta(-z)+\eta_{1}z)^{\nu}. 
\en
{}From the relation $-\eta(z)+\eta(-z)=\xi(z)\xi(-z)$ we have 
\be 
\eta_1=0\quad\hbox{and}\quad\xi(z)\eta(-z)^\nu
=\xi(z)\eta(z)^\nu\hbox{ in }\bar{Z}_N.
\en  
Hence (\ref{pf:FeiginFeigin-to-prove1}) is equivalent to 
\be 
[\xi(z)\eta(z)^{\nu-1}]_{\ge N}=0 \quad {\rm and} \quad 
[\eta(z)^{\nu}]_{\ge N}=0. 
\en 

We have $[J_{\mu}(z)]_{\ge N-\mu+1}=0$ in $\bar{Z}_{N}$.  
Note that 
\be
J_{2\nu}(z)-\nu\xi_{0}J_{2\nu-1}(z)=(\eta_{0}+\eta(z))^{\nu}. 
\en 
Hence we find 
\bea 
[(\eta_{0}+\eta(z))^{\nu}]_{\ge N-2\nu+2}=0\hbox{ in }\bar Z_N.
\label{pf:FeiginFeigin-rel1} 
\ena 
{}From this we can prove that $[\eta(z)^{\nu}]_{\ge N}=0$ 
by induction on $\nu$. 
This result and 
$[J_{2\nu-1}(z)]_{\ge N-2\nu+2}=0$ in $\bar{Z}_{N}$ imply that 
$[\xi(z)\eta(z)^{\nu-1}]_{\ge N}=0$. 
\end{proof} 
We denote by $Z'$ the subalgebra of $Z$ generated by $\xi_n,\eta_n$ $(n\geq1)$.
\begin{lem}\label{IMAGE}
The image of the subalgebra $Z'$ in $\bar{Z}_{N}$ 
belongs to $\iota_N(\bar Z_{N-1})$. 
\end{lem}
\begin{proof}
Since $\iota_N$ is an algebra homomorphism, it is enough to prove that
$\eta(z)$ and $\xi(z)$ belong to $\iota_N(\bar Z_{N-1})$.
The latter is obvious from the definition (\ref{IOTA1}).
Therefore, $\xi(z)\xi_1$ belongs to $\iota_N(\bar Z_{N-1})$.
Note that $\eta_1=0$ in $\bar Z_N$. Therefore, from (\ref{IOTA2})
we have $\eta(z)\in\iota_N(\bar Z_{N-1})$.
\end{proof}

\begin{lem}\label{FeiginFeigin2}
Let $\varphi: Z \longrightarrow Z$ be 
the $Z'$-linear map defined by 
\be 
\varphi(\eta_0^k)=\xi_0\eta_0^k,\quad 
\varphi(\xi_0\eta_0^k)=\frac{\eta_0^{k+1}}{k+1}. 
\en 
This map induces a surjection
$\varphi_N:\bar{Z}_{N-1}\longrightarrow\bar{Z}_N/\iota_N(\bar{Z}_{N-1})$. 
\end{lem} 
\begin{proof} 
It is enough to prove that 
\bea 
&& 
\varphi(\xi_{0}^{\delta}\eta_{0}^{k}(\xi(z)\xi(-z)+\eta(z)-\eta(-z)))=0 
\quad {\rm and} 
\label{pf:FeiginFeigin-to-prove3} \\ 
&& 
\varphi(\xi_{0}^{\delta}\eta_{0}^{k}[J_{\mu}(z)]_{\ge N-\mu})=0 \quad 
{\rm in} \,\,  \bar{Z}_{N}/\iota_N(\bar{Z}_{N-1})
\label{pf:FeiginFeigin-to-prove4} 
\ena 
for $\delta=0, 1$ and $k \ge 0$. 

Since $\xi(z)\xi(-z)+\eta(z)-\eta(-z)\in Z'$, it is clear that
(\ref{pf:FeiginFeigin-to-prove3}) holds. Here we prove
(\ref{pf:FeiginFeigin-to-prove4}) for odd $\mu$. The proof for even $\mu$
is similar. 

{}First consider the case $\delta=0$.
Expanding the factor $(\eta_{0}+\eta(z))^{\nu-1}$ in $J_{2\nu-1}(z)$ 
we have 
\be 
\eta_{0}^{k}J_{2\nu-1}(z)=\sum_{s=0}^{\nu-1}\binom{\nu-1}{s}\left( 
\xi_0\eta_0^{k+s}\eta(z)^{\nu-s-1}+\eta_0^{k+s}\xi(z)\eta(z)^{\nu-s-1}\right). 
\en
Therefore, we have
\be 
\varphi(\eta_{0}^{k}J_{2\nu-1}(z))&=&\sum_{s=0}^{\nu-1}\binom{\nu-1}{s}\left( 
\frac{\eta_{0}^{k+s+1}}{k+s+1}\eta(z)^{\nu-s-1}+
\xi_{0}\eta_{0}^{k+s}\xi(z)\eta(z)^{\nu-s-1} \right) \\ 
&=&\sum_{s=0}^{\nu-1}\binom{\nu-1}{s} 
\frac{\eta_{0}^{k+s+1}}{k+s+1}\eta(z)^{\nu-s-1}+
\xi_{0}\eta_{0}^{k}\xi(z)(\eta_{0}+\eta(z))^{\nu-1}. 
\en
Now we apply the following identity to the first term.
\be 
\sum_{a=0}^{n}\binom{n}{a}\frac{x^{a+b}}{a+b}y^{n-a}&=&  
\sum_{j=0}^{b-1}(-1)^j
\frac{(b-1) \cdots (b-j)}{(n+1) \cdots (n+j+1)}x^{b-j-1}(x+y)^{n+j+1} \nn \\ 
&& {}+
(-1)^{b}\frac{(b-1)!}{(n+1) \cdots (n+b)}y^{n+b},  
\en
where $x$ and $y$ are commuting variables and $b=1, 2, \ldots $. 
Then we get 
\be 
&&\varphi(\eta_{0}^{k}J_{2\nu-1}(z)) \nn \\ 
&&=\frac{1}{\nu}\eta_0^kJ_{2\nu}(z)+\sum_{j=1}^{k}(-1)^j
\frac{k(k-1)\cdots (k-j+1)}{\nu(\nu+1) \cdots (\nu+j)}
\eta_{0}^{k-j}(\eta_{0}+\eta(z))^{\nu+j} \nn \\ 
&& {}+
(-1)^{k+1}\frac{k!}{\nu(\nu+1) \cdots (\nu+k)}\eta(z)^{\nu+k}. 
\en
We have (\ref{pf:FeiginFeigin-rel1}) and also
$[J_{2\nu}(z)]_{\ge N-2\nu+1}=0$ in $\bar{Z}_{N}$.
Moreover the coefficients of $\eta(z)^{\nu+k}$ are in $\iota(\bar{Z}_{N-1})$
by Lemma \ref{IMAGE}. Therefore we find 
(\ref{pf:FeiginFeigin-to-prove4}) with $\delta=0$ and odd $\mu$. 

The proof for $\delta=1$ is similar. Using $(\xi_0+\xi(z))J_{2\nu-1}(z)=0$,
we have
\be 
&&-\varphi(\xi_{0}\eta_{0}^{k}J_{2\nu-1}(z)) \\ 
&& {}=\sum_{j=0}^{k}(-1)^j
\frac{k(k-1)\cdots (k-j+1)}{\nu(\nu+1) \cdots (\nu+j)}
\eta_{0}^{k-j}\xi(z)(\eta_{0}+\eta(z))^{\nu+j} \nn \\ 
&& {}+
(-1)^{k+1}\frac{k!}{\nu(\nu+1) \cdots (\nu+k)}\xi(z)\eta(z)^{\nu+k}. 
\en
Note that 
\be 
\xi(z)(\eta_{0}+\eta(z))^{\nu+j}=J_{2(\nu+j)+1}(z)-\xi_{0}J_{2(\nu+j)}(z).
\en 
{}From this equality and $[J_{\mu}(z)]_{\ge N-\mu+1}=0$ in $\bar{Z}_{N}$, 
we see that 
\be 
[\xi(z)(\eta_{0}+\eta(z))^{\nu+j}]_{\ge N-2\nu+1}=0. 
\en 
The coefficients of $\xi(z)\eta(z)^{\nu+k}$ are in $\iota(\bar{Z}_{N-1})$
by Lemma \ref{IMAGE}. Thus we get (\ref{pf:FeiginFeigin-to-prove4}) with 
$\delta=1$ and odd $\mu$. 
\end{proof} 

{}From Lemma \ref{FeiginFeigin1} and Lemma \ref{FeiginFeigin2} 
we get the following diagram: 
\be 
\begin{CD} 
& & \bar{Z}_{N-1} & & & & \bar{Z}_{N-1} & &\\ 
& & @V{\iota_N}VV & & @V{\varphi_{N}}VV & &\\ 
0 @>>> \iota_N(\bar{Z}_{N-1}) @>>> \bar{Z}_{N} @>>> 
\bar{Z}_{N}/\iota_N(\bar{Z}_{N-1}) @>>> 0 \\ 
& & @VVV & & @VVV & &\\ 
& & 0 & & & & 0 & &. 
\end{CD} 
\en 
Here and after, for formal series 
\be 
f=\sum_{\alpha_{1}, \ldots , \alpha_{\nu}}f_{\alpha_{1}, \ldots , \alpha_{\nu}} 
z_{1}^{\alpha_{1}} \cdots z_{\nu}^{\alpha_{\nu}}, \quad 
g=\sum_{\alpha_{1}, \ldots , \alpha_{\nu}}g_{\alpha_{1}, \ldots , \alpha_{\nu}} 
z_{1}^{\alpha_{1}} \cdots z_{\nu}^{\alpha_{\nu}} 
\en 
with integer coefficients, 
we write $f \le g$ if 
$f_{\alpha_{1}, \ldots , \alpha_{\nu}} \le g_{\alpha_{1}, \ldots , \alpha_{\nu}}$ 
holds for all $\alpha_{1}, \ldots , \alpha_{\nu}$. 
{}For a homogeneous element $b \in Z$ we have 
\be 
&& 
\deg{\iota(b)}=\deg{b}+\wt{b}, \quad 
\wt{\iota(b)}=\wt{b}, \\ 
&& 
\deg{\varphi(b)}=\deg{b}, \quad \wt{\varphi(b)}=\wt{b}+1. 
\en 
Hence we find 
\bea 
\chi_{\scriptscriptstyle N}(q,z)\le
\chi_{\scriptscriptstyle N-1}(q,qz)+z\chi_{\scriptscriptstyle N-1}(q,z).
\label{FeiginFeigin:ineq} 
\ena 

It is easy to see that $\chi_{1}(q, z)=1+z$. 
Starting from this and using (\ref{FeiginFeigin:ineq}) repeatedly, 
we get the following upper bound for the character $\chi_\sN$.
\begin{prop}\label{FeiginFeigin:upper}
We have
\bea 
\chi_\sN(q,z)\le\sum_{l=0}^N\left[N\atop l\right]z^l.
\label{FeiginFeigin:upper-ineq}
\ena 
\end{prop} 

Let us prove the equality in (\ref{FeiginFeigin:upper-ineq}).
It is enough to prove that 
\bea 
\chi_\sN(1, 1)=\dim_{\mathbb{C}}{\bar{Z}_{N}} \ge 
\sum_{l=0}^{N} {N \atopwithdelims() l}=2^{N}. 
\label{FeiginFeigin:lower-ineq} 
\ena 

{}For $c=(c_{1}, \ldots , c_{N}) \in \mathbb{C}^{N}$, 
introduce the evaluation map 
\bea 
e_{c}:  \mathbb{C}[z_{1}, \ldots , z_{N}] 
\longrightarrow \mathbb{C}, \quad 
P(z_{1}, \ldots , z_{N}) \mapsto P(c_{1}, \ldots , c_{N}). 
\label{ec}
\ena 
This map induces 
\be
&&\bar{Z}_N\otimes R_N\longrightarrow\bar{Z}_N,\\
&&\Lambda_N\otimes\C[z_1,\ldots,z_N]\longrightarrow\Lambda_N,\\
&&A_N\otimes R_N\longrightarrow A_N.
\en 
We denote  these induced maps by the same letter $e_{c}$. 

The space $\vpi(\Rc)$ contains the coefficients of 
\be
\prod_{j=1}^{N}(1-z_{j}z)\cdot\vpi(\Xc(z))= 
\sum_{a=1}^{N}\psi_{a}\prod_{j=1}^{a-1}(1-z_{j}z) 
\cdot z_{a}z \cdot \prod_{j=a+1}^{N}(1+z_{j}z). 
\en 
Hence if $c=(c_{1}, \ldots , c_{N})$ satisfies 
\bea 
\prod_{j=1}^Nc_j\prod_{1 \le j<j' \le N}(c_{j}+c_{j'}) \not= 0, 
\label{cond:generic} 
\ena
we have 
\bea 
e_{c}(\vpi(\Rc))=\Lambda_N. 
\label{image:e_{c}} 
\ena 

We have the surjections (\ref{SURJ}) and 
\be 
\mathcal{Z}_N \twoheadrightarrow\vpi(\Rc) \subset 
\Lambda_N\otimes\C[z_1,\ldots,z_N]. 
\en 
Evaluating these maps at $c$ satisfying (\ref{cond:generic}), we get 
the exact sequences
\be 
\begin{CD}
\bar{Z}_N\otimes R_N@>>>\gr\Zc_N@>>>0\\ 
@V{e_c}VV@V{e_c}VV&&\\ 
\bar{Z}_N@>>>e_c(\gr\Zc_N)@>>>0\\
\end{CD} 
\en
and
\be  
\begin{CD} 
\Zc_N@>>>\vpi(\Rc)@>>>0\\ 
@V{e_c}VV @V{e_c}VV \\ 
e_c(\Zc_N) @>>>\Lambda_N@>>>0,\\
\end{CD}
\en
where the vertical arrows are surjective. 

The space $e_c(\Zc_N)$ is given by $e_c(\Zc_N)=Z/e_c(\Ic_N)$. 
Introduce the filtration $\{F_{s}(Z/e_c(\Ic_N))\}$ on $Z/e_c(\Ic_N)$ 
in the same way as (\ref{def:filtration}), that is 
$F_{s}(Z/e_c(\Ic_N))=Z_{\le s}/(Z_{\le s} \cap e_{c}(\Ic_N))$. 
Then the associated graded space 
$\gr(Z/e_c(\Ic_N))=\gr(e_c(\Zc_N))$ 
is isomorphic to $e_{c}(\gr\Zc_N)$. 
Hence we get 
\be 
\dim_\C\bar{Z}_{N}\ge\dim_\C\gr e_c(\Zc_N)= 
\dim_\C e_c(\Zc_N)\ge\dim_\C\Lambda_N=2^N. 
\en 
We have proved (\ref{FeiginFeigin:lower-ineq}). Thus we obtain
\begin{thm}\label{thm:pre-supernomial} 
The character $\chi_\sN(q,z)$ of the bi-graded algebra $\bar Z_N$ is 
given by 
\be 
\chi_\sN(q,z)=\sum_{l=0}^N\left[N\atop l\right]z^{l}. 
\en 
\end{thm} 
\begin{cor}\label{cor:grZN}
We have the following isomorphism as $\C$-algebras: 
\be 
\bar Z_N\simeq\gr(e_c(\Zc_N))\quad{\rm and}\quad e_c(\Zc_N)\simeq\Lambda_N,
\en 
at any point $c$ satisfying \eqref{cond:generic}. 
\end{cor}
\begin{prop}\label{prop:isom} 
We have the isomorphism of $R_N$-modules:
\bea 
\bar Z_N\otimes_\C R_N\simeq\gr\Zc_N.
\label{isom:gr}
\ena 
\end{prop}
\begin{proof} 
We have the following exact sequences.
\bea 
\begin{CD} 
&&\bar Z_N\otimes_\C R_{N}@>{\theta}>>\gr\Zc_N@>>>0\\
&&@V{e_c}VV @V{e_c}VV\\
0@>>>\bar Z_N@>{\theta'}>>\gr(e_c(\Zc_N))@>>>0.
\end{CD}\label{CD:Z/I} 
\ena 
Take a $\C$-basis $\{Q_i\}_{i=1,\ldots,2^N}$ of $\bar Z_N$. 
Then the set $\{\theta(Q_i\otimes1)\}_i$ generates $\gr\Zc_N$ 
over $R_N$. 
Let us prove that it is linearly independent over $R_N$. 

Suppose that $\sum_ir_i\theta(Q_i\otimes1)=0$ for some $r_i\in R_N$. 
Evaluating this equality at $c$ we get 
\be 
&&e_c(\sum_ir_i\theta(Q_i\otimes1))
=\sum_ie_c(r_i)(e_c\circ\theta)(Q_i\otimes1)\\ 
&&=\sum_ie_c(r_i) (\theta'\circ e_c)(Q_i\otimes1)=\sum_ie_c(r_i)\theta'(Q_i)=0.
\en 
Since $\theta'$ is an isomorphism and 
$\{Q_i\}$ is a basis, we have $e_c(r_i)=0$ for all $i$. 
Thus $r_{i}$ satisfies $e_{c}(r_{i})=0$ for any generic point $c$, 
and this implies $r_{i}=0$. 
\end{proof}  

\subsection{Proof of Theorems \ref{thm:WN}, \ref{thm:chWN}}
\label{subsec:4.4}
We conclude this section by proving 
Theorems \ref{thm:WN} and \ref{thm:chWN}. 
Recall that we fix $N,l$ satisfying $0\le l\le N$. We set
\be
\Delta_+=\prod_{1\le i<j\le N}(z_i+z_j).
\en
The following two Lemmas are proved in \cite{N}.
\begin{lem}\label{lem:N1}
Let $P_a$ $(1\le a\le\binom{N}{l})$ be arbitrary elements
of $W_{N,l}$, and set
\bea
P_a=\sum_{J}P_{aJ}\,
X_1^{j_1}\wedge\cdots\wedge X_l^{j_l},
\label{Pa}
\ena
where $J=(j_1,\cdots,j_l)$
runs over the set of indices satisfying 
$0\le j_1<\cdots<j_l\le N-1$. 
Then $\det\bigl(P_{aJ}\bigr)$ is
divisible by $\Delta_+^{\binom{N-1}{l-1}+\binom{N-2}{l-1}}$.
\end{lem}
\begin{proof}
If $N=1$, there is nothing to prove. 
Suppose $N\ge 2$, and 
consider the matrix $M=\bigl(P_{aJ}\bigr)$. 
It is sufficient to show that 
$\det M$ is divisible by $(z_1+z_2)^m$ with 
$m=\binom{N-1}{l-1}+\binom{N-2}{l-1}$. 
Let $L$ be the $N$-dimensional space of column vectors
over the field $\C(z_2,\cdots,z_N)$. 
Set
${\bf a}={}^t(1,z_2^{-1},\cdots,z_2^{-N+1})$, 
${\bf b}={}^t(1,(-z_2)^{-1},\cdots,(-z_2)^{-N+1})$. 
By the minimality condition \eqref{MIN}, the subspace 
$L'={\bf a}\wedge\bigl(\wedge^{l-1}L\bigr)
+{\bf b}\wedge\bigl(\wedge^{l-1}L\bigr)$
of $\wedge^l L$ 
belongs to the kernel of $M\bigl|_{z_1=-z_2}$.
Since ${\bf a}\wedge{\bf b}\neq 0$, 
we have $\dim L'=m$.
The assertion follows from the fact that, 
if $X(z)$ is a square matrix whose entries are 
polynomials in $z$ and if ${\rm corank }\,X(0)=k$, 
then $\det X(z)$ is divisible by $z^k$.
\end{proof}

\begin{lem}\label{lem:N2}
Let $\{Q_a\}_{1\le a\le\binom{N}{l}}$ be 
a set of homogeneous elements of $W_{N,l}$ with the 
following properties:
\bea
&&\sum_{a=1}^{\binom{N}{l}}q^{d_a}=\qbin{N}{l}
\qquad (d_a=\deg Q_a),
\label{da}\\
&&\mbox{$\{e_c(Q_a)\}$ is linearly independent for some 
$c\in\C^N$ satisfying \eqref{cond:generic}.}
\label{db}
\ena
In the second line, $e_c$ denotes the evaluation map 
\eqref{ec}.
Then $\{Q_a\}_{1\le a\le\binom{N}{l}}$ is a 
free $R_N$-basis of $W_{N,l}$. 
\end{lem}
\begin{proof}
As in \eqref{Pa}, denote by $Q_{aJ}$ 
the transition coefficients between the $\{Q_a\}$ 
and the monomials 
$X^J=X_1^{j_1}\wedge\cdots\wedge X_l^{j_l}$, 
$J=(j_1,\cdots,j_l)$. 
We have $\deg Q_{aJ}=d_a+|J|$
where $|J|=j_1+\cdots+j_l$. 
Using \eqref{da} one computes easily
\be
&&\sum_{a=1}^{\binom{N}{l}}d_a=\binom{N}{l}\frac{l(N-l)}{2},
\\
&&\sum_{J}|J|=\binom{N}{l}\bigl(\binom{l}{2}+\frac{l(N-l)}{2}\bigr),
\en
and hence
\be
\deg\det\bigl(Q_{aJ}\bigr)
&=&\sum_{a=1}^{\binom{N}{l}}d_a+\sum_J|J|
\\
&=&\binom{N}{2}\bigl(\binom{N-1}{l-1}+\binom{N-2}{l-1}\bigr).
\en
By Lemma \ref{lem:N1}, it follows that
$\det\bigl(Q_{aJ}\bigr)=C\cdot
\Delta_+^{\binom{N-1}{l-1}+\binom{N-2}{l-1}}$ 
for some $C\in\C$. 
Evaluating both sides at $z=c$ and using \eqref{db} we find
that $C\neq 0$. 
In particular, $\{Q_a\}$ is a linearly independent set
over the field $\C(z_1,\cdots,z_N)^{\mathfrak{S}_N}$.  
Let us show that it is a basis of $W_{N,l}$ over $R_N$.   
Take $P=\sum_J P_J X^J \in W_{N,l}$. 
By the linear independence of $\{Q_a\}$, there exist unique
elements $p_a\in\C(z_1,\cdots,z_N)^{\mathfrak{S}_N}$ such that 
$P=\sum_{a=1}^{\binom{N}{l}}p_a\,Q_a$ holds. 
Use Cramer's rule to solve
the equation $P_J=\sum_{a=1}^{\binom{N}{l}}p_a\,Q_{aJ}$ 
for $p_a$. By Lemma \ref{lem:N1}, 
we see that $p_a$ are polynomials. 
This proves the Lemma. 
\end{proof}
\medskip

\noindent{\it Proof of
Theorems \ref{thm:WN} and \ref{thm:chWN}.}\quad 

According to Theorem \ref{thm:pre-supernomial}, there exist
homogeneous elements $\zeta_a$ ($1\le a\le \binom{N}{l}$) 
of $Z$, satisfying $\wt\zeta_a=l$, 
$\sum_a q^{\deg \zeta_a}=\qbin{N}{l}$, 
and 
whose equivalence classes give a basis of $\bar{Z}_N$.  
By Corollary \ref{cor:grZN}, we may assume that 
$\{e_c(\zeta_a)\}$ is a linearly independent set for 
some $c\in \C^N$. 
Proposition \ref{prop:isom} shows that 
$\{\zeta_a\otimes 1\}$ span $\Zc_N$ over $R_N$. 
Let $\bar{\rho}_{N}$ denote the composition map 
\begin{eqnarray}
\Zc_N\longrightarrow \Rc/\Ii_N\longrightarrow
\vpi(\Rc)\longrightarrow W_N.
\label{pi}
\end{eqnarray}
Let $Q_a=\pi(\zeta_a\otimes 1)$. 
Then \eqref{da} is satisfied. 
{}From the remark below \eqref{Cc0}, 
the map 
\be
\psi_{m_1}\cdots\psi_{m_l}\mapsto 
e_c\Bigl(\Skew\bigl(G_{m_1}(X_1)\cdots G_{m_l}(X_l)\bigr)
\Bigr)
\en
is injective, so that 
$\{e_c(Q_a)\}$ is linearly independent over $\C$. 
Therefore Lemma \ref{lem:N2} applies, and  
$W_{N,l}$ is a free $R_N$-module with free basis $\{Q_a\}$. 
This completes the proof of Theorems \ref{thm:WN} and 
\ref{thm:chWN}.
\qed

{}From the proof we obtain 
\begin{cor}\label{cor:iso-rho-N} 
The natural map 
\bea
\bar{\rho}_{N}:\Zc_N\overset{\sim}{\longrightarrow} W_N 
\label{rho}
\ena
is an isomorphism. 
\end{cor}


\section{The character of $M_{N,l}$}\label{sec:Gordon}

Recall that the space $M_{N,l}$ is defined as 
the quotient of $W_{N, l}$ by the subspace 
generated by $\Sigma_1(X)$ and $\Sigma_2(X_1,X_2)$. 
The goal of this section is the following result. 
\begin{thm}\label{thm:Nakayashiki}
The space $M_{N,l}$ is a free $R_N$-module with the 
character
\bea
\ch_q M_{N,l}=
\frac{1}{(q)_N}\left( \qbin{N}{l}-\qbin{N}{l-1}\right).
\label{eq:character-Nakayashiki} 
\ena
\end{thm}
{}Formula \eqref{eq:character-Nakayashiki} 
was obtained by Nakayashiki \cite{N}.
We will give here an alternative proof 
based on the results of the previous section. 

\subsection{Estimate from both ends}
\label{subsec:above-Nakayashiki}
In the previous section, we proved the identity in 
Theorem \ref{thm:pre-supernomial} 
by showing two inequalities:
estimate from above \eqref{FeiginFeigin:upper-ineq} 
and estimate from below
\eqref{FeiginFeigin:lower-ineq}. 
We prove Theorem \ref{thm:Nakayashiki} 
in the same way. 

In order to obtain an estimate from above, 
we make use of the isomorphism \eqref{rho}, 
\be
\bar{\rho}_{N}:
\Zc_N=(Z\otimes R_N)/\Ic_N \overset{\sim}{\rightarrow}
W_N,
\en
where $Z$ is defined in (\ref{LC}), and 
$\Ic_N$ is given in Definition \ref{def:JN}. 
Let $\pi:W_{N,l}\rightarrow M_{N,l}$ be the canonical projection. 
The filtration \eqref{def:filtration} on $\Zc_N$ induces  
a filtration $\{\pi\bigl(\bar{\rho}_{N}(F_s)\bigr)\}$ on $M_{N,l}$.

\begin{definition}
Let $\Jb_N$ be the ideal of $Z$ 
generated by $J_N$ and $\xi_0,\eta_0$, 
where  $J_N$ is the ideal given in Definition \ref{def:J_N}.
We define 
\be
\Zb_{N,l}=Z_{l}/(Z_{l}\cap \Jb_{N}), 
\en
where $Z_{l}$ denotes the subspace of weight $l$. 
\end{definition}

{}From the correspondence in Proposition \ref{prop:current-cycle-correspondence},  
we have a surjection 
\bea
\Zb_{N,l}\otimes R_N \rightarrow \gr M_{N,l}\rightarrow 0.
\label{surj2}
\ena
In the following subsections we will prove that  
\begin{prop}\label{prop:Mnlup}
\bea
\ch_q\Zb_{N,l}\le 
\qbin{N}{l}-\qbin{N}{l-1}.
\label{Wbc}
\ena
\end{prop}

{}For the estimate from below, we 
use the following result of Tarasov (\cite{T}, in the proof of Theorem 4.4).
It can also be derived using the knowledge of crystal basis
(see Section \ref{subsec:crystal}).
\begin{prop}\label{prop:below-Nakayashiki}$($\cite{T}$)$
Let $e_c:W_{N,l}\rightarrow A_{N,l}$ 
$(c\in \C^N)$ be the evaluation map, and  
denote the map induced on the quotient space $M_{N,l}$ 
by the same letter $e_c$.
Then for a generic $c$ we have 
\be 
\dim_\C e_c(M_{N,l})=\binom{N}{l}-\binom{N}{l-1}. 
\en
\end{prop}

Theorem \ref{thm:Nakayashiki} follows from 
Proposition \ref{prop:Mnlup} 
and Proposition \ref{prop:below-Nakayashiki}, 
by reasoning in a similar manner as in Subsection 
\ref{subsec:FeiginFeigin}. 

\subsection{Kostka polynomial for type $A_{1}$} 
Before proceeding to the proof
of Proposition \ref{prop:Mnlup}, 
it is useful to rephrase \eqref{Wbc}
in terms of Kostka polynomials. 
The Kostka polynomial  
for type $A_{1}$ is a polynomial $K_{m,\nu}(q)$ in $q$, 
depending on an integer $m$  
and an array $\nu=(\nu_1,\cdots,\nu_N)$ of non-negative
integers. 
The following fermionic formula is available. 
\bea
&&
K_{m,\nu}(q)
=
\sum_{\bf{n}}
q^{c(\bf{n})}
\prod_{j} 
\qbin{P_j+n_j}{n_j}.
\label{eq:Kos-N}
\ena
Here we set $m_{a}=\sharp\{j\mid \nu_j=a\}$ and 
the sum ranges over all the sequences 
${\bf n}=(n_{j})_{j=1, 2, \ldots}$ of 
non-negative integers $n_j\in \Z_{\ge 0}$ such that
$2\sum_{j\ge 1} jn_j=\sum_{j\ge 1} jm_j-m$. 
The quantities $c(\bf{n})$ and $P_{j}$ are defined by 
\bea
&&c({\bf n})=\sum_{j,j'\ge 1} A_{jj'}n_jn_{j'}, 
\label{def:gordon-c-N} 
\\
&&
A_{jj'}=\min(j,j'), \quad 
P_j=\sum_{j'\ge 1} A_{jj'}(m_{j'}-2n_{j'}). 
\label{def:Pj}
\ena

In the simplest case where $\nu=(1^N)=(1,\cdots,1)$, 
the Kostka polynomial $K_{m, (1^{N})}(q)$ is given by
a simple formula:
\bea
&&K_{m,(1^N)}(q)=
\begin{cases}
\qbin{N}{\frac{N-m}{2}}-\qbin{N}{\frac{N-m-2}{2}}
& (m\equiv N\bmod 2),\\
0& (\mbox{otherwise}).
\end{cases}
\label{eq:formula-Kostka} 
\ena
Hence \eqref{Wbc} is equivalently stated as 
\bea 
\ch_q\Zb_{N,l}\le K_{N-2l, (1^{N})}(q).
\label{Wbcc}
\ena 

\subsection{Dual space}
To show \eqref{Wbcc}, 
it is convenient to pass to the dual space. 
Let $\C[x_1,\cdots,x_s]^{{\rm Skew}}$ 
be the space of skew-symmetric polynomials in $x_1,\cdots,x_s$, 
and $\C[y_1,\cdots,y_t]^{{\rm Symm}}$
the space of symmetric polynomials in $y_1,\cdots,y_t$. 
Denote by $\Lambda(\xi_0,\xi_1,\xi_2,\ldots)_t$ or
$\C[\eta_0,\eta_1,\eta_2,\ldots]_t$,
the homogeneous component of weight $t$ with respect to the weight
defined by (\ref{DEGWEI}).
Set 
\be 
\bar{\xi}(z)=\xi_{0}+\xi(z)=\sum_{n=0}^{\infty}\xi_{n}z^{n}, \quad 
\bar{\eta}(z)=\eta_{0}+\eta(z)=\sum_{n=0}^{\infty}\eta_{n}z^{n}. 
\en 
There are non-degenerate bilinear pairings
\bea
&&\Lambda(\xi_0,\xi_1,\xi_2,\ldots)_s\times
\C[x_1,\cdots,x_s]^{{\rm Skew}}\longrightarrow \C,
\label{pair1}\\
&&\C[\eta_0,\eta_1,\eta_2,\ldots]_t\times
\C[y_1,\cdots,y_t]^{{\rm Symm}}\longrightarrow \C,
\label{pair2}
\ena
characterized by 
\bea
&&\langle \bar{\xi}(x_1)\cdots\bar{\xi}(x_s),f\rangle=f(x_1,\cdots,x_s),
\label{coup1}\\
&&\langle \bar{\eta}(y_1)\cdots\bar{\eta}(y_t),g\rangle=g(y_1,\cdots,y_t),
\label{coup2}
\ena
for $f\in \C[x_1,\cdots,x_s]^{{\rm Skew}}$,  
$g\in \C[y_1,\cdots,y_t]^{{\rm Symm}}$.   
The pairings \eqref{pair1},\eqref{pair2} respect degrees 
with the assignment $\deg x_a=1,\deg y_b=1$. 

We have thus a non-degenerate pairing
\bea
&& 
\label{def:pairing} \\ 
&& 
\bigl(\Lambda[\xi]\otimes\C[\eta]\bigr)_l
\times
\bigoplus_{s+2t=l}(\C[x_1,\cdots,x_s]^{{\rm Skew}}\otimes
\C[y_1,\cdots,y_t]^{{\rm Symm}})
\longrightarrow \C. \nn 
\ena
The dual space of $\Zb_{N,l}$ is identified 
via \eqref{def:pairing} with the space $\F_{N,l}$, which  consists
of sets of polynomials $\f=(f_{s,t})_{s+2t=l}$ 
orthogonal to the ideal $\Jb_{N,l}$. 
Explicitly the generators of  $\Jb_{N,l}$ are
\bea
&&\xi_0,~~\eta_0, 
\label{J4}\\
&&\xi(z)\xi(-z)+\eta(z)-\eta(-z),
\label{J1}\\
&&[\eta(z)^\nu]_{\ge N-2\nu+1}\qquad (\nu\ge 1),
\label{J2}\\
&&[\xi(z)\eta(z)^{\nu-1}]_{\ge N-2\nu+2}\qquad (\nu\ge 1).
\nn
\ena
{}From the rules \eqref{coup1}, \eqref{coup2} of the pairing,
we find the following 
conditions defining $\F_{N,l}$.

\begin{lem}\label{lem:orth}
A set of polynomials $\f=(f_{s,t})_{s+2t=l}$ belongs to 
$\F_{N,l}$ if and only if the following conditions
are satisfied. 
\begin{enumerate}
\item $f_{s,t}(x_1,\cdots,x_s;y_1,\cdots,y_t)$ 
is skew-symmetric with respect to $x_1,\ldots,x_s$ 
and symmetric with respect to $y_1,\ldots,y_t$.
\item 
There exists a polynomial $g_{s,t}$ such that
\be
f_{s,t}=x_1\cdots x_s(y_1\cdots y_t)^2g_{s,t}.
\en
\item The polynomials $f_{s,t}$ satisfy the relations
\be
&&f_{s+2,t}(z,-z,x_1,\ldots,x_s;y_1,\ldots,y_t)
+f_{s,t+1}(x_1,\ldots,x_s;z,y_1,\ldots,y_t)\\
&&-f_{s,t+1}(x_1,\ldots,x_s;-z,y_1,\ldots,y_t)=0,
\en
\item For all $\nu\ge 1$ we have 
\be
&&\deg_zf_{s,t}(x_1,\ldots,x_s;
\underbrace{z,\ldots,z}_\nu,y_{\nu+1},\ldots,y_t)\leq N-2\nu,
\\
&&
\deg_zf_{s,t}(z,x_2,\ldots,x_s;
\underbrace{z,\ldots,z}_{\nu-1},
y_{\nu},\ldots,y_t)\leq N-2\nu+1.
\en
\end{enumerate}
\end{lem}

The inequality \eqref{Wbcc} is equivalent to 
\begin{prop}\label{prop:above2-Nakayashiki}
We have
\be
\ch_q\F_{N,l}\le K_{N-2l,(1^N)}(q).
\en
\end{prop}
In the next subsection
we will prove Proposition \ref{prop:above2-Nakayashiki}, 
using the fermionic formula of $K_{N-2l, (1^{N})}(q)$.

\subsection{Proof of Proposition \ref{prop:above2-Nakayashiki}}
\label{subsec:Gordon}

Let ${\mathcal P_l}$ be the set of partitions of $l$. 
For $\lambda=(\lambda_1,\ldots,\lambda_n)
=(1^{n_1},2^{n_2},\ldots,k^{n_k}) \in {\mathcal P_{l}}$, 
we have $\lambda_1\geq\lambda_2\geq\cdots\geq\lambda_n>0$, 
$n_\alpha\geq0$ and
$\sum_{i=1}^n\lambda_i=\sum_{\alpha=1}^k\alpha n_\alpha=l$.
We define a total order of
${\mathcal P_l}$: $\lambda>\lambda'$ if and only if for some $j$ we have
$\lambda_i=\lambda'_i$ if $1\leq i<j$ and $\lambda_j>\lambda'_j$.

We define a polynomial $\varphi_\lambda({\bf f})$
of the variables $v_1,\ldots,v_n$. We set $s=\sum_{\alpha:{\rm odd}}n_\alpha$
and $t=(l-s)/2$. Note that $l-s$ is even. Let $i_a$ $(1\leq a\leq s)$
be the set of indices such that $i_1<i_2<\cdots<i_s$ and
$\lambda_{i_a}\in 2\Z+1$. We set
\begin{equation}\label{SPECIALIZATION}
\varphi_\lambda({\bf f})=f_{s,t}(v_{i_1},\ldots,v_{i_s};
\underbrace{v_1,\ldots,v_1}_{\nu_1},\ldots,
\underbrace{v_n,\ldots,v_n}_{\nu_n}),
\end{equation}
where
\begin{eqnarray}
\lambda_i=
\begin{cases}
2\nu_i+1&\hbox{if }i=i_1,\ldots,i_s;\\
2\nu_i&\hbox{otherwise}.
\end{cases}
\end{eqnarray}
We define a filtration $\Gamma_\lambda$ of the vector space 
${\mathcal D}_{N,l}$:
\bea
\Gamma_\lambda=\bigcap_{\lambda'>\lambda}{\rm Ker}\,\varphi_{\lambda'}.
\label{def:Gordon-filtration} 
\ena
The associated graded space is defined as
$\Gamma_\lambda/\Gamma'_\lambda$ where
$\Gamma'_\lambda=\Gamma_\lambda\cap{\rm Ker}\,\varphi_\lambda$.
Since $\varphi_\lambda({\bf f})=0$
if ${\bf f}\in{\rm Ker}\,\varphi_\lambda$,
the specialization ${\bf f}\mapsto\varphi_\lambda({\bf f})$
is an injective mapping defined on the graded component
$\Gamma_\lambda/\Gamma'_\lambda$. Our aim is to determine
the image of this mapping.

\begin{prop}\label{FACTOR}
Let ${\bf f}\in\Gamma_\lambda/\Gamma'_\lambda$ and
$F=\varphi_\lambda({\bf f})$. 
The polynomial $F(v_1,\ldots,v_n)$ is divisible by
\begin{equation}
\prod_av_a^{\lambda_a}
\prod_{a>b}(v_a^2-v_b^2)^{\lambda_a}.
\end{equation}
\end{prop}

Before the proof, 
let us see that Proposition \ref{prop:above2-Nakayashiki} 
follows from this proposition. 
Set 
\be 
\varphi_{\lambda}({\bf f})= 
\left( \prod_av_a^{\lambda_a}
\prod_{a>b}(v_a^2-v_b^2)^{\lambda_a} \right) f_{\lambda}. 
\en 
Then $f_{\lambda}$ is symmetric with respect to $v_{a}$ and $v_{b}$ 
such that $\lambda_{a}=\lambda_{b}$. 
The degree of $f_{\lambda}$ with respect to $v_{a}$ 
is at most $P_{\lambda_{a}}$, where $P_{\lambda_{a}}$ is given by \eqref{def:Pj} 
with $m=N-2l$ and $\nu=(1^{N})$. 
Hence we have 
\be 
{\rm ch}_{q}\varphi_{\lambda}(\Gamma_{\lambda}/\Gamma_{\lambda}') \le 
q^{c({\bf n})}
\prod_{j}
\qbin{P_j+n_j}{n_j}, 
\en 
where $c({\bf n})$ is given by \eqref{def:gordon-c-N}, 
and ${\bf n}=(n_{j})_{j=1, 2, \ldots }$ is 
determined from $\lambda$ by $\lambda=(1^{n_{1}}, 2^{n_{2}}, \ldots)$. 
Thus we get the upper estimate in Proposition \ref{prop:above2-Nakayashiki}. 

The proof of Proposition \ref{FACTOR} is divided into several lemmas.
\begin{lem}\label{BASIC}
Let $f_{s,t}$ be a member of ${\bf f}\in \F_{N,l}$.
We have the equality
\[
f_{s,t}(v,\ldots;v,\ldots)=f_{s,t}(v,\ldots;-v,\ldots).
\]
\end{lem}
\begin{proof}
This is a consequence of the skew-symmetry of $f_{s+2,t-1}$
in the first set of variables and the relation (iii) in Lemma \ref{lem:orth}.
\end{proof}
\begin{lem}\label{1EVEN}
Let $\lambda=(\lambda_1,\ldots,\lambda_n)$ be a partition such that
for some $a>b$ we have $\lambda_a=2m+1,\lambda_b=2\nu$. We denote by
$\kappa$ the
partition 
obtained from $\lambda$ by splitting $\lambda_a$ to $(1,2^m)$.
We use the variables $v_1,\ldots,v_n$ and $w_1,\ldots,w_m$
to represent $\varphi_\kappa({\bf f});$ in particular, the variables
$v_a$ and $v_b$ correspond to the parts $1$ and $2\nu$.
The polynomial $\varphi_\kappa({\bf f})$ where ${\bf f}\in\Gamma_\lambda$
is divisible by $v_a^2-v_b^2$.
\end{lem}
\begin{proof}
Let $\kappa'$ be the 
partition obtained from $\kappa$ by splitting
the part $2\nu$ to $2^\nu$. 
We introduce the variables $y_1,\ldots,y_\nu$
in place of $v_b$ to represent 
$\varphi_{\kappa'}({\bf f})$. 
Let us denote
this polynomial by $f(v_a;y_1,\ldots,y_\nu)$ forgetting the dependence
on the other variables. The assertion follows if we show two equalities
$f(v;v,\ldots,v)=0$ and $f(v;-v,\ldots,-v)=0$. Let $\kappa''$ be
a partition obtained from $\kappa$ by merging the parts $1,2\nu$
in a new part $2\nu+1$. Then, we have $\kappa''>\lambda$, and therefore
\[
f(v;v,\ldots,v)=\varphi_{\kappa''}({\bf f})=0.
\]
The second equality follows from this and Lemma \ref{BASIC}.
\end{proof}
\begin{lem}\label{CARTAN}
Let
$f^{(0)}(x_1,x_2,x_3,x_4;y_1,\ldots,y_{\nu-2})$,
$f^{(1)}(x_1,x_2;y_1,\ldots,y_{\nu-1})$
and\\
$f^{(2)}(y_1,\ldots,y_\nu)$ be polynomials, which are skew-symmetric
in $\{x_a\}$, symmetric in $\{y_b\}$, and satisfy the relations
\begin{eqnarray}
&&f^{(0)}(x_1,x_2,v,-v;y_1,\ldots,y_{\nu-2})\label{KEY1}\\
&&=-f^{(1)}(x_1,x_2,;v,y_1,\ldots,y_{\nu-2})
+f^{(1)}(x_1,x_2,;-v,y_1,\ldots,y_{\nu-2}),\nonumber\\
&&f^{(1)}(v,-v;y_1,\ldots,y_{\nu-1})\label{KEY}\\
&&=-f^{(2)}(v,y_1,\ldots,y_{\nu-1})+f^{(2)}(-v,y_1,\ldots,y_{\nu-1}).\nonumber
\end{eqnarray}
If the polynomial $f^{(2)}(v,\ldots,v)$ of $v$ is identically $0$,
then for all $0\leq\nu'\leq\nu$ we have
$f^{(\nu)}_{\nu'}=f^{(2)}(\underbrace{v,\ldots,v}_{\nu'},
\underbrace{-v,\ldots,-v}_{\nu-\nu'})=0$.
\end{lem}
\begin{proof}
Lemma \ref{BASIC} implies that $f^{(1)}(v,-v;\underbrace{v,\ldots,v}_{\nu'},
\underbrace{-v,\ldots,-v}_{\nu-\nu'-1})$ are independent of
$0\leq \nu'\leq \nu-1$. Using (\ref{KEY}), we write these $f^{(1)}$
in terms of $f^{(2)}$. Let
\[
A^{(\nu)}_{a,b}=
\begin{cases}
2&\hbox{if }a=b;\\
{}-1&\hbox{if }a=b\pm1;\\
0&\hbox{otherwise}.
\end{cases}
\]
Under the condition $f^{(2)}(v,\ldots,v)=f^{(2)}(-v,\ldots,-v)=0$, we have
\[
\sum_{\nu'=1}^{\nu-1}A^{(\nu)}_{a,\nu'}f^{(\nu)}_{\nu'}=0
\]
for all $1\leq a\leq\nu-1$. Since $A^{(\nu)}$ is non-degenerate,
the assertion follows.
\end{proof}
\begin{lem}\label{1ODD}
Let $\lambda=(\lambda_1,\ldots,\lambda_n)$ be a partition such that
for some $a>b$ we have $\lambda_a=2\nu'+1,\lambda_b=2\nu-1$. We denote by $\kappa$ a
partition obtained from $\lambda$ by splitting $\lambda_a$ to $(1,2^{\nu'})$.
We use the variables $v_1,\ldots,v_n$ and $w_1,\ldots,w_{\nu'}$
to represent 
$\varphi_\kappa({\bf f});$ in particular, the variables
$v_a$ and $v_b$ correspond to the parts $1$ and $2\nu-1$.
The polynomial $\varphi_\kappa({\bf f})$ where ${\bf f}\in\Gamma_\lambda$
is divisible by $v_a^2-v_b^2$.
\end{lem}
\begin{proof}
Let $\lambda^{(0)},\lambda^{(1)},\lambda^{(2)}$ be partitions obtained from
$\kappa$ by replacing the parts $1,2\nu-1$
with new parts $(1^4,2^{\nu-2}),(1^2,2^{\nu-1}),(2^\nu)$, respectively.
We write
\begin{eqnarray*}
&&\varphi_{\lambda^{(0)}}({\bf f})=f^{(0)}(x_1,x_2,x_3,x_4;y_1,\ldots,y_{\nu-2}),\\
&&\varphi_{\lambda^{(1)}}({\bf f})=f^{(1)}(x_1,x_2;y_1,\ldots,y_{\nu-1}),\\
&&\varphi_{\lambda^{(2)}}({\bf f})=f^{(2)}(y_1,\ldots,y_\nu),
\end{eqnarray*}
forgetting the variables other than those which correspond to the new parts.
It is enough to show two equations
$f^{(1)}(v,v;v,\ldots,v)=0$ and $f^{(1)}(v,-v;-v,\ldots,-v)=0$.
The former follows from the skew-symmetry of $f_{s,t}$
in the first set of variables. Let $\kappa'$ be a partition obtained from $\kappa$
by merging $1,2\nu-1$ in $2\nu$.
Then, we have $\kappa'>\lambda$, and therefore,
$f^{(2)}(v,v,\ldots,v)=\varphi_{\kappa'}({\bf f})=0$.
By Lemma \ref{CARTAN}, it then follows that
$f^{(2)}(v,-v,\ldots,-v)=f^{(2)}(-v,-v,\ldots,-v)=0$.
Because of (\ref{KEY}), these equalities imply $f^{(1)}(v,-v;-v,\ldots,-v)=0$.
\end{proof}
\begin{lem}\label{2}
Let $\lambda=(\lambda_1,\ldots,\lambda_n)$ be a partition.
Take a pair of indices $a>b$ such that $\lambda_a\geq2$.
We denote by $\kappa$ a partition obtained from $\lambda$ by splitting
$\lambda_a$ to $(1,2^{\nu'})$ if $\lambda_a=2\nu'+1$, or to $2^{\nu'}$
if $\lambda_a=2\nu'$. If $\lambda_a=2\nu'+1$,
we use the variables $v_1,\ldots,v_n$ and $w_1,\ldots,w_{\nu'}$,
to represent $\varphi_\kappa({\bf f});$ 
or if $\lambda_a=2\nu'$, 
we use $v_1,\ldots,v_n$ 
except for $v_a$ and $w_1,\ldots,w_{\nu'}$.
In particular, the variables
$w_1,\ldots,w_{\nu'}$ and $v_b$ correspond to the parts $2^{\nu'}$ and $\lambda_{b}$.
The polynomial $\varphi_\kappa({\bf f})$ where ${\bf f}\in\Gamma_\lambda$
is divisible by $(w_i^2-v_b^2)^2$.
\end{lem}
\begin{proof}{\it Case $\lambda_b=2\nu-2$}.

\noindent We follow the same steps as in the proof of Lemma \ref{1ODD}
with the parts $1,2\nu-1$ replaced with $2,2\nu-2$,
and define $f^{(0)},f^{(1)},f^{(2)}$, and $\kappa'$. Note that $\kappa'>\lambda$.
We now want to prove that both $f^{(2)}(v,\underbrace{w,\ldots,w}_{\nu-1})$
and $f^{(2)}(-v,\underbrace{w,\ldots,w}_{\nu-1})$ are divisible by $(v-w)^2$.
It is enough to show $f^{(2)}(v,v\ldots,v)$ $=0$,
$\frac{\partial f^{(2)}}{\partial y_1}(v,v,\ldots,v)=0$,
$f^{(2)}(-v,v\ldots,v)=0$ and
$\frac{\partial f^{(2)}}{\partial y_1}(-v,v,\ldots,v)=0$.
The first equality follows from the condition
${\bf f}\in\Gamma_\lambda\subset{\rm Ker}\,\varphi_{\kappa'}$,
and the second, then follows from the symmetry of $f^{(2)}$ in the variables
$y_1,\ldots,y_\nu$. The third follows from Lemma \ref{CARTAN}.

Let $\kappa''$ be a partition obtained from $\kappa$ by replacing two parts
$2$ and $2\nu-2$ by new parts $1$ and $2\nu-1$. We have $\kappa''>\lambda$.
Because of (\ref{KEY}) and the skew-symmetry of $f^{(1)}$
in $x_1,x_2$, the last equality follows from
$\frac{\partial f^{(1)}}{\partial x_2}(v,-v;v,\ldots,v)=0$,
which is a consequence of
${\bf f}\in\Gamma_\lambda\subset{\rm Ker}\,\varphi_{\kappa''}$.

{\it Case $\lambda_b=2\nu-1$}.

\noindent Let $\kappa'$ be a partition obtained from $\kappa$ by splitting $2\nu-1$
to $(1,2^{\nu-1})$. Along with a part $2$ coming from the splitting of
$\lambda_a\geq2$, it contains the parts $(1,2^\nu)$.
We introduce the variables $x$ and $y_1,\ldots,y_\nu$
corresponding to these parts, and 
write $\varphi_{\kappa'}({\bf f})=
f(x_1;y_1,\ldots,y_\nu)$ forgetting the other variables. Set
$h(v,w)=f(v;w,v,\ldots,v)$. We want to show that $h(v,w)$ is divisible by
$(v-w)^2$ and $(v+w)^2$, or equivalently,
$f(v;v,v,\ldots,v)=0$, $\frac{\partial f}{\partial y_1}(v;v,v,\ldots,v)=0$,
$f(v;-v,v,\ldots,v)=0$ and
$\frac{\partial f}{\partial y_1}(v;-v,v,\ldots,v)=0$.
Let $\kappa''$ be a partition obtained from $\kappa$ by replacing
two parts $2$ and $2\nu-1$ by new parts $1$ and $2\nu$.
We have $\kappa''>\lambda$.
The first equality follows from $f(v;w,\ldots,w)=0$, which is a consequence of
${\bf f}\in\Gamma_\lambda\subset{\rm Ker}\,\varphi_{\kappa''}$.
The second then follows from the symmetry of
$f(x_1;y_1,\ldots,y_\nu)$ in the variables $y_1,\ldots,y_\nu$.
By Lemma \ref{BASIC} we have the third, $f(v;-v,v,\ldots,v)=0$,
and also $f(v;w,v,\ldots,v)=f(v;w,-v,\ldots,-v)$.
In particular, we have
$\frac{\partial f}{\partial y_1}(v;-v,v,\ldots,v)=
\frac{\partial f}{\partial y_1}(v;-v,\ldots,-v)$. This is in 
fact zero because $f(v;w,\ldots,w)=0$.
\end{proof}

{\it Proof of Proposition \ref{FACTOR}}.
We use Lemmas \ref{1EVEN}, \ref{1ODD} and \ref{2} to show that
$F$ has the factor $(v_a^2-v_b^2)^{\lambda_a}$.
Let $\kappa$ be a partition defined in Lemma \ref{2}. 
Since the number of odd parts, i.e., $s$, in $\lambda$ and in $\kappa$
are the same, the same function $f_{s,t}$ is used both in
$\varphi_\kappa({\bf f})$ and $\varphi_\lambda({\bf f})$.
Therefore, the latter is obtained by specialization of the former.
The assertion follows from this observation.\qed


\newcommand{\vvpi}{\Pi_{n,l}}
\newcommand{\Bc}{\mathcal{B}}

\section{Form factors of the restricted sine-Gordon model}
\label{sec:RSGdef}

\subsection{Formulation}\label{subsec:RSGformulation}

When the coupling constant $\xi$ becomes rational, 
a `reduction' of the SG theory takes place \cite{RS}. 
In this subsection, we formulate 
form factors of the restricted sine Gordon (RSG) model 
in the case where $\xi$ is an integer $r\ge 3$. 
Set $\epsilon=e^{-\pi i/r}$, and denote by $\Ure(\slt)$ the subalgebra 
of $\Ure$ generated by $E=e_1,F=f_1,T^{\pm 1}=t_1^{\pm 1}$.  
We use the {\it opposite} coproduct 
\bea
&&\Delta'(E)=E\otimes T+1\otimes E,\quad
\Delta'(F)=F\otimes 1+T^{-1}\otimes F,\quad\Delta'(T)=T\otimes T.
\label{OPP} 
\ena
We introduce the gauge transformation
\bea
\tilde f_n(\beta_{1}, \ldots , \beta_{n})=
G(\beta_{1}, \ldots , \beta_{n}) f_n(\beta_{1}, \ldots , \beta_{n}), 
\ena 
where
\be 
G(\beta_{1}, \ldots , \beta_{n}):=e^{\frac1{2r}\sum_j\beta_j\sigma^z_j},
\en
which makes the action of $\Ure(\slt)$ on $V^{\otimes N}$
to be the standard one, i.e., without spectral parameters.
We will consider the decomposition of the space of
highest weight vectors in this action. This procedure leads us to the
$r$-restricted paths.

{}First let us summarize some facts about representations of $\Ure(\slt)$. 
The details are given in appendix \ref{app:goodbad}. 
Besides $V=\C v_+\oplus \C v_-$,   
we use three types of $\Ure(\slt)$-modules \cite{RT}, 
\bea
&&V^s,~~X^s(\alpha)\quad (0\le s\le r-2, \alpha=\pm 1),
\label{module}\\
&&W^s(\alpha)\quad (0\le s\le r-1,\alpha=\pm 1). 
\nn
\ena
The modules $V^s$ ($0\le s\le r-2$) are 
specializations of irreducible modules at $q=\epsilon$. 
They are irreducible and $\dim V^s=s+1$. In particular, $V^1=V$. 
The modules $X^s(\alpha)$ and $W^s(\alpha)$ 
are indecomposable, and have dimensions $2r$ and $r$, respectively. 

The tensor product $V^{\otimes n}$ is a direct sum 
of `good' and `bad' subspaces (see Definition \ref{def:goodbad}) 
\bea
V^{\otimes n}=\G^{(r)}_n\oplus \Bc^{(r)}_n.
\label{goodbad}
\ena
The `good' subspace $\G^{(r)}_n$ is a direct sum of 
$V^s$ ($0\le s\le r-2$), 
while the `bad' subspace $\Bc^{(r)}_n$ 
is a direct sum of $X^s(\alpha)$ ($0\le s\le r-2$)
and $W^{r-1}(\alpha)$. 
The decomposition 
\eqref{goodbad} is orthogonal relative to 
the standard 
symmetric bilinear form $(~,~)$ on $V^{\otimes n}$.

In the case of generic $\xi$, we considered  
form factors $f=(f_n)_{n=0}^\infty$ taking values in 
$\Omega_{n,l}=\Ker e_1\cap (V^{\otimes n})_l$ in the action
$\pi_{\zeta_{1}^{-1}} \otimes \cdots \otimes \pi_{\zeta_{N}^{-1}}$.
This space is invariant under the action of
the operators which enter the axioms for form factors of the SG model: 
\be 
&& 
P_{j,j+1}S_{j,j+1}(\beta_{j}-\beta_{j+1}) \\ 
&& {}=G(\ldots , \beta_{j+1}, \beta_{j}, \ldots )^{-1}\tilde{S}_{j,j+1}(\beta_{j}-\beta_{j+1})
G(\ldots , \beta_{j}, \beta_{j+1}, \ldots )  
\en 
and 
\be 
&& 
e^{\frac{(n-2l)\pi i}{2\xi}\sigma_n^z}P_{n,n-1}\cdots P_{2,1} \\ 
&& {}=
G(\beta_{1}, \ldots , \beta_{n-1}, \beta_{n}+2\pi i)^{-1}\Pi_{n,l} 
G(\beta_{n}, \beta_{1}, \ldots , \beta_{n-1}). 
\en 
Here the operator $\tilde{S}(\beta)$ is given by 
\be 
\tilde{S}(\beta)=S_{0}(\beta) 
\frac{R^{+}\zeta-(R^{+})^{-1}\zeta^{-1}}{q\zeta-q^{-1}\zeta^{-1}}, \quad 
\zeta=e^{\frac{\beta}{\xi}}.  
\en 
See \eqref{def:Rplus} and \eqref{def:vpi} for the definition 
of $R^{+}$ and $\Pi_{n, l}$. 

In the restricted case, we consider the gauge transformed action.
Though the subspaces $\mathcal{G}^{(r)}_n,\mathcal{B}^{(r)}_n$
are not invariant under the actions of $\tilde{S}_{j,j+1}(\beta)$ and 
$\Pi_{n, l}$, 
it can be shown that the space $\Omega_{n,l}\cap\mathcal{B}^{(r)}_n$ is invariant 
(see Lemma \ref{lem:Om}). This observation makes the following definition
of the form factors of the RSG model well defined.

We define 
\bea
\Omega^{(r)}_{n,l}=\Omega_{n,l}/\Omega_{n,l}\cap\Bc^{(r)}_n.
\label{Orest}
\ena
The operator $\tilde{S}_{j,j+1}(\beta)$ and $\vvpi$ are well defined on 
$\Omega^{(r)}_{n,l}$. Let 
\be
\Pc:\Omega_{n,l}\rightarrow \Omega^{(r)}_{n,l}
\en
be the projection. 

\begin{definition} 
{}For a form factor $f=(f_{n})_{n=0}^{\infty}$ 
of the SG model satisfying the axioms ${\rm (A0)-(A3)}$,  
we call the tower of the $\Omega^{(r)}_{n,l}$-valued 
functions $(\Pc \tilde{f}_{n})_{n=0}^{\infty}$ 
a form factor of the RSG model.
\end{definition} 

A form factor $(\Pc \tilde{f}_{n} )_{n=0}^{\infty}$ is called 
{\it $N$-minimal} if 
\be 
\Pc \tilde{f}_{n} =0 \quad {\rm for} \,\, n<N. 
\en 
{}From the axiom (A3) the $N$-particle minimal form factor satisfies  
\bea 
{\rm res}_{\beta_{N}=\beta_{N-1}+\pi i}\Pc \tilde{f}_{N}=0. 
\label{cond:N-minimal-formfactor} 
\ena 

\subsection{The qKZ equation of face type}\label{sect:faceqKZ} 
In this section we introduce the basis of the tensor product $V^{\otimes N}$ 
parametrized by certain combinatorial objects called paths. 
For the time being, let the parameter $\xi\in\C$ be generic and set
\bea\label{QI}
\qi{n}=(q^n-q^{-n})/(q-q^{-1}),
\ena
where $q=e^{-\pi i/\xi}$. We consider the action of $U_q(\slt)$ 
as given in Section \ref{NCC}.
For a non-negative half integer $j\in\frac12\Z$, 
we take a basis $\bar v^j_m$ $(-j\leq m\leq j,j-m\in\Z)$
of the irreducible $(2j+1)$-dimensional representation such that 
\begin{eqnarray}
E\bar v^j_m&=&q^{-m+\frac12}\sqrt{\qi{j+m}\qi{j-m+1}}
\,\bar{v}^j_{m-1},\\
F\bar v^j_m&=&q^{m+\frac12}\sqrt{\qi{j-m}\qi{j+m+1}}
\,\bar{v}^j_{m+1}. 
\end{eqnarray}
(The basis $\{\bar v^j_m\}$ is related to $\{v^{2j}_{j-m}\}$ used 
in Appendix \ref{app:goodbad} by a scalar multiple.)
We set
\begin{eqnarray}
\left[\begin{matrix}j&1/2&j+1/2\\m&\pm1/2&m\pm1/2\end{matrix}\right]
&=&q^{\frac{\pm j-m}2}\sqrt{\frac{\qi{j\pm m+1}}{\qi{2j+1}}},\label{3J1}\\
\left[\begin{matrix}j&1/2&j-1/2\\m&\pm1/2&m\pm1/2\end{matrix}\right]
&=&\pm q^{\frac{\mp(j+1)-m}2}\sqrt{\frac{\qi{j\mp m}}{\qi{2j+1}}}.\label{3J2}
\end{eqnarray}
These are called the $q$-deformed $3j$ symbols.
If $j,m$ do not satisfy the conditions stated above, 
they are defined to be zero.  
The decomposition of the tensor product 
$\C^{2j+1}\otimes\C^2=\C^{2j+2}\oplus\C^{2j}$
is given by
\begin{eqnarray}
&&\bar v^{j\pm\frac12}_{m+\frac12}=\\
&&\left[\begin{matrix}j&1/2&j\pm1/2\\m&1/2&m+1/2\end{matrix}\right]
\bar v^{j}_m\otimes\bar v^{\frac12}_{\frac12}
+\left[\begin{matrix}j&1/2&j\pm1/2\\m+1&-1/2&m+1/2\end{matrix}\right]
\bar v^{j}_{m+1}\otimes\bar v^{\frac12}_{-\frac12}.\nonumber
\end{eqnarray}

A path of length $N$ is a sequence $J=(j_1,\ldots,j_N)$ 
satisfying $j_n\in(1/2)\Z$, $j_1=1/2$
and $j_{n+1}=j_n\pm\frac12$. The integer $2j_N$ is called the weight of
the path $J$. A path $J$ is called {\it classically restricted} 
if $j_n\geq0$ for all $0\leq n\leq N$. 
For a classically restricted path $J$
and $m_N\in \Z+j_N$ with $-j_N\leq m_N\leq j_N$, 
we define a vector $u_{J,m_N}\in\C^2\otimes\cdots\otimes\C^2$ by
\[
u_{J,m_N}=\sum_{\varepsilon_1,\ldots,\varepsilon_N=\pm1\atop
\varepsilon_1+\cdots+\varepsilon_n=2m_n}
\left[\begin{matrix}j_1&1/2&j_2\\m_1&\varepsilon_2/2&m_2
\end{matrix}\right]
\cdots
\left[\begin{matrix}j_{N-1}&1/2&j_N\\m_{N-1}&\varepsilon_N/2&m_N
\end{matrix}\right]
v_M.
\]
Here the vector $v_M$ is specified by the subset 
$M=\{n\mid\varepsilon_n=-1\}$, 
and the sum is restricted by the condition 
$\varepsilon_1+\cdots+\varepsilon_N=2m_N$.  
For $n<N$ the equation
$\varepsilon_1+\cdots+\varepsilon_n=2m_n$ defines $m_n$ in the summand.

The vectors $u_{J,m}$ constitute an
orthonormal basis of the tensor product $V^{\otimes N}$. 
In particular, the vectors $u_J=u_{J,j_N}$ 
constitute a basis of ${\rm Ker}\,E$. 
Since the integral $I(G(v_{N,l}),P)$ belongs to 
$\Omega_{N,l}={\rm Ker}\,E\cap(V^{\otimes N})_l$, 
it is a linear combination of $u_{J}$ such that $j_N=(N-2l)/2$:
\bea
&&I(G(v_{N,l}),P)=\sum_{J,j_N=(N-2l)/2}\tilde{\psi}_{P,J}u_J,
\label{rI}\\
&&\tilde{\psi}_{P,J}=\sum_{M,\sharp(M)=l}q^{\nu(M)}
e^{\frac{1}{2\xi}(\sum_{k \not\in M}\beta_{k}-\sum_{m \in M}\beta_{m})} 
C_{J,M}I(w_M,P),
\label{rPJ}\\
&&C_{J,M}=
\left[\begin{matrix}j_1&1/2&j_2\\m_1&\varepsilon_2/2&m_2
\end{matrix}\right]
\cdots
\left[\begin{matrix}j_{N-1}&1/2&j_N\\m_{N-1}&\varepsilon_N/2&j_N
\end{matrix}\right],
\label{rCJ}
\ena
where
$\varepsilon_1+\cdots+\varepsilon_n=2m_n$ for $1\leq n\leq N-1$,
and $M=\{n\mid \varepsilon_n=-1\}$.

By a standard argument \cite{Pas}, it follows 
from Theorem \ref{QKZ} and Corollary \ref{INV} 
that $\tilde{\psi}_{P,J}(\beta_1,\ldots,\beta_N)$
satisfies the qKZ equation of face type:
\be
&&\tilde{\psi}_{P,(\ldots,j_{i-1},j_i,j_{i+1},\ldots)}
(\ldots,\beta_{i+1},\beta_{i},\ldots)\label{BOL}\\
&&=\sum_{j_i'}
\tilde{\psi}_{P,(\ldots,j_{i-1},j'_i,j_{i+1},\ldots)}
(\ldots,\beta_{i},\beta_{i+1},\ldots)
\BW{j_{i-1}}{j_i}{j_i'}{j_{i+1}}{\beta_i-\beta_{i+1}},\nonumber
\en
where the coefficients are the Boltzmann weights 
of the RSOS model \cite{ABF}
\be
&&\BW{j\mp1/2}{j}{j}{j\pm1/2}{\beta}=1,\\
&&\BW{j}{j\pm1/2}{j\pm1/2}{j}{\beta}=
\frac{\qi{2j+1\mp u}}{\qi{2j+1}}
\frac{1}{\qi{1+u}},\\
&&\BW{j}{j\pm1/2}{j\mp1/2}{j}{\beta}=
\frac{\sqrt{\qi{2j}\qi{2j+2}}}{\qi{2j+1}}
\frac{\qi{u}}{\qi{1+u}},\\
\en
and $u=-\frac\beta{\pi i}$. 

Now let us consider the case where $\xi=r\geq3$ is an integer.  
Accordingly $q$ is specialized to the root of unity $\epsilon=e^{-\pi i/r}$. 
A classically restricted path $J$ is called
{\it $r$-restricted} if $2j_n\leq r-2$ for all $1\leq n\leq N$. 
It is known that the set of the vectors $\{\Pc u_{J}\}$ 
associated with $r$-restricted paths give a basis of $\Omega_{N}^{(r)}$. 

The $q$ integer
$\qi{j}=\frac{\epsilon^{j}-\epsilon^{-j}}{\epsilon-\epsilon^{-1}}$ 
vanishes if $j$ is a multiple of $r$. 
However, the Boltzmann weights appearing in the qKZ equation of face type
(\ref{BOL}) are non-zero and finite if the paths in the equations are
$r$-restricted. 
Consider the special $r$-restricted path 
\begin{equation}
J_{N,l}=(\underbrace{\frac12,0,\ldots,\frac12,0}_{2l},
\frac12,1,\frac32,\ldots,\frac{N-2l}2).
\end{equation}
Solving the qKZ equation of face type, one can obtain
$\tilde{\psi}_{P,J}$ for an arbitrary $r$-restricted path $J$ 
in terms of the one for $J_{N,l}$. 
Therefore, if $\tilde{\psi}_{P,J_{N,l}}=0$
then $\tilde{\psi}_{P,J}=0$ for all  paths $J$ of length $N$ and weight $N-2l$. 
In the next section, we construct minimal deformed cycles with this property.

\subsection{Restricted cocycles and restricted null cycles}\label{sect:restricted-null-cycle}
We fix $N,l$. Let $J$ be an $r$-restricted path of length $N$ and weight
$N-2l$. We call a linear combination of the deformed cocycles 
$\tilde{w}_{J}={\rm Skew}_{a_{1}, \ldots , a_{l}}\tilde{g}_{J}$, where 
\bea 
\tilde{g}_J=
\sum_{\# M=l} 
e^{\frac{1}{2\xi}(\sum_{k \not\in M}\beta_{k}-\sum_{m \in M}\beta_{m})} 
\epsilon^{\nu(M)}C_{J,M}g_M, 
\label{def:g-tilde} 
\ena 
an $r$-restricted cocycle. 
Here $C_{J,M}$ are defined in \eqref{rCJ}. 
A deformed cycle $P$ is called an $r$-restricted
null cycle if $I(\tilde{w},P)=0$ for all $r$-restricted cocycles. 
As discussed
at the end of the previous section, this is equivalent to 
$I(\tilde{w}_{J_{N,l}},P)=0$.
\begin{lem}\label{RDEG}
Let $Q$ be the polynomial corresponding to $\tilde{w}_{J_{N,l}}$ 
in $(\ref{WQ})$. Then, we have
\begin{equation}
{\rm deg}_{a_p}Q\leq N+l-2.\label{RES}
\end{equation}
\end{lem}
\begin{proof}
For the path $J_{N,l}$, 
the coefficient $C_{J_{N,l},M}$ is non-zero if and only if 
\[
M=(\varepsilon_1,-\varepsilon_1,\ldots,\varepsilon_l,-\varepsilon_l,
\underbrace{+,\ldots,+}_{N-2l}).
\]
{}From (\ref{3J1}) and (\ref{3J2}), we have
\[
C_{J_{N,l},M}=(-1)^{\frac12\sum_{p=1}^l(\varepsilon_p+1)}
\frac{\epsilon^{\frac12\sum_{p=1}^l\varepsilon_p}}{\qi{2}^{\frac l2}}.
\]
Taking the summation over each $\varepsilon_p$
in the right hand side of \eqref{def:g-tilde}, we obtain (\ref{RES}).
\end{proof}

Set
\begin{equation}
\mu=r-1-(N-2l).
\end{equation}
We assume that $\mu\geq1$. This is equivalent to
\[
N-2l\leq r-2.
\]
Namely, unless $\mu\geq1$, there exists no $r$-restricted path of length $N$
and weight $N-2l$. We also assume that $\mu\leq l$. This is equivalent to
\[
N-l\geq r-1.
\]
Namely, unless $\mu\leq l$, all the classically restricted paths of length
$N$ and weight $N-2l$ are $r$-restricted. The aim of this section is to find
as many minimal $r$-restricted null cycles as 
there are classically restricted 
paths that are not $r$-restricted.

We define minimal cycles $\Gamma_1$ (when $r\equiv N\bmod2$) and $\Gamma_2$ by
\begin{eqnarray}
&&\Gamma_1(X)
=X^{-1}\left(\Theta(X)-(-1)^{N+r}\Theta(-X)\right), \label{def:gamma1} \\
&&\Gamma_2(X_1,X_2)=X_1^{-1}X_2^{-1}\left(\frac{X_1-X_2}{X_1+X_2}
\Theta(X_{1}, X_{2})-\Theta(X_{1}, -X_{2}) \right), \label{def:gamma2}
\end{eqnarray}
where $\Theta(X)$ and $\Theta(X_{1}, X_{2})$ are given by 
\eqref{def:thetaX} and \eqref{def:sigmaX}, respectively. 
Observe that $\Gamma_1$ is a polynomial of $X$
of degree less than $N$ if $r \equiv N ({\rm mod} 2)$, 
and $\Gamma_2$ is a skew-symmetric polynomial of
$X_1,X_2$ of degree less than $N$ in each variable. Moreover, they
satisfy the condition of minimality (\ref{MIN}).
\begin{prop}
{\rm (i)} Suppose that $\mu=2\nu+1$.
For $\bar P\in W_{N, l-\mu}$, we set
\[
P=\Gamma_1\wedge(\wedge^\nu\Gamma_2)\wedge\bar P.
\]
This is an $r$-restricted null cycle.

{\rm (ii)} Suppose in addition that $\mu+1\leq l$.
For $\bar P\in W_{N, l-\mu-1}$, we set
\[
P=(\wedge^{\nu+1}\Gamma_2)\wedge\bar P.
\]
This is an $r$-restricted null cycle.

{\rm (iii)} Suppose that $\mu=2\nu$.
For $\bar P\in W_{N, l-\mu}$, we set
\[
P=(\wedge^\nu\Gamma_2)\wedge\bar P.
\]
This is an $r$-restricted null cycle.
\end{prop}
\begin{proof}
We prove (i). 
Note that $r\equiv N\bmod2$. 
We have the following 
equalities.
\be
&&\nabla_{\alpha,-r\pi i}\left(X^{-1}\Theta(X)\right)
=\Gamma_1(X)\hbox{ if }r\equiv N\bmod2,\\
&&\nabla_{\alpha_1,\alpha_2,-r\pi i}
\left(\frac{X_2^{-1}-X_1^{-1}}{X_1+X_2}\Theta(X_{1})\Theta(X_{2})\right)
=\frac{X_2^{-1}-X_1^{-1}}{X_1+X_2}\Theta(X_{1}, X_{2}), \\
&&X_1^{-1}X_2^{-1}\Theta(X_{1}, -X_{2})=\Gamma_1(X_1)X_2^{-1}\prod_{j=1}^N(z_jX_2+1)
{}-X_1^{-1}\prod_{j=1}^N(z_jX_1+1)\Gamma_1(X_2).
\en
If the convergence of the integrals is assured, we can show that $I(w,P)=0$
for each of (i),(ii),(iii) by repeating a similar argument
as in the proof of Proposition \ref{SIGMA}. The present case is different from
that by the factor $X^{-1}$ (or $X_1^{-1}X_2^{-1}$). Therefore, the
convergence of the integral when $\alpha\rightarrow\infty$
($\alpha_1,\alpha_2\rightarrow\infty$) only matters.

Consider the integral
\[
I(w,P)=\int_Cd\alpha\phi(\alpha)w(a)P(X).
\]
Here $P=X^{-1}\Theta(X)$ and $w$ is of the form (\ref{WQ})
with $l=1$, where $Q$ is a polynomial of $a$ satisfying
\[
{\rm deg}_aQ\leq\kappa.
\]
The integral is convergent if
\begin{equation}
r-3N+2\kappa<0.\label{CONV}
\end{equation}
The same estimate holds for the double integral $I(w,P)$ with
\[
P=\frac{X_2^{-1}-X_1^{-1}}{X_1+X_2}\Theta(X_{1})\Theta(X_{2}).
\]

We show that $I(w,P)=0$ for the case (i). 
Note that (\ref{CONV}) is equivalent to
\begin{equation}
\kappa<N+l-\nu-1.\label{KAPPA}
\end{equation}
We take
\[
Q(a_1,\ldots,a_l)=a_1^{\lambda_1}\cdots a_l^{\lambda_l}.
\]
Because of the skew-symmetry of $P$ one can assume that
$\lambda_1>\cdots>\lambda_l$. Because of Lemma \ref{RDEG},
we can also assume that $\lambda_1\leq N+l-2$. It is enough to show that
for any permutation of the variables $a_1,\ldots,a_l$ by $\sigma\in{\frak S}_l$
we have $I(\sigma(Q),P_0)=0$ where
\[
P_0(X_1,\ldots,X_l)=\Gamma_1(X_1)\Gamma_2(X_2,X_3)
\cdots\Gamma_2(X_{\mu-1},X_\mu)\bar P(X_{\mu+1},\ldots,X_l).
\]

The integral is zero if we can
apply the twisted difference method to one of the factors 
$\Gamma_1(X_1),\Gamma_2(X_2,X_3),\cdots,\Gamma_2(X_{\mu-1},X_\mu)$.
It is the case if (\ref{KAPPA}) is valid for
\[
\kappa={\rm min}(\lambda_{\sigma(1)},{\rm max}(\lambda_{\sigma(2)},
\lambda_{\sigma(3)}),\ldots,
{\rm max}(\lambda_{\sigma(\rho-1)},\lambda_{\sigma(\rho)})).
\]
If this is not the case, we have $\lambda_{\nu+1}\geq N+l-\nu-1$. This implies
\[
\lambda_1\geq\lambda_{\nu+1}+\nu\geq N+l-1.
\]
This is a contradiction.

The proofs for the cases (ii) and (iii) are similar.
\end{proof}

Now we define the space of $r$-restricted null cycles $D_{N, l}^{(r)}$ 
as follows. 
If $\mu=2\nu$ we set 
\be 
D_{N, l}^{(r)}=\Sigma_{1}\wedge W_{N, l-1}+\Sigma_{2}\wedge W_{N, l-2}+
(\wedge^{\nu}\Gamma_{2}) \wedge W_{N, l-2\nu}, 
\en 
and if $\mu=2\nu+1$ then 
\be 
&& 
D_{N, l}^{(r)}=\Sigma_{1}\wedge W_{N, l-1}+\Sigma_{2}\wedge W_{N, l-2} \\ 
&& \qquad \quad 
{}+\Gamma_{1} \wedge (\wedge^{\nu} \Gamma_{2}) \wedge W_{N, l-2\nu-1}+
(\wedge^{\nu+1}\Gamma_{2}) \wedge W_{N, l-2\nu-2}. 
\en 

Set 
\be 
M_{N, l}^{(r)}=W_{N, l}/D_{N, l}^{(r)}. 
\en 
We identify $M_{N, l}^{(r)}$ with the space of $N$-particle 
minimal form factors of the RSG model at $\xi=r$ by the map 
$P \mapsto \Pc \tilde{f}_{P}$. 

Recall the definition \eqref{def:degree-onW} 
of the degree on $W_{N, l}$. 
The generators $\Sigma_{1}, \Sigma_{2}, \Gamma_{1}$ and $\Gamma_{2}$
of $D_{N, l}^{(r)}$ are homogeneous elements. 
Hence $M_{N, l}^{(r)}$ is graded. Consider the character
\be 
\ch_{q}M_{N, l}^{(r)}=\sum_{d}q^{d}\dim_{\C}(M_{N, l}^{(r)})_{d}, 
\en 
where $(M_{N, l}^{(r)})_{d}$ is the homogeneous component 
of degree $d$. 
We will see in the next subsection (Theorem \ref{thm:4.1}) that 
it is represented in terms of the restricted Kostka polynomial. 

\subsection{Restricted Kostka polynomial}\label{subsec:rKostka}
The level-restricted Kostka polynomial 
$K^{(k)}_{m,\nu}(q)$ 
is given by the same type of fermionic formula as 
in \eqref{eq:Kos-N}: 
\bea
&&
K^{(k)}_{m,\nu}(q)
=
\sum_{n_1,\cdots,n_k}
q^{c(n_1,\cdots,n_k)}
\prod_{j=1}^k
\qbin{P_j+n_j}{n_j}.
\label{resKos}
\ena
Here we set $m_{a}=\sharp\{j\mid \nu_j=a\}$, 
\bea
&&c(n_1,\cdots,n_k)=\sum_{j,j'=1}^kA_{jj'}n_jn_{j'}
+\sum_{j=1}^kv_jn_j, 
\label{def:gordon-c} 
\\
&&
A_{jj'}=\min(j,j'),\quad v_j=\max(0,j-k+m), \\
&&
P_j=\sum_{j'=1}^kA_{jj'}(m_{j'}-2n_{j'})-v_j, 
\ena
and the sum ranges over 
$n_j\in \Z_{\ge 0}$ such that
$2\sum_{j=1}^k jn_j=\sum_{j=1}^kjm_j-m$. 
We will need the simplest case where $\nu=(1^N)=(1,\cdots,1)$. 
Note that $K^{(k)}_{m,(1^N)}(q)=0$ if  
$m\not \equiv N \bmod 2$. 

We are now in a position to state the main result of this paper.
\begin{thm}\label{thm:4.1}
The space $M^{(r)}_{N,l}$ is a free $R_N$-module with the 
character
\be
\ch_q M^{(r)}_{N,l}=\frac{1}{(q)_N}K^{(r-2)}_{N-2l,(1^N)}(q).
\en
\end{thm}

Theorem \ref{thm:4.1}
is reduced to the following two statements. 
\begin{prop}\label{prop:4.2}
There exist homogeneous elements 
$Q_j\in M^{(r)}_{N,l}$ $(1\le j\le d)$ 
such that
$M^{(r)}_{N,l}=\sum_{j=1}^dR_NQ_j$ and 
\be
\sum_{j=1}^d q^{\deg Q_j}\le K^{(r-2)}_{N-2l,(1^N)}(q).
\en
\end{prop}

\begin{prop}\label{prop:4.3}
Let $e_c:W_{N,l}\rightarrow A_{N,l}$ 
$(c\in \C^N)$ be the evaluation map, and  
denote the map induced on the quotient space $M^{(r)}_{N,l}$ 
by the same letter $e_c$.
Then for a generic $c$ we have 
\be 
\dim_\C e_c(M^{(r)}_{N,l})
\ge  K^{(r-2)}_{N-2l,(1^N)}(1). 
\en
\end{prop}
Theorem \ref{thm:4.1} follows from these propositions
by the same reasoning as given in section 
\ref{subsec:FeiginFeigin}. 
Below we prove Proposition \ref{prop:4.2} 
in section \ref{subsec:rGordon}
and Proposition \ref{prop:4.3} 
in section \ref{subsec:crystal}.

\subsection{Estimate from above}\label{subsec:rGordon}
The proof of Proposition \ref{prop:4.2} is quite similar to 
that of Proposition \ref{prop:Mnlup}.  
We make use of the isomorphism 
$\bar{\rho}_{N}:\mathcal{Z}_N \overset{\sim}{\rightarrow} W_N$. 
Let $\pi^{(r)}:W_{N,l}\rightarrow M^{(r)}_{N,l}$ be the 
canonical projection. 
The filtration \eqref{def:filtration} on $\Zc_N$ induces  
a filtration $\{\pi^{(r)}\bigl(\bar{\rho}_{N}(F_s)\bigr)\}$ 
on $M^{(r)}_{N,l}$.

We denote by $\Jb^{(r)}_{N,l}$ the ideal of 
$Z=\Lambda[\xi]\otimes\C[\eta]$ 
generated by the elements \eqref{J4}--\eqref{J2} and  
\bea
\begin{cases}
\eta_2^\nu & (\mbox{ if $\mu=2\nu$ is even}),\\
\xi_1\eta_2^\nu,~~\eta_2^{\nu+1} 
& (\mbox{ if $\mu=2\nu+1$ is odd}).\\
\end{cases}
\label{J5}
\ena
Set 
\be
\Zb^{(r)}_{N,l}=Z_{l}/(Z_{l}\cap \Jb^{(r)}_{N,l}). 
\en

Recall the correspondence in Proposition \ref{prop:current-cycle-correspondence}. 
In particular, we have 
\bea 
x_{1}^{-} \mapsto -\frac12\Gamma_1(X), \quad 
{}-4i(x_1^-)^{(2)} \mapsto \Gamma_2(X_1,X_2), 
\label{eq:fermion-null2} 
\ena 
where $\Gamma_1(X)$ and $\Gamma_2(X_1,X_2)$ are given by \eqref{def:gamma1} 
and \eqref{def:gamma2}, respectively. 
{}From this fact and the definition of $\Jb^{(r)}_{N,l}$, 
we have a surjection
\bea
\Zb^{(r)}_{N,l}\otimes R_N \rightarrow 
\gr M^{(r)}_{N,l}\rightarrow 0.
\ena
Therefore, Proposition \ref{prop:4.2} will follow if we show 
the estimate 
\bea
\ch_q\Zb^{(r)}_{N,l}\le K^{(r-2)}_{N-2l,(1^N)}(q).
\label{Wbc-res}
\ena

To show \eqref{Wbc-res} we consider the dual space of $\Zb^{(r)}_{N,l}$ 
determined by the pairing \eqref{def:pairing}. 
Then the dual space $\F^{(r)}_{N,l}$ of $\Zb^{(r)}_{N,l}$ 
is described as follows: 

\begin{lem}\label{lem:orth-res}
A set of polynomials $\f=(f_{st})_{s+2t=l}$ belongs to 
$\F^{(r)}_{N,l}$ if and only if the conditions
${\rm (i)-(iv)}$ in Lemma \ref{lem:orth} and the following condition are satisfied:

\noindent 
${\rm (v)}$\,\, If $\mu=2\nu$ is even, then  
\be
g_{s,t}(x_1,\ldots,x_s;
\underbrace{0,\ldots,0}_\nu,y_{\nu+1},\ldots,y_t)=0.
\en
If $\mu=2\nu+1$ is odd, then
\be
&&g_{s,t}(0,x_2,\ldots,x_s;
\underbrace{0,\ldots,0}_\nu,y_{\nu+1},\ldots,y_t)=0,\\
&&g_{s,t}(x_1,\ldots,x_s;
\underbrace{0,\ldots,0}_{\nu+1},y_{\nu+2},\ldots,y_t)=0.
\en
\end{lem}

The statement \eqref{Wbc-res} is equivalent to 
\begin{prop}\label{prop:4.4}
We have
\be
\ch_q\F^{(r)}_{N,l}\le K^{(r-2)}_{N-2l,(1^N)}(q).
\en
\end{prop}
This result is a super-symmetric version of a result in \cite{FJKLM}.

We prove Proposition \ref{prop:4.4} in the same way as 
in the proof of Proposition \ref{prop:above2-Nakayashiki}. 
For a partition $\lambda$ of $l$, 
we define the map $\varphi_{\lambda}$ by \eqref{SPECIALIZATION}. 
Introduce the filtration $\Gamma_\lambda$ of the vector space
${\mathcal D}^{(r)}_{N,l}$ 
by \eqref{def:Gordon-filtration} and consider the associated graded space 
$\Gamma_\lambda/\Gamma'_\lambda$, where
$\Gamma'_\lambda=\Gamma_\lambda\cap{\rm Ker}\,\varphi_\lambda$.
Then the map ${\bf f}\mapsto\varphi_\lambda({\bf f})$
is an injective mapping defined on the graded component
$\Gamma_\lambda/\Gamma'_\lambda$. 

We use $(x)_+={\rm max}(x,0)$.
Proposition \ref{prop:4.4} follows form the following proposition and 
the fermionic formula \eqref{resKos} for the restricted Kostka polynomial. 
\begin{prop}\label{FACTOR-res}
Let ${\bf f}\in\Gamma_\lambda/\Gamma'_\lambda$ and
$F=\varphi_\lambda({\bf f})$. 
The polynomial $F(v_1,\ldots,v_n)$ is divisible by
\begin{equation}
\prod_a v_a^{\lambda_a+(\lambda_a-\mu+1)_+}
\prod_{a>b}(v_a^2-v_b^2)^{\lambda_a}.
\end{equation}
\end{prop}

\begin{proof} 
{}From Proposition \ref{FACTOR} it is enough to prove that 
$F$ has the factor $v_a^{\alpha+(\alpha-\mu+1)_+}$
where $\alpha=\lambda_a$.
Because of (ii) in Lemma \ref{lem:orth}, it is enough to show that the function
\[
h(v)=\begin{cases}
g_{s,t}(v,\ldots;\underbrace{v,\ldots,v}_{\nu-1},\ldots)&\hbox{if }\alpha=2\nu-1;\\
g_{s,t}(\ldots;\underbrace{v,\ldots,v}_\nu,\ldots)&\hbox{if }\alpha=2\nu,
\end{cases}
\]
satisfies 
\bea 
\partial_v^jh(0)=0 \quad {\rm for} \,\,  0\leq j\leq\alpha-\mu.
\label{state:zero-factor} 
\ena 
If $\alpha<\mu$, there is nothing to show, and if $\alpha=\mu$,
the assertion follows from (v) in Lemma \ref{lem:orth-res}.
We consider the case $\alpha>\mu$ in the following. 

The following lemmas are straightforward.
\begin{lem}\label{2DIFF}
Let $g^{(0)}(x_1,x_2,x_3,x_4),g^{(1)}(x_1,x_2;y_1)$ and $g^{(2)}(y_1,y_2)$
be polynomials which are skew-symmetric in the variables $x_i$
and satisfy the relations
\begin{eqnarray*}
&&g^{(0)}(v_1,-v_1,v_2,-v_2)=g^{(1)}(v_2,-v_2;v_1)-g^{(1)}(v_2,-v_2;-v_1),\\
&&g^{(1)}(v_2,-v_2;v_1)=g^{(2)}(v_1,v_2)-g^{(2)}(v_1,-v_2).
\end{eqnarray*}
We have
\[
\frac{\partial^2g^{(2)}}{\partial v_1\partial v_2}(0,0)=0.
\]
\end{lem}
\begin{lem}\label{1DIFF}
Let $g^{(0)}(x_1,x_2,x_3)$ and $g^{(1)}(x_1;y_1)$
be polynomials which are skew-symmetric in the variables $x_i$
and satisfy the relation
\[
g^{(0)}(v_1,-v_1,v_2)=g^{(1)}(v_2;v_1)-g^{(1)}(v_2;-v_1).
\]
We have
\[
\frac{\partial g^{(1)}}{\partial v_1}(0;0)=0.
\]
\end{lem}
\begin{lem}\label{1ZERO}
Let $g^{(0)}(x_1,x_2)$ and $g^{(1)}(y_1)$
be polynomials satisfying the relations
\begin{eqnarray*}
g^{(0)}(v,-v)=g^{(1)}(v)-g^{(1)}(-v),\\
g^{(0)}(0,v)=g^{(0)}(v,0)=0
\end{eqnarray*}
We have
\[
\frac{\partial g^{(1)}}{\partial v}(0)=0.
\]
\end{lem}

Now let us prove \eqref{state:zero-factor}. 
Recall that $\alpha>\mu$. 
We divide the proof in four cases.
\medskip

{\it Case} $\alpha=2\nu,\mu=2\nu-2\kappa$.

\noindent We have
\begin{equation}\label{HEVEN}
h(v)=g_{s,t}(\ldots;\underbrace{v,\ldots,v}_\nu,\ldots).
\end{equation}
We use
\[
g_{s,t}(\ldots;y_1,\ldots,y_\kappa,
\underbrace{0,\ldots,0}_{\nu-\kappa}\ldots)=0.
\]
The derivative $\partial_v^jh(0)$ is a linear combination of
\begin{equation}\label{DER}
{\partial^jg_{s,t}}/{\partial y_1^{\delta_1}\cdots\partial y_\nu
^{\delta_\nu}}\mid_{y_1=\ldots=y_\nu=0},
\end{equation}
where $\delta_1\geq\delta_2\geq\cdots\geq\delta_\nu\geq0$.
By using $j\leq2\kappa$, one can check the following.\\
(i) Unless there are two or more $1$ in $\{\delta_1,\ldots,\delta_\nu\}$,
we have $\delta_i=0$ for $i>\kappa$, and therefore, (\ref{DER}) is zero;\\
(ii)If there are two or more $1$ in $\{\delta_1,\ldots,\delta_\nu\}$,
then by Lemma \ref{2DIFF} we can deduce that (\ref{DER}) is zero.
\medskip

{{\it Case}} $\alpha=2\nu+1,\mu=2\nu+1-2\kappa$.

\noindent We have
\begin{equation}\label{HODD}
h(v)=g_{s,t}(v,\ldots;\underbrace{v,\ldots,v}_\nu,\ldots).
\end{equation}
We use
\begin{eqnarray}
g_{s,t}(0,\ldots;y_1,\ldots,y_\kappa,
\underbrace{0,\ldots,0}_{\nu-\kappa}\ldots)=0,\label{G1}\\
g_{s,t}(x_1,\ldots;y_1,\ldots,y_{\kappa-1},
\underbrace{0,\ldots,0}_{\nu-\kappa+1}\ldots)=0.\label{G2}
\end{eqnarray}
The derivative $\partial_v^j h(0)$ is a linear combination of
\begin{equation}\label{DER2}
{\partial^jg_{s,t}}/\partial x_1^{\gamma_1}
\partial y_1^{\delta_1}\cdots\partial y_\nu
^{\delta_\nu}\mid_{x_1=y_1=\ldots=y_\nu=0},
\end{equation}
where $\delta_1\geq\delta_2\geq\cdots\geq\delta_\nu\geq0$.
By using $j\leq 2\kappa$, one can check the following.\\
(i) If $\gamma_1=0$, unless there are two or more $1$ in
$\{\delta_1,\ldots,\delta_\nu\}$, we have $\delta_i=0$ for $i>\kappa$,
and, therefore, because of (\ref{G1}), the derivative (\ref{DER2}) is zero;\\
(ii) If $\gamma_1=1$, unless there are one or more $1$ in
$\{\delta_1,\ldots,\delta_\nu\}$, we have $\delta_i=0$ for $i>\kappa-1$,
and, therefore, because of (\ref{G2}), the derivative (\ref{DER2}) is zero;\\
(iii) if $\gamma_1\geq2$, unless there are two or more $1$ in
$\{\delta_1,\ldots,\delta_\nu\}$, we have $\delta_i=0$ for $i>\kappa-1$,
and, therefore, because of (\ref{G2}), the derivative (\ref{DER2}) is zero;\\
(iv) If there are two or more $1$ in $\{\delta_1,\ldots,\delta_\nu\}$,
or if $\gamma_1=1$ and there are one or more $1$ in
$\{\delta_1,\ldots,\delta_\nu\}$, then by Lemmas \ref{2DIFF} and \ref{1DIFF}
we can deduce that the derivative (\ref{DER2}) is zero.
\medskip

{{\it Case}} $\alpha=2\nu,\mu=2\nu-(2\kappa-1)$.

\noindent We have (\ref{HEVEN}).
We use (\ref{G1}) and (\ref{G2}).
The derivative $\partial_v^j h(0)$ is a linear combination of (\ref{DER}).
By using $j\leq 2\kappa-1$, one can check the following.\\
(i) Unless there are one or more $1$ in
$\{\delta_1,\ldots,\delta_\nu\}$, we have $\delta_i=0$ for $i>\kappa-1$,
and, therefore, because of (\ref{G2}), the derivative (\ref{DER}) is zero;\\
(ii) If there are one or more $1$ in $\{\delta_1,\ldots,\delta_\nu\}$, then
because of (\ref{G1}) we can apply Lemma \ref{1ZERO} and deduce that
the derivative (\ref{DER}) is zero.
\medskip

{{\it Case}} $\alpha=2\nu+1,\mu=2(\nu-\kappa)$.

\noindent We have (\ref{HODD}). We use
\begin{equation}\label{LAST}
g_{s,t}(x_1,\ldots;y_1,\ldots,y_\kappa,
\underbrace{0,\ldots,0}_{\nu-\kappa}\ldots)=0.
\end{equation}
The derivative $\partial_v^j h(0)$ is a linear combination of (\ref{DER2}).
By using $j\leq 2\kappa+1$, one can check the following.\\
(i) If $\gamma_1=0$, unless there are one or more $1$ in
$\{\delta_1,\ldots,\delta_\nu\}$, we have $\delta_i=0$ for $i>\kappa$,
and, therefore, because of (\ref{LAST}), the derivative (\ref{DER2}) is zero;\\
(ii) If $\gamma_1\geq1$, unless there are two or more $1$ in
$\{\delta_1,\ldots,\delta_\nu\}$, we have $\delta_i=0$ for $i>\kappa$,
and, therefore, because of (\ref{LAST}), the derivative (\ref{DER2}) is zero;\\
(iii) If there are two or more $1$ in $\{\delta_1,\ldots,\delta_\nu\}$,
or if $\gamma_1=0$ and there are one or more $1$ in
$\{\delta_1,\ldots,\delta_\nu\}$, then by Lemmas \ref{2DIFF} and \ref{1DIFF},
we can deduce that the derivative (\ref{DER2}) is zero. 
\end{proof}

\subsection{Estimate from below}\label{subsec:crystal}
For the proof of Proposition \ref{prop:4.3}, 
we use the results of Kashiwara \cite{Kas} 
on the existence of global basis for level zero representations.
In this subsection we 
fix non-negative integers $m,k$ with $0\le m\le k$.

First we note a simple fact.
Take $a=(a_1,\cdots,a_N)\in (\C^{\times})^N$.  
Set $A=\mathbb{C}[q, q^{-1}]$, $\epsilon=\sqrt{-1}$, 
and 
\be
&&(\mathbb{V}_{K})_a=(\Vq)_{a_1}\otimes\cdots \otimes
(\Vq)_{a_N}, 
\\
&&
(\mathbb{V}_A)_a=(V_A)_{a_1}\otimes\cdots\otimes (V_A)_{a_N},
\\
&&(\mathbb{V}_\epsilon)_a
=(V_\epsilon)_{a_1}\otimes\cdots\otimes
 (V_\epsilon)_{a_N}.
\en

\begin{lem}\label{lem:4.2}
Let $Y_{k,m}$ denote the right ideal of $U_A$ generated by 
$f_1,(e_0^{(2)})^{\nu}$ if $k-m+1=2\nu$, 
$f_1,e_0(e_0^{(2)})^{\nu}, (e_0^{(2)})^{\nu+1}$ if $k-m+1=2\nu+1$.
Then we have 
\be
&&\dim_{\C} \bigl((\mathbb{V}_\epsilon)_a
/\bigl(Y_{k,m}(\mathbb{V}_\epsilon)_a\bigr)_m
\\
&&\ge 
\dim_{K} \bigl((\mathbb{V}_{K})_a
/\bigl(f_1(\mathbb{V}_{K})_a+e_0^{k-m+1}(\mathbb{V}_{K})_a\bigr)_m.
\en
\end{lem}
\begin{proof}
This follows from the specialization argument. 
In fact, 
\be
\dim_{\C}\bigl(Y_{k,m}(\mathbb{V}_\epsilon)_a\bigr)_m
&\le&
{\rm rank}_{A} \bigl(Y_{k,m}(\mathbb{V}_{A})_a\bigr)_m
\\
&=&
\dim_{K}\bigl(Y_{k,m}(\mathbb{V}_{K})_a\bigr)_m.
\en
Since $Y_{k,m}(\mathbb{V}_{K})_a=
f_1(\mathbb{V}_{K})_a+e_0^{k-m+1}(\mathbb{V}_{K})_a$, 
the assertion follows. 
\end{proof}

By Lemma \ref{lem:4.2}, the proof of Proposition \ref{prop:4.3}
is reduced to the following statement. 
\begin{prop}\label{prop:4.5}
\bea
&&\dim_{K} \bigl((\mathbb{V}_{K})_a
/\bigl(f_1 
(\mathbb{V}_{K})_a
+e_0^{k-m+1} (\mathbb{V}_{K})_a\bigr)\bigr)_m
=K^{(k)}_{m,(1^N)}(1).
\label{gencoinv}
\ena
\end{prop}
\begin{proof}
Let $\A$ be the subring of $K=\C(q)$ consisting of 
rational functions which are regular at $q=0$. 
The two-dimensional module $V_K$ is 
a level $0$ fundamental representation  
of $\U$ in the sense of \cite{Kas}. 
It has the crystal base $(L,B)$ with
$L=\A v_+\oplus \A v_-$ and $B=\{v_+,v_-\}\subset L/qL$. 
Let $(L^\aff, B^\aff)$ be the affinization,   
and set 
\be
&&\mathbb{L}=L^{\otimes N}, \quad \mathbb{B}=B^{\otimes N},
\\
&&\mathbb{L}^\aff=(L^\aff)^{\otimes N}, 
\quad \mathbb{B}^\aff=(B^\aff)^{\otimes N}.
\en
Let further $\cnorm$ be the involution 
introduced in \cite{Kas}(eq.(8.9) and a few lines above). 
In the below, 
we use the standard notation 
in crystal theory such as $\tilde{e}_i,\tilde{f}_i$,
$\varepsilon_i(b)=\max\{n\mid \tilde{e}_i^n(b)\neq 0\}$
and $\varphi_i(b)=\max\{n\mid \tilde{f}_i^n(b)\neq 0\}$
($i=0,1$). 
The following is a special case of the general results
proved in \cite{Kas} (Theorem 8.5 and Theorem 6.2).
\begin{enumerate}
\item 
For any $b\in{ \mathbb{L}^\aff}/q {\mathbb{L}^\aff}$, 
there is a unique element
$G(b)\in \mathbb{L}^\aff \cap (\mathbb{V}_A)^\aff$ 
such that $\cnorm G(b)=G(b)$ and
$G(b)\equiv b~\bmod q\mathbb{L}^\aff$. 
The set 
$\{G(b)\}_{b\in \mathbb{B}^\aff}$ is a 
$K$-basis of $\mathbb{V}^\aff$. 
\item For any $i=0,1$ and $n\ge 0$, we have 
\be
f_i^n \mathbb{V}^\aff=
\bigoplus_{b\in \mathbb{B}^\aff 
\atop \varepsilon_i(b)\ge n}\,K G(b),
\quad
e_i^n \mathbb{V}^\aff=
\bigoplus_{b\in \mathbb{B}^\aff 
\atop \varphi_i(b)\ge n}\,K G(b).
\en
\end{enumerate}

{}From the uniqueness of $G$, we have 
\be
G(z_\nu b)=z_\nu G(b)
\qquad (b\in \mathbb{B}^\aff).
\en
Let $a_1,\cdots,a_N\in K^{\times}$ 
be such that $a_\nu/a_{\nu+1}\in \A$ for 
$1\le\nu\le N-1$. 
Let 
$\psi:\mathbb{V}^\aff\longrightarrow \mathbb{V}_a$
be the canonical surjection. 
{}By the argument \cite{Kas} (in the proof of Theorem 9.1), 
$\{\psi(G(b))\}_{b\in \mathbb{B}}$ is a $K$-basis of 
$\mathbb{V}_a$. 
Since $\psi$ is $\U$-linear and 
$\psi(G(z_\nu b))=a_\nu\psi(G(b))$, 
we have for any $i=0,1$ and $n\ge 0$ 
\be
f_i^n \mathbb{V}_a
=\bigoplus_{b\in \mathbb{B} \atop \varepsilon_i(b)\ge n}
\,K \psi(G(b)),
\qquad
e_i^n \mathbb{V}_a
=\bigoplus_{b\in \mathbb{B} 
\atop \varphi_i(b)\ge n}\,K \psi(G(b)).
\en
Consider the subspace 
\be
f_1\mathbb{V}_a+e_0^{k-m+1}\mathbb{V}_a.
\en
{}From the reasoning above, it has the set 
$\{\psi(G(b))\mid b\in \mathbb{B},
~~ \epsilon_1(b)\ge 1~~\mbox{ or }~~
\varphi_0(b)\ge k-m+1\}$ as a basis.
Therefore the left hand side of \eqref{gencoinv}
is equal to 
\be
\sharp\{b\in \mathbb{B}\mid \tilde{e}_1b=0,~~
\tilde{f}_{0}^{k-m+1}b=0,~~
\mathop{\rm wt} b=m\}.
\en
It is well known (see e.g. \cite{HKKOTY}) 
that the last line
is given by the restricted Kostka number $K^{(k)}_{m,(1^N)}(1)$.
This proves
Proposition \ref{prop:4.5}.
\end{proof}


\section{Virasoro characters}\label{sec:Virch}

So far we have dealt with 
the space of minimal deformed cycles for each fixed $N$, 
assuming that they are polynomials in $z_1,\cdots,z_N$. 
In this section we consider the total space 
\be
&&M^{(r)}_m
=\bigoplus_{N\ge 0\atop N-2l=m}
M_{N,l}^{(r)}\otimes
\C[z_1^{\pm 1},\cdots,z_N^{\pm 1}]^{{\mathfrak S}_N}. 
\en
This space is $\Z$-graded 
with the assignment $\deg X_i=-1,\deg z_j=1$. 
However its character does not make sense
because each graded component is infinite dimensional.  
Following \cite{Kou}, we consider instead a family of 
subspaces for which the character is well-defined.
Recall that the axioms (A2),(A3) contain the parameter 
$\varepsilon=0,1$. 
For each $L=0,1,2,\cdots$ with $L\equiv \varepsilon\bmod 2$,  
define 
\bea
&&M^{(r)}_m[L]
=\{P\in M^{(r)}_m
\mid \bigl(\prod_{j=1}^Nz_j\bigr)^{\frac{L-\varepsilon}{2}}P
~\mbox{ is a polynomial }\},
\label{Mre}
\ena
so that $M^{(r)}_m =\cup_{L\equiv \varepsilon~\bmod 2}
M^{(r)}_m[L]$.
We use the grading defined for 
$P\in 
M_{N,l}^{(r)}\otimes
\C[z_1^{\pm 1},\cdots,z_N^{\pm 1}]^{{\mathfrak S}_N}$
as
\be
\deg' P=\deg P+\frac{N^2}{4}-\frac{\varepsilon N}{2}
+\frac{m(m+2)}{4r}
\qquad (m=N-2l).
\en
This is nothing but the degree 
of the $N$-particle minimal form factor $\Pc \tilde{f}_{P}$ 
of the RSG model, that is 
\be 
\tilde{f}_{P}(\beta_{1}+\Lambda , \ldots , \beta_{N}+\Lambda)= 
e^{\Lambda \deg'{P}}\tilde{f}_{P}(\beta_{1}, \ldots , \beta_{N}). 
\en 
The corresponding character is 
\be
\ch_{q} M^{(r)}_m[L]=
\sum_{N\ge 0}
\frac{q^{\frac{N^2}{4}+\frac{m(m+2)}{4r}-\frac{LN}{2}}}{(q)_N}
K^{(r-2)}_{m,(1^N)}(q).
\en
We show below (Theoerm \ref{thm:vir}) 
that it is expressible in terms of 
characters of the Virasoro algebra. 

Consider the irreducible characters of the Virasoro 
minimal unitary series 
\be
\chi^{(r,r+1)}_{b, a}(q)=q^{-\frac{c}{24}+h_{b, a}}
\chih^{(r,r+1)}_{b, a}(q).
\en
Here $1\le b\le r-1$, $1\le a\le r$, 
\be
&&c=1-\frac{6}{r(r+1)},
\quad h_{b, a}=\frac{((r+1)b-ra)^2-1}{4r(r+1)},
\en
and $\chih^{(r,r+1)}_{b,a}(q)=1+O(q)$.
Explicitly we have 
\bea
&&\chih^{(r,r+1)}_{b,a}(q)=
\frac{1}{(q)_\infty}
\label{virch}\\
&&\times
\Bigl(\sum_{n\in \Z}q^{r(r+1)n^2+((r+1)b-ra)n}
{}-\sum_{n\in \Z}q^{r(r+1)n^2+((r+1)b+ra)n+ba}\Bigr), 
\nn
\ena
where $(q)_{\infty}=\prod_{j=1}^{\infty}(1-q^{j})$. 
We shall use also their finitization due to \cite{ABF}. 
For $L\in\Z_{\ge 0}$ such that $L\equiv b-a~\bmod 2$,  
define a polynomial
\bea
\chih^{(r,r+1)}_{b,a}(q;L)
&=&
\sum_{n\in \Z}q^{r(r+1)n^2+((r+1)b-ra)n}
\qbin{L}{\displaystyle{\frac{L-b+a}{2}-(r+1)n}}
\nn\\
&&-\sum_{n\in \Z}q^{r(r+1)n^2+((r+1)b+ra)n+ba}
\qbin{L}{\displaystyle{\frac{L-b-a}{2}-(r+1)n}}.
\label{reg3}
\ena
We set $\chih^{(r,r+1)}_{b,a}(q;L)=0$ if 
$L\not\equiv b-a~\bmod 2$. 
In the notation of \cite{ABF} we have
$\chih^{(r,r+1)}_{b,a}(q;L)=q^{-(a-b)(a-b-1)/4}X_L(a,b,b+1)$, 
$r=r_{ABF}-1$, where $r_{ABF}$ signifies the parameter 
$r$ used in \cite{ABF}.
With the definition
\be
\chi^{(r,r+1)}_{b, a}(q;L)=q^{-\frac{c}{24}+h_{b, a}}
\chih^{(r,r+1)}_{b, a}(q;L),
\en
it is obvious that
\be
\lim_{L\rightarrow\infty\atop L\equiv a-b~\bmod 2}
\chi^{(r,r+1)}_{b,a}(q;L)=\chi^{(r,r+1)}_{b,a}(q).
\en

\begin{thm}\label{thm:vir}
Let $0\le m\le r-2$, $L\in\Z_{\ge 0}$. 
Then the following identity holds. 
\bea
&&\ch_{q} M^{(r)}_m[L]
=
\sum_{1\le a\le r\atop a\equiv L-1~\bmod 2}
\chi^{(r,r+1)}_{m+1,a}(q)
\chi^{(r,r+1)}_{1,a}(q^{-1};L).
\label{chiral}
\ena
\end{thm}
Note that 
\be
\chi^{(r,r+1)}_{1,a}(q^{-1};L)
=q^{\frac{c}{24}-h_{1a}-\left(\frac{L}{2}\right)^2+\left(
\frac{1-a}{2}\right)^2}
\times
K^{(r-1)}_{a-1,(1^L)}(q).
\en

For the proof we need
\begin{lem}\label{lem:vir}
For any $N\in\Z_{\ge 0}$ we have
\bea
\sum_{p\ge 0}\frac{(zq^{p+1};q)_\infty}{(q)_p}
z^pq^{p(p-N)}
=\sum_{s\ge 0}\qbin{N}{s}_{q^{-1}}z^s\,
\label{iden}
\ena
where $(z;q)_{n}=\prod_{j=1}^{n}(1-q^{j-1}z)$ and 
$\qbin{N}{s}_{q^{-1}}$ signifies $\qbin{N}{s}$ with $q$ replaced by $q^{-1}$. 
\end{lem}
\begin{proof}
Let $L_N(z)$ (resp. $R_N(z)$) stand for the left (resp. right) 
hand side.
It is straightforward to check that 
$L_N(z)=zL_{N-1}(z)+L_{N-1}(q^{-1}z)$ 
and the same relation for $R_N(z)$. 
Clearly $R_0(z)=1$.
That $L_0(z)=1$ follows from the identity
$\sum_{n\ge 0}q^{n(n-1)}z^n/((q)_n(z)_n)=1/(z)_\infty$. 
\end{proof}

\noindent
{\it Proof of Theorem \ref{thm:vir}.}\quad
We prove the assertion in the equivalent form
\bea
&&\sum_{N\ge 0}
\frac{q^{\frac{N^2-m^{2}}{4}-\frac{LN}{2}}}{(q)_N}
K^{(r-2)}_{m,(1^N)}(q)
\nn\\
&&
=\sum_{1\le a\le r\atop a\equiv L-1~\bmod 2}
\chih^{(r,r+1)}_{m+1,a}(q)q^{-\frac{(a-1)m}{2}}
\chih^{(r,r+1)}_{1,a}(q^{-1};L).
\label{id2}
\ena
Let us start from the right hand side. 
Substituting \eqref{reg3} and using the relations
\be
&&
\widehat{\chi}^{(r,r+1)}_{b,-a}(q)=-q^{-ba}\widehat{\chi}^{(r,r+1)}_{b,a}(q), 
\\
&&\widehat{\chi}^{(r,r+1)}_{b,a-2(r+1)n}(q)
=q^{-r(r+1)n^2-((r+1)b-ra)n}\widehat{\chi}^{(r,r+1)}_{b,a}(q), 
\\
&&\widehat{\chi}^{(r,r+1)}_{b,a}(q)=0\quad (a\in (r+1)\Z),
\en
we obtain 
\be
\sum_{a\in \Z\atop a\equiv L-1\bmod 2}
\widehat{\chi}^{(r,r+1)}_{m+1,a}(q)q^{-\frac{(a-1)m}{2}}
\qbin{L}{\frac{L+a-1}{2}}_{q^{-1}}.
\en
Applying \eqref{iden} we transform this expression into 
\be
&&
\sum_{n\in \Z\atop p\ge 0}\Bigl(
q^{rn^2+(m+1)n+(p-rn-\frac{m}{2})^2-\frac{m^2}{4}-L(p-rn-\frac{m}{2})}
\frac{1}{(q)_{p-2rn-m}(q)_p}
\\
&&-
q^{rn^2+(m+1)n+m+1+
(p+rn+\frac{m+2}{2})^2-\frac{(m+2)^2}{4}-L(p+rn+\frac{m+2}{2})}
\frac{1}{(q)_{p+2rn+m+2}(q)_p}
\Bigr).
\en
Here we set $1/(q)_a=0$ for a negative integer $a$.
Changing the variable $p$ to $N$ where
$p=\frac{N+m}{2}+rn$ for the first term
and $p=\frac{N-m-2}{2}-rn$ for the second, we obtain
\bea
\sum_{N\ge 0\atop N\equiv m\bmod 2}
\frac{q^{\frac{N^2-m^2}{4}-\frac{LN}{2}}}{(q)_N}
\sum_{n\in\Z}
q^{rn^2+(m+1)n}
K_{m+2rn, (1^{N})}(q), 
\label{aaa} 
\ena
where $K_{m+2rn, (1^{N})}(q)$ is the (non-restricted) Kostka polynomial 
given by \eqref{eq:formula-Kostka}. 
We now make use of the following alternating sum formula
for the restricted Kostka polynomial (\cite{SS}, eq.(6.8))
\bea
K^{(r-2)}_{m,\nu}(q) 
&=&\sum_{i\ge 0}q^{ri^2+(m+1)i}K_{2r i+m,\nu}(q) 
\label{alter} \\
&& 
{}-\sum_{i> 0}q^{r i^2-(m+1)i}
K_{2r i-m-2,\nu}(q) 
\nn 
\ena
and the property 
\be 
K_{m, (1^{N})}(q)=-K_{-m-2, (1^{N})}(q).  
\en 
Then \eqref{aaa} becomes 
the left hand side of \eqref{id2} and the proof is over.
\qed
\medskip

For small $L$, the right hand side of \eqref{chiral}
is given by
\be
L=0&:&q^{\frac{c}{24}} \chi^{(r,r+1)}_{m+1,1}(q),
\\
L=1&:&q^{\frac{c}{24}-h_{12}} \chi^{(r,r+1)}_{m+1,2}(q),
\\
L=2&:&q^{\frac{c}{24}} \chi^{(r,r+1)}_{m+1,1}(q)+
q^{\frac{c}{24}-h_{13}}\chi^{(r,r+1)}_{m+1,3}(q),
\\
L=3&:&q^{\frac{c}{24}-h_{12}} (1+q^{-2})\chi^{(r,r+1)}_{m+1,2}(q)+
q^{\frac{c}{24}-h_{14}}\chi^{(r,r+1)}_{m+1,4}(q),
\en
and so forth. 
As long as $0\le L \le r-1$, each time $L$ is increased, 
a new term $\chi^{(r,r+1)}_{m+1,L+1}(q)$ appears.
This agrees with the formula for
the two-particle form factors of 
the exponential operator \cite{L}.

If the argument $q^{-1}$ in the second factor
in \eqref{chiral} were an independent variable $\bar{q}$, then 
in the limit $L \rightarrow\infty$ we would obtain
\be
\sum_{1\le a\le r\atop a\not\equiv \varepsilon~\bmod 2}
\chi^{(r,r+1)}_{m+1,a}(q)
\chi^{(r,r+1)}_{1,a}(\bar{q})
\qquad (\varepsilon =0, 1),
\en
a structure reminiscent of a 
modular invariant partition function of CFT.   
In the massive theory the two chiralities are not separated. 
This point was first observed in \cite{Kou}
in a few simple cases including the non-unitary models
$(p,p')=(2,5),(3,5)$.

\appendix 

\section{Quantum affine algebra}\label{app:aff} 
We summarize here our convention 
concerning the 
quantum loop algebra $\U_q(\widetilde{\mathfrak{sl}}_2)$.  
Let $K=\C(q)$ be the field of rational functions 
in indeterminate $q$. 
The 
quantum loop algebra $\U=\U_q(\widetilde{\mathfrak{sl}}_2)$ 
is a Hopf algebra over $K$ generated by $e_i,f_i,t_i$ ($i=0,1$) 
under the following defining relations. 
\be
&&t_0t_1=t_1t_0=1,
\\
&&
t_ie_jt_i^{-1}=q^{a_{ij}}e_j,
\quad
t_if_jt_i^{-1}=q^{-a_{ij}}f_j,
\\
&&[e_i,f_j]=\delta_{ij}\frac{t_i-t_i^{-1}}{q-q^{-1}},
\\
&&e_i^{(3)}e_j-e_i^{(2)}e_je_i+e_ie_je_i^{(2)}-e_je_i^{(3)}=0
\qquad(i\neq j),
\\
&&f_i^{(3)}f_j-f_i^{(2)}f_jf_i+f_if_jf_i^{(2)}-f_jf_i^{(3)}=0
\qquad(i\neq j).
\en
Here $a_{ii}=2$, $a_{ij}=-2$ ($i\neq j$), and we have set  
\be
&&x^{(n)}=\frac{x^n}{\qi{n}!},
\quad 
\qi{n}!=\prod_{j=1}^n\qi{j},
\quad 
\qi{j}=\frac{q^j-q^{-j}}{q-q^{-1}}.
\en
We choose the coproduct
\bea
&&
\Delta(e_i)=e_i\otimes 1+t_i\otimes e_i,
\quad
\Delta(f_i)=f_i\otimes t_i^{-1}+1\otimes f_i,
\quad \Delta(t_i)=t_i\otimes t_i.
\label{copro} 
\ena

The algebra $\U$ has an alternative presentation in terms 
of the Drinfeld generators
$x^{\pm}_k$ ($k\in\Z$), $a_n$ ($n\in\Z\backslash\{0\}$) 
and $t_1^{\pm 1}$. 
They satisfy the relations 
\bea
&&[t_1,a_n]=0,\quad [a_m,a_n]=0,
\label{Dr1}\\
&&
t_1x^{\pm}_kt_1^{-1}=q^{\pm 2}x^{\pm}_k,
\label{Dr2}\\
&&[a_n,x_k^{\pm}]=\pm\frac{\qi{2n}}{n}x^{\pm}_{k+n},
\label{Dr3}\\
&&x^{\pm}_{k+1}x^{\pm}_l-q^{\pm 2}x_l^{\pm}x_{k+1}^\pm
=q^{\pm 2}x^{\pm}_{k}x^{\pm}_{l+1}-x_{l+1}^{\pm}x_{k}^\pm,
\label{Dr4}\\
&&[x^+_k,x_l^-]=
\frac{\varphi^+_{k+l}-\varphi^{-}_{k+l}}{q-q^{-1}},
\label{Dr5}
\ena
where
\be
\sum_{k\ge 0}\varphi^{\pm}_{\pm k}z^k
=t_1^{\pm 1}\exp\left(\pm(q-q^{-1})\sum_{n=1}^\infty
a_{\pm n}z^n\right)
\en
and $\varphi^{\pm}_{\pm k}=0$ for $k<0$.
The two sets of generators are related by
\be
x_0^+=e_1,~~
x_1^-=e_0t_1,~~
x_0^-=f_1,~~
x_{-1}^+=t_1^{-1}f_0.
\en

We shall deal also with the algebra with $q$ 
specialized to a complex number. 
Let $\Ur$ be the subalgebra of $\U$ generated by the elements 
$e_i^{(s)}$, $f_i^{(s)}$ ($i=0,1$, $s\in\Z_{\ge 0}$) 
and $t_1^{\pm 1}$ over $A=\C[q,q^{-1}]$.  
$\Ur$ is also generated by 
the elements $(x^{\pm}_k)^{(s)}$ ($k\in \Z$, $s\in\Z_{\ge 0}$)
and $t_1^{\pm 1}$ \cite{CP}. 
For a non-zero complex number $\epsilon\in\C$, 
consider the ring homomorphism $A\rightarrow\C$ 
which sends $q$ to $\epsilon$. 
We define $\Ure=\Ur\otimes_{A}\C$. 
Specialization of modules is defined in a similar manner. 
Let $W$ be a $\U$-module equipped with a 
free $A$-submodule $W_A$ satisfying 
$\Ur W_A\subset W_A$ and $W=W_A\otimes_{A}K$.   
We say that $W$ is defined over $A$, 
and that the $\Ure$-module $W_A\otimes_{A}\C$ 
is the specialization of $W$ to $\epsilon$. 

In this paper we shall mainly consider the modules
\be
&&\Vq=K v_+\oplus K v_-, 
\quad \Vq^\aff=\Vq\otimes K[z^{\pm 1}], 
\\
&&V_A=A v_+\oplus A v_-,
\quad V_A^\aff=V_A\otimes A[z^{\pm 1}], 
\\
&&V=\C v_+\oplus \C v_-,
\quad V^\aff=V\otimes \C[z^{\pm 1}]. 
\en
The space $\Vq^\aff$ is a module 
over $\U\otimes K[z^{\pm 1}]$ 
defined by the assignment 
\be
e_0\mapsto z\sigma^-,~~e_1\mapsto\sigma^+,~~
f_0\mapsto z^{-1}\sigma^+,~~f_1\mapsto \sigma^-,~~
t_1\mapsto \tau,  
\en
where $\sigma^{a}$ ($a=\pm, z$), $\tau$ denote 
linear operators given in the above basis by 
\be
\sigma^+=\begin{pmatrix}0&1\\0&0\\\end{pmatrix},
~~
\sigma^-=\begin{pmatrix}0&0\\1&0\\\end{pmatrix},
~~
\sigma^z=\begin{pmatrix}1&0\\0&-1\\\end{pmatrix},
~~
\tau=\begin{pmatrix}q&0\\0&q^{-1}\\\end{pmatrix}.
\en
We identify the $N$-fold tensor product 
module $(\Vq^\aff)^{\otimes N}$ with 
$\Vq^{\otimes N}\otimes K[z_1^{\pm1},\cdots,z_N^{\pm1}]$. 
Similarly $(V_A^\aff)^{\otimes N}$ 
(resp. $(V^\aff)^{\otimes N}$) is a module over
$\Ur\otimes A[z_1^{\pm1},\cdots,z_N^{\pm1}]$ 
(resp. $\Ure\otimes \C[z_1^{\pm1},\cdots,z_N^{\pm1}]$). 
For a $\U$-module $W$, 
we write $W[m]=\{v\in W\mid t_1v=q^hv\}$ and call it  
the subspace of weight $m$.  
This gives rise to a grading on 
$(V^\aff)^{\otimes N}$ 
and $(V_\epsilon)^{\otimes N}$ for any $\epsilon\in\C^\times$. 
Fixing $N$, we often write 
$(V_\epsilon)^{\otimes N}[N-2l]$
as 
$((V_\epsilon)^{\otimes N})_l$.


\section{Fermionic realization}\label{app:JW}

In this appendix we give the details about 
Proposition \ref{prop:fermi}.

Introduce the Jordan-Wigner fermions 
\be
&&\psi^{*}_a=\sigma_a^+(-i\sigma^z_{a+1})\cdots(-i\sigma^z_N),
\\
&&
\psi_a=\sigma_a^-(i\sigma^z_{a+1})\cdots(i\sigma^z_N).
\en
Here the subscript $a=1,\cdots,N$ of 
$\sigma^{\bullet}_a$ indicates that it acts 
on the $a$-th tensor component of $V^{\otimes N}$.
We have $[\psi_a,\psi_b]_+=[\psi^*_a,\psi^*_b]_+=0$, 
$[\psi^*_a,\psi_b]_+=\delta_{ab}$ ($1\le a,b\le N$)
and $\sigma^z_a=-2\psi_a\psi^{*}_a+1$. 
We make an identification 
$\Lambda_N \simeq V^{\otimes N}$.  
\medskip

\noindent{\it Proof of Proposition \ref{prop:fermi}.}\quad
Set $\varepsilon=-i(q-i)+(q-i)^2/2$, so that 
$q^{\pm 1}\equiv\pm i(1\pm\varepsilon+\varepsilon^2/2)$ 
$\bmod (q-i)^3$. 
Using the same letter $\vpi$ to denote 
the representation $\Ur\rightarrow 
\End_A((V_A^\aff)^{\otimes N})$, we set 
\be
&&\vpi(iq^{-1}a_1)\equiv
\alpha_0+\varepsilon\alpha_1+\varepsilon^2\alpha_2
~~\bmod~~\varepsilon^3,
\\
&&\vpi(x^{-}_0)\equiv
\beta_0+\varepsilon\beta_1~~\bmod~~\varepsilon^2,
\\
&&\vpi(\Xc(z))\equiv
\gamma_0(z)+\varepsilon\gamma_1(z)
~~\bmod~~\varepsilon^2.
\en
Explicitly we have 
\be
&&
\alpha_0=\sum_{a=1}^Nz_a,
\quad
\alpha_1=-\sum_{a=1}^Nz_a\sigma_a^z+
4\sum_{a<b}z_a\psi_a\psi^{*}_b,
\\
&&
\alpha_2-\frac{1}{2}\alpha_0
=4\sum_{a<b<c}z_a\psi_a\sigma^z_b\psi^{*}_c,
\\
&&\beta_0=\sum_{a=1}^N(-1)^{N-a}\psi_a,
\quad
\beta_1=
\sum_{a<b}(-1)^{N-a+1}\psi_a\sigma^z_b.
\en
Substituting these into the relation 
\be
x^-_{k+1}=-\frac{1}{\qi{2}}[a_1,x^-_k],
\en
we obtain 
\bea
&&2\gamma_0(z)=z[\alpha_1,\beta_0+\gamma_0(z)],
\label{gam1}\\
&&2\gamma_1(z)=z\left([\alpha_1,\beta_1+\gamma_1(z)]
+[\alpha_2,\beta_0+\gamma_0(z)]\right).  
\label{gam2}
\ena
The formal series $\gamma_0(z)$, $\gamma_1(z)$ are uniquely 
determined by these relations.
We seek them in the form 
\be
&&\gamma_0(z)=\sum_{a=1}^NA_a(z)\psi_a,
\\
&&\gamma_1(z)=\sum_{a<b}B_{ab}(z)\psi_a\sigma^z_b
+\sum_{a<b<c}C_{abc}(z)\psi_a\psi_b\psi^*_c.
\en
Then \eqref{gam1},\eqref{gam2} are rewritten as the relations
for the coefficients:
\be
&&(1-z_az)A_a(z)=z_az(1+2\sum_{b=a+1}^NA_b(z)),
\\
&&
(1-z_az)B_{ab}(z)=2z_az\sum_{a<c<b}B_{cb}(z)
+\frac{z_a(1-z_bz)}{z_b}A_b(z) \qquad (a<b), 
\\
&&(1-z_az-z_bz+z_cz)C_{abc}(z)=
4z_az(A_b(z)+B_{bc}(z))
+4z_bz(B_{ab}(z)-B_{ac}(z))
\\
&&\qquad+2z_az\sum_{a<p<b}C_{pbc}(z)-2z_az\sum_{b<p<c}C_{bpc}(z)
\\
&&\qquad +
2z_bz\sum_{b<p<c}C_{apc}(z)-\sum_{b<p<c}2z_pzC_{abp}(z) \qquad (a<b<c). 
\en
A direct computation shows that the following 
is the solution. 
\be
&&
A_a(z)=\frac{z_az}{1-z_az}\prod_{j=a+1}^{N}\frac{1+z_jz}{1-z_jz},
\\
&&
B_{ab}(z)=\frac{z_az}{1-z_az}
\prod_{a<j\le N\atop j\neq b}\frac{1+z_jz}{1-z_jz} \quad (a<b), 
\\
&&C_{abc}(z)=8\frac{z_az}{1-z_az}
\Bigl(\prod_{a<j<b}\frac{1+z_jz}{1-z_jz}\Bigr)\frac{z_bz}{1-z_bz}
\frac{1}{1-z_cz}
\prod_{c<j\le N}\frac{1+z_jz}{1-z_jz} \quad (a<b<c). 
\en
Calculating 
\be
&&
\vpi((x^-_0)^{(2)})
\equiv-\frac{i}{2}[\beta_0,\beta_1]_+~~\bmod\varepsilon, 
\\ 
&&
\vpi((\Xc^-(z))^{(2)})
\equiv-\frac{i}{2}[\gamma_0(z),\gamma_1(z)]_+~~\bmod\varepsilon,
\en
we arrive at the desired formulas. 
\qed

\medskip

\noindent{\it Remark.}\quad
In a similar manner, one can show that 
the elements $a_n$
 act as multiplication by a scalar,  
\be
a_n=
\begin{cases}
\frac{i^{n-1}}{n}\sum_{j=1}^Nz_j^n & (\mbox{$n$: odd}),\\
0& (\mbox{$n$: even}).\\
\end{cases}
\en
The action of $a_n$ with odd $n$ coincide with 
that of integrals of motion of the SG model. 
\qed


\section{Trigonometric hypergeometric space}\label{app:TrigHGS}
In this appendix we give an account of the connection between
the space of polynomials $W_N=\oplus_{l=0}^{N}W_{N,l}$ 
and the tensor product 
$\Vq^{\otimes N}\otimes K[z_1^{\pm 1},\cdots,z_N^{\pm 1}]$
specialized to $q=\sqrt{-1}$. 
Our exposition is based on \cite{TV} with a slight modification.  

Let us recall a well-known construction in the algebraic 
Bethe Ansatz.
Consider the $R$ matrix
\be
R(z)=\begin{pmatrix}
1      &                       &                     & \\
       &\ds{\frac{(1-z)q}{1-q^2z}}&\ds{\frac{1-q^2}{1-q^2z}} & \\
       &\ds{\frac{(1-q^2)z}{1-q^2z}}&\ds{\frac{(1-z)q}{1-q^2z}}& \\
       &                       &                     &1 \\
\end{pmatrix}.
\en
Let $R_{ij}(z)$ stand for the matrix 
$R(z)$ acting on the ($i,j$)-th tensor 
component of $\Vq^{\otimes (N+1)}$ ($0\le i,j\le N$). 
Define operators $A(z),B(z),C(z),D(z)$ by 
\be
\begin{pmatrix}
A(z)&B(z)\\
C(z)&D(z)\\
\end{pmatrix}
=\frac{\Theta(q^{-2}z^{-1})}{1-q^{-2}}
R_{0N}(z/z_N)\cdots R_{01}(z/z_1),
\en
where $\Theta(X)=\prod_{j=1}^N(1-z_jX)$ is given by \eqref{def:thetaX}. 
Then $C(z)$ is a polynomial in $z^{-1}$ of degree $N-1$. 
It is also a polynomial in $z_1,\cdots,z_N$ and 
a Laurent polynomial in $q$. 
We take the basis $\{v_+^*,v_-^*\}\subset V_{K}^*$ 
dual to $\{v_+,v_-\}$,  
and set for $v\in \bigl(V_{K}^{\otimes N}\bigr)_l$ 
\bea
\Cc'(v)=
\langle v_+^*\otimes\cdots\otimes v_+^*,
C(X_1^{-1})\cdots C(X_l^{-1})v\rangle.
\label{Cc}
\ena
Since $C(z)C(w)=C(w)C(z)$, 
the right hand side is symmetric in $X_1,\cdots,X_l$. 
For the vectors \eqref{vM} we have
\bea
\Cc'(v_M)=\Sym\left(
G'_{m_1}(X_1)\cdots G'_{m_l}(X_{l})
\prod_{j<j'}\frac{q^{-1}X_j-qX_{j'}}{X_j-X_{j'}}\right),
\label{CvM}
\ena
where
\be
G'_m(X)=q^{m-N}
\prod_{j=1}^{m-1}(1-q^{-2}z_jX)\prod_{j=m+1}^N(1-z_jX).
\en

Let $F_{N,l}$ denote the space of symmetric
polynomials $P$ in $X_1,\cdots,X_l$ with coefficients in
$K[z_1^{\pm 1},\cdots,z_N^{\pm 1}]$,
which have degree at most $N-1$ in each $X_i$ and
satisfies the condition 
\be
P(z_{k}^{-1}, q^{2}z_{k}^{-1}, X_{3}, \cdots , X_{l})=0 \quad 
\mbox{for} \quad k=1, \ldots , N. 
\en
{}From \eqref{CvM}, it is easy to see that $\Cc'(v_M)\in F_{N,l}$.
Set $F_N=\oplus_{l=0}^N F_{N,l}$. 
The space $F_N$ is a version of the trigonometric
hypergeometric space introduced in \cite{TV}.

Extending the definition \eqref{Cc} by linearity,
we obtain a map
\bea
\Cc':\Vq^{\otimes N}\otimes K[z_1^{\pm 1},\cdots,z_N^{\pm 1}]
\longrightarrow  F_N.
\label{Cc2}
\ena

\begin{prop}$($cf. \cite{TV,T}$)$
The space $F_N$ is endowed with the structure of a $\U$-module
such that \eqref{Cc2} is an intertwiner.
The action of $\U$ on $P\in F_{N,l}$ is given as follows:
\be
&&(e_1 P)(X_1,\cdots,X_{l-1})=q^{N-l}P(X_1,\cdots,X_{l-1},0),
\\
&&
(f_0 P)(X_1,\cdots,X_{l-1})=(-1)^{N-1}
q^{2N-l}\prod_{j=1}^Nz_j^{-1}
\\
&&\quad
\times
(z^{N-1}P(X_1,\cdots,X_{l-1},z^{-1}))\bigl|_{z=0},
\\
&&
(f_1P)(X_1,\cdots,X_{l+1})=
\frac{q^{-N+l}}{1-q^{-2}}
\sum_{\nu=1}^{l+1}P(X_1,\overset{\overset{\nu}{\smallfrown}}{\cdots},X_{l+1})
\\
&&\times
\Bigl(q^{N}
\Theta(q^{-2}X_\nu)\prod_{1\le k\le l+1\atop k\neq \nu}
\frac{q^{-2}X_k-X_\nu}{X_k-X_\nu}
{}-q^{-N}
\Theta(X_\nu)\prod_{1\le k\le l+1\atop k\neq \nu}
\frac{q^{2}X_k- X_\nu}{X_k-X_\nu}\Bigr),
\\
&&
(e_0 P)(X_1,\cdots,X_{l+1})=
\frac{q^{-N+l}}{1-q^{-2}}\sum_{\nu=1}^{l+1}
P(X_1,\overset{\overset{\nu}{\smallfrown}}{\cdots},X_{l+1})
\\
&&
\times
X_\nu^{-1}
\Bigl(\Theta(q^{-2}X_\nu)\prod_{1\le k\le l+1\atop k\neq \nu}
\frac{X_k-q^2X_\nu}{X_k-X_\nu}
{}-\Theta(X_\nu)\prod_{1\le k\le l+1\atop k\neq \nu}
\frac{X_k-q^{-2} X_\nu}{X_k-X_\nu}\Bigr),
\\
&&
(t_1P)(X_1,\cdots,X_l)=q^{N-2l}P(X_1,\cdots,X_l),
\en
and $e_1P=f_0P=0$ for $l=0$.
\end{prop}

\begin{prop}\label{prop:divpow}
Set
\be
F_{A,N,l}:=F_{N,l}\cap
A[z_1^{\pm 1},\cdots,z_N^{\pm 1}][X_1,\cdots,X_l],
\en
and $F_{A,N}:=\oplus_{l=0}^NF_{A,N,l}$.
Then for $i=0,1$ and $s\ge 0$ we have
\be
e_i^{(s)}F_{A,N}\subset F_{A,N},\quad
f_i^{(s)}F_{A,N}\subset F_{A,N}.
\en
\end{prop}
\begin{proof}
Take $P\in F_{A,N,l}$.
We are to show that,  for any $s\ge 0$ and $x=e_i,f_i$,
$x^sP$ is divisible by
$\prod_{j=1}^s(q^j-q^{-j})/(q-q^{-1})$.
For $x=f_1$ or $e_0$, this follows from the
formula for the action given above and the identity
\be
\Sym\prod_{1\le j<j'\le s}\frac{q^{-1}X_j-qX_{j'}}{X_j-X_{j'}}
=\prod_{j=1}^s\frac{q^j-q^{-j}}{q-q^{-1}}.
\en
The assertion for $x=e_1$ or $f_0$
can be shown from the Lemma below by choosing $t=q^2$ and
specializing $X=0$ or $X=\infty$.
\end{proof}

\begin{lem}
Let $Q(X_1,\cdots,X_l)$ be a symmetric polynomial in
$X_1,\cdots,X_l$ with coefficients in $\C[z_1,\cdots,z_N,t]$,
which has degree at most $N-1$ in each $X_i$.
Assume further that
$Q(z_{k}^{-1}, tz_{k}^{-1}, X_{3}, \cdots , X_{l})=0$ 
for $k=1, \ldots , N$. 
Then, for any $2\le m\le l$,
$Q(X,tX,\cdots,t^{m-1}X,X_{m+1},\cdots,X_l)$ is divisible by
$\prod_{j=1}^m(t^j-1)/(t-1)$.
\end{lem}
\begin{proof}
Let $\epsilon$ be a primitive $d$-th root of unity 
with $2\le d\le m$, and consider
\bea
Q(X,tX,\cdots,t^{d-1}X,X_{d+1},\cdots,X_l)\bigl|_{t=\epsilon}.
\label{Qt}
\ena
Viewed as a polynomial of $X$,
\eqref{Qt} has degree at most $d(N-1)$.
By the assumption it has $dN$ distinct zeroes
$\epsilon^{j}z_{k}^{-1}$ ($0\le j\le d-1$, $1\le k\le N$).
Hence \eqref{Qt} vanishes identically.

More generally,
\be
Q(X_1,tX_1,\cdots,t^{d-1}X_1,\cdots,
X_r,tX_r,\cdots,t^{d-1}X_r,X_{rd+1},\cdots,X_l)
\en
has $r$-fold zeroes at $t=\epsilon$,
as shown by differentiation with respect to $t$.
Therefore, if $rd\le m$, then 
\be
Q(X,tX,\cdots,t^{m-1}X,X_{m+1},\cdots,X_l)
\en
is divisible by $\varphi_d(t)^r$, 
where $\varphi_d(t)$
stands for the $d$-th cyclotomic polynomial.
Noting that
\be
\prod_{j=1}^m\frac{t^j-1}{t-1}=\prod_{2\le j\le m}
\prod_{d|j\atop 2\le d\le m}\varphi_d(t)
=\prod_{2\le d\le m}\varphi_d(t)^{[m/d]},
\en
where $[x]$ is the greatest integer not exceeding $x$,
we obtain the assertion.
\end{proof}

Let $\Ur^+$ be the subalgebra of $\Ur$
generated by $e_0^{(s)},e_1^{(s)},f_1^{(s)}$ ($s\ge 0$)
and $t_i^{\pm 1}$.
Let further $W'_{A,N}$ denote the subspace
of $F_{A,N}$ consisting of elements $P$ which are
symmetric polynomials in $z_1,\cdots,z_N$
and satisfy the condition
\bea
P=0\quad\mbox {if}\quad X_1^{-1}=q^{-2}z_{N-1}=z_{N}.
\label{xzz}
\ena

\begin{lem}\label{lem:a1}
\be
\Cc'(\Ur^+ v_+^{\otimes N})\subset W'_{A,N}.
\en
\end{lem}
\begin{proof}
By Proposition \ref{prop:divpow},
we have $\Ur^+ F_{A,N}\subset F_{A,N}$.
Since $\Cc'(v_+^{\otimes N})=1\in F_{A,N}$,
$x\cdot v_+^{\otimes N}$ ($x\in \Ur^+$)
is a polynomial in
$z_1,\cdots,z_N$ with coefficients in $A$.
If we write
\be
C(z)=C(z|z_1,\cdots,z_N) \quad {\rm and} \quad 
Y_i=P_{i\,i+1}R_{i\,i+1}(z_i/z_{i+1}),
\en
then we have
\be
Y_iC(z|\cdots,z_i,z_{i+1},\cdots)Y_i^{-1}
=C(z|\cdots,z_{i+1},z_i,\cdots).
\en
The operator $Y_i:\Vq[z_i, z_i^{-1}]\otimes\Vq[z_{i+1},
z_{i+1}^{-1}]
\rightarrow
\Vq[z_{i+1},z_{i+1}^{-1}]\otimes\Vq[z_{i},z_i^{-1}]$
commutes with the action of $\U$, 
and
leaves $v_+^{*\otimes N}$ and $v^{\otimes N}_+$ invariant.
The symmetry in $z_1,\cdots,z_N$
follows from these properties.
The condition \eqref{xzz} is a consequence of the property
$\langle v_+^*\otimes\cdots\otimes v_+^*,
C(z|\cdots,q^2z,z)v\rangle=0$
for any $v$, which can be verified
easily from the definition.
\end{proof}

When $q$ is specialized to $\sqrt{-1}$,
the image of $\Cc'$ becomes divisible by
$\prod_{j<j'}(X_j+X_{j'})$, as is seen from \eqref{CvM}.
Redefining
\be
\Cc(v)=\prod_{j<j'}\frac{X_j-X_{j'}}{i(X_j+X_{j'})}
\cdot \Cc'(v)
\qquad(v\in (\Vq^{\otimes N})_l)
\en
we obtain the map \eqref{Cc0}.
Proposition \ref{prop:2.2} follows from Lemma \ref{lem:a1}.


\setcounter{equation}{0}
\section{Representations of $U_\epsilon(\slt)$ at roots of $1$}
\label{app:goodbad}
We collect here some facts 
about representations of $U_\epsilon(\slt)$ used
in the text. 
Recall that $\epsilon=e^{-\pi i /r}$ with $r\ge 3$. 

The $U_\epsilon(\slt)$-modules 
$V^s(\alpha),X^s(\alpha)$ ($0\le s\le r-2$, $\alpha=\pm 1$)
and $W^s(\alpha)$ ($0\le s\le r-1$, $\alpha=\pm 1$)
are defined as follows. 
The module $V^s(\alpha)$ has basis $\{v^s_k(\alpha)\}_{0\le k\le s}$
with the action of $\Ure(\slt)$ given by 
\be
&&E v^s_k(\alpha)=\alpha
\qi{k}\qi{s+1-k}v^s_{k-1}(\alpha), 
\\
&&F v^s_k(\alpha)=v^s_{k+1}(\alpha), \quad
T v^s_k(\alpha)=\alpha \epsilon^{s-2k}v^s_k(\alpha),
\en
where $\qi{n}=\frac{\epsilon^{n}-\epsilon^{-n}}{\epsilon-\epsilon^{-1}}$ 
and $v^s_{-1}(\alpha)=v^s_{s+1}(\alpha)=0$. 
We abbreviate $V^s(1)$ to $V^s$ and $V^{1}$ to $V$.

The module $W^s(\alpha)$ has basis $\{w^s_k(\alpha)\}_{0\le k\le r-1}$
with the action of $\Ure(\slt)$ given by 
\be
&&E w^s_k(\alpha)=\alpha
\qi{k}\qi{s+1-k} w^s_{k-1}(\alpha), 
\\
&&F w^s_k(\alpha)=w^s_{k+1}(\alpha), \quad
T w^s_k(\alpha)=\alpha \epsilon^{s-2k}w^s_k(\alpha),
\en
where $w^s_{-1}(\alpha)=w^s_{r}(\alpha)=0$. 
The module $X^s(\alpha)$ has basis 
\be 
\{x^s_k(\alpha),y^s_k(\alpha)\}_{0\le k\le s}\cup
\{a^s_k(\alpha),b^s_k(\alpha)\}_{0\le k\le r-2-s}, 
\en 
and the action of $\Ure(\slt)$ is given as follows. 
\be
&&
E x^s_k(\alpha)=\alpha
\qi{k}\qi{s+1-k} x^s_{k-1}(\alpha)
\quad (0\le k\le s), 
\\
&&E y^s_k(\alpha)=
\begin{cases}
\alpha
\qi{k}\qi{s+1-k} y^s_{k-1}(\alpha)&\quad (1\le k\le s), \\
a^s_{r-2-s}(\alpha) & (k=0),\\
\end{cases}
\\
&&E a^s_k(\alpha)=-\alpha
\qi{k}\qi{r-1-s-k} a^s_{k-1}(\alpha)
\quad (0\le k\le r-2-s),  
\\
&&E b^s_k(\alpha)=
\begin{cases}
{}-\alpha
\qi{k}\qi{r-1-s-k} b^s_{k-1}(\alpha)
+a^s_{k-1}(\alpha) & (1\le k\le r-2-s),\\
x^s_s(\alpha) & (k=0),\\
\end{cases} 
\\
&&F x^s_k(\alpha)=x^s_{k+1}(\alpha),
 \quad
F y^s_k(\alpha)=y^s_{k+1}(\alpha)
\quad (0\le k\le s), 
\\
&&F a^s_k(\alpha)=a^s_{k+1}(\alpha),
 \quad
F b^s_k(\alpha)=b^s_{k+1}(\alpha)
\quad (0\le k\le r-2-s),  
\\
&&
T x^s_k(\alpha)=\alpha \epsilon^{s-2k}x^s_k(\alpha),
\quad
T y^s_k(\alpha)=\alpha \epsilon^{s-2k}y^s_k(\alpha)
\quad (0\le k\le s), 
\\
&&
T a^s_k(\alpha)=-\alpha \epsilon^{r-2-s-2k}a^s_k(\alpha),
\quad
T b^s_k(\alpha)=-\alpha \epsilon^{r-2-s-2k}b^s_k(\alpha)
\quad (0\le k\le r-2-s).
\en
Here we have set 
$x^s_{-1}(\alpha)=a^s_{-1}(\alpha)=0$, 
$x^s_{s+1}(\alpha)=a^s_{0}(\alpha)$, 
$y^s_{s+1}(\alpha)=0$, 
$a^s_{r-1-s}(\alpha)=0$, 
$b^s_{r-1-s}(\alpha)=y^s_0(\alpha)$.  

The modules $V^s(\alpha)$ ($0\le s\le r-2$) 
and $W^{r-1}(\alpha)$ are irreducible.
The others are indecomposable and we have 
\be
&&0\rightarrow V^{r-2-s}(-\alpha)\rightarrow
W^s(\alpha)\rightarrow V^s(\alpha)\rightarrow 0
\qquad (0\le s\le r-2),
\\
&&0\rightarrow W^{s}(\alpha)\rightarrow
X^s(\alpha)\rightarrow W^{r-2-s}(-\alpha)\rightarrow 0 
\qquad (0\le s\le r-2).
\en

Upon tensoring with $V$, 
these modules decompose as \cite{RT} 
\bea
&&V^s(\alpha)\otimes V=V^{s+1}(\alpha)\oplus V^{s-1}(\alpha),
\quad (0\le s\le r-2),
\label{dcp1}\\
&&W^s(\alpha)\otimes V=W^{s+1}(\alpha)\oplus W^{s-1}(\alpha),
\quad (0\le s\le r-2),
\label{dcp2}\\
&&W^{r-1}(\alpha)\otimes V=X^0(\alpha),
\label{dcp3}\\
&&X^s(\alpha)\otimes V=X^{s+1}(\alpha)\oplus X^{s-1}(\alpha),
\quad (0\le s\le r-2).
\label{dcp4}
\ena
In the above, 
we set 
\be
&&V^{-1}(\alpha)=0,\quad V^{r-1}(\alpha)=W^{r-1}(\alpha), 
\\
&&W^{-1}(\alpha)=W^{r-1}(-\alpha),  
\\
&&
X^{-1}(\alpha)=W^{r-1}(-\alpha)\oplus W^{r-1}(-\alpha),
\quad
X^{r-1}(\alpha)=W^{r-1}(\alpha)\oplus W^{r-1}(\alpha).
\en
Applying the above rule repeatedly, we see that 
$V^{\otimes n}$ decomposes as a direct sum of 
$V^s,X^s(\alpha)$ ($0\le s \le r-2$) and $W^{r-1}(\alpha)$, $\alpha=\pm 1$.

Define subspaces $\G^{(r)}_n,\Bc^{(r)}_n$ of $V^{\otimes n}$ 
inductively as follows. 

\begin{definition}\label{def:goodbad}
We set $\G^{(r)}_1=V=V^{1}$, $\Bc^{(r)}_1=0$. 
{}For $n\ge 2$, 
$\G^{(r)}_n$ is the direct sum of the $V^s$ $(0\le s\le r-2)$
appearing in the decomposition of $\G^{(r)}_{n-1}\otimes V$. 
$\Bc^{(r)}_n$ is the sum of $\Bc^{(r)}_{n-1}\otimes V$ and 
the direct sum of $W^{r-1}(1)$'s appearing in 
$\G^{(r)}_{n-1}\otimes V$.  
\end{definition}

We have
\be
V^{\otimes n}=\G^{(r)}_n\oplus \Bc^{(r)}_n.
\en
The decomposition \eqref{dcp1} is orthogonal  with respect to 
the standard symmetric bilinear form $(~,~)$ on $V^{\otimes N}$. 
Hence $\G^{(r)}_n$ and $\Bc^{(r)}_n$ are orthogonal. 
Note that $F^{r-1}\G^{(r)}_n=0$. 

Let 
\be 
\Omega_{n, l}=\Ker E \cap (V^{\otimes n})_{l}. 
\en 

\begin{lem}\label{lem:fr-1}
If $u\in \Omega_{n,l}\cap\Bc^{(r)}_n$ and $F^{r-1}v=0$, then 
$(u,v)=0$.
\end{lem}
\begin{proof}
We may assume that $v$ belongs to one of the subspaces 
isomorphic to \eqref{module}.
{}From the structure of these modules, we see that
the condition $F^{r-1}v=0$ implies $v\in \Im F$ or $v\in V^s$.
If $v=f v'$ for some $v'$, then
we have $(u,v)=(u,Fv')=(ET^{-1}u,v')=0$. 
Otherwise $v\in \G^{(r)}_n$,     
and the assertion follows from the orthogonality of 
$\G^{(r)}_n$ and $\Bc^{(r)}_n$.  
\end{proof}

We denote the basis of $V=V^{1}$ by $v_{+}=v_{0}^{1}(1)$ and
$v_{-}=v_{1}^{1}(1)$. 
Let $R^{+}$ be the linear operator on $V \otimes V$ given by
\bea
&&R^{+} v_\pm \otimes v_\pm =\epsilon v_\pm \otimes v_\pm,
\nn \\
&&R^{+} v_+\otimes v_-= v_-\otimes v_+, 
\label{def:Rplus} \\
&&R^{+} v_-\otimes v_+= (\epsilon-\epsilon^{-1})v_-\otimes v_+
+v_+\otimes v_-. \nn
\ena
Denote by $R^{+}_{i\,i+1}\in \End(V^{\otimes n})$ ($1\le i\le n-1$) the operator 
acting as $R^{+}$ on the $(i,i+1)$ tensor factor and as identity elsewhere. 
They commute with the action of $\Ure(\slt)$ on $V^{\otimes n}$ 
defined by the opposite coproduct $\Delta'$ \eqref{OPP}. 
Define further
\bea
\Pi_{n, l}=P_{n-1\,n}\cdots P_{23}P_{12}
\cdot D_1^{l-n/2-1},
\label{def:vpi} 
\ena
where $D\in\End(V)$, $D^{1/2} v_{\pm}=e^{\mp \pi i/2r}v_{\pm}$.
It is easy to check that
\be
R^{+}_{i\,i+1}\Omega_{n,l}\subset \Omega_{n,l},\quad 
\Pi_{n, l}\Omega_{n,l}\subset \Omega_{n,l}.
\en
The subspaces $\mathcal{G}_{n}^{(r)},\mathcal{B}_{n}^{(r)}$
are not invariant under the actions of $R^{+}_{i\,i+1}$ and $\Pi_{n, l}$.
Nevertheless we have
\begin{lem}\label{lem:Om}
The space $\Omega_{n,l}\cap \Bc^{(r)}_n$ is invariant by 
the operators 
$R^{+}_{i\,i+1}$ $(1\le i\le n-1)$, $\vvpi$.
\end{lem}
\begin{proof}
Let $u\in \Omega_{n,l}\cap \Bc^{(r)}_n$. 
Since $\Bc^{(r)}_n$ is the orthogonal complement of $\G^{(r)}_n$, 
the assertion will follow if we show that 
\bea
&&(R^{+}_{i\,i+1}u,\G^{(r)}_n)=0,
\label{ort1}
\\
&&(\vvpi u,\G^{(r)}_n)=0.
\label{ort}
\ena
The equation \eqref{ort1} is a consequence of 
the relation $(R^{+}_{i\,i+1}u,v)=(u,R^{+}_{i\,i+1}v)$, 
$F^{r-1}R^{+}_{i\,i+1}\G^{(r)}_n=R^{+}_{i\,i+1}F^{r-1}\G^{(r)}_{n}=0$ and Lemma \ref{lem:fr-1}.

Let us verify \eqref{ort}. 
Take $v\in \G^{(r)}_n$ and set 
$\tilde{v}=D_1^{l-n/2-1}P_{12}P_{23}\cdots P_{n-1\,n}v$, 
so that $(\vvpi u,v)=(u,\tilde{v})$. 
We have $v\in V^s\otimes V$ for some $0\le s\le r-2$,  
and $\tilde{v}\in V\otimes V^s$. 
Therefore we have either $F^{r-1}\tilde{v}=0$, or else 
$s=r-2$ and $\tilde{v}=v_+\otimes v'$, $Ev'=0$. 
The latter does not take place. 
Indeed, since $v_+$ is an eigenvector of $D$, 
it would mean that 
$v= P_{n-1\,n}\cdots P_{12}D_1^{-l+n/2+1}\tilde{v}$ 
is proportional to $v'\otimes v_+$,  
which belong to the irreducible component 
$W^{r-1}(1)$ of $V^{r-2}\otimes V$.
This is a contradiction. 
Hence Lemma \ref{lem:fr-1} implies \eqref{ort}.
\end{proof}

\bigskip 
\noindent
{\it Acknowledgments.}\quad
JM is partially supported by 
the Grant-in-Aid for Scientific Research (B2) no.12440039, 
and TM is partially supported by 
(A1) no.13304010, Japan Society for the Promotion of Science.
YT is supported by the Japan Society for the Promotion of Science.


\begin{thebibliography}{[FJKLM]}

\bibitem{ABF}
G.~Andrews, R.~Baxter and P.~Forrester,
\newblock Eight-vertex SOS model and generalized
Rogers-Ramanujan-type identities,
\newblock {\em J. Stat. Phys.} {\bf 35} (1984), 193--266.

\bibitem{BBS} 
O.~Babelon, D.~Bernard and F.~Smirnov, 
\newblock Null vectors in integrable field theory, 
\newblock {\em Commun. Math. Phys.} 
{\bf 186} (1997), 601--648.


\bibitem{CM}
J.~Cardy and G.~Mussardo, 
\newblock Form factors of descendent operators in perturbed 
conformal field theories,
\newblock {\em Nucl. Phys.}{\bf B340} (1990), 387--402.

\bibitem{CP}
V.~Chari and A.~Pressley,
\newblock Quantum affine algebras at roots of unity,
\newblock {\em Representation Theory} (electronic), 
{\bf 1} (1997), 280--382.


\bibitem{FF}
B.~L. Feigin and E.~Feigin,
\newblock $q$-characters of the tensor products in
$\slt$-case,
\newblock math.QA/0201111.

\bibitem{FJKLM}
B.~Feigin, M.~Jimbo, R.~Kedem, S.~Loktev and T.~Miwa,
Spaces of coinvariants and fusion product I. 
{}From equivalence theorem to Kostka polynomials, 
\newblock math.QA/0205324.

\bibitem{FS}
B.~L. Feigin and A.~V. Stoyanovsky,
\newblock Functional models 
for representations of current algebras and 
  semi-imfinite {Schubert} cells, 
\newblock {\em Funct.~Anal.~and~Its~Appl.}
{\bf 28} (1993) 55--72.

\bibitem{HKKOTY}
G.~Hatayama, A.~N.~Kirillov, A.~Kuniba,
M.~Okado, T.~Takagi and Y.~Yamada,
\newblock Character formulae of $\widehat{sl}_n$-modules
and inhomogeneous paths,
\newblock {\em Nucl. Phys. B} {\bf 536} (1999), 575--616.

\bibitem{JM}
M.~Jimbo and T.~Miwa,
\newblock Quantum KZ equation with $|q|=1$ and correlation functions 
of the XXZ model in the gapless regime, 
\newblock {\em J. Phys. A: Math. Gen.} {\bf 29} (1996), 2923--2958.


\bibitem{Kas}
M.~Kashiwara,
\newblock On level zero representations of quantized enveloping
algebras,
\newblock {\em Duke Math. J. } {\bf 112} (2002), 117--195.

\bibitem{Kou}
A.~Koubek,
\newblock The space of local operators in perturbed conformal 
field theories,
\newblock {\em Nucl. Phys.} {\bf B435}[FS] (1995), 703--734.

\bibitem{L}
S.~Lukyanov,
\newblock Form-factors of exponential fields 
in the sine-Gordon model,
\newblock {\em Mod.Phys.Lett.} {\bf A12} (1997), 2543-2550.

\bibitem{N}
A.~Nakayashiki,
\newblock Residues of $q$-hypergeometric integrals and 
characters of affine Lie algebras,
\newblock math.QA/0210168.

\bibitem{NT} 
A.~Nakayashiki and Y.~Takeyama, 
\newblock On form factors of the $SU(2)$ invariant Thirring model, 
\newblock in MathPhys Odyssey 2001, Integrable Models and Beyond- 
in honor of Barry M. McCoy, ed. M.~Kashiwara and T.~Miwa, 
\newblock Progr. in Math. Phys., Birk\"auser, 2002, 357--390.  

\bibitem{NPT}
A.~Nakayashiki, V.~Tarasov and S.~Pakulyak,
\newblock On solutions of the KZ and qKZ equations at
level $0$,
\newblock {\em Ann. Inst. Henri Poincar\'e} {\bf 71}
(1999), 459--496.

\bibitem{Pas} 
V.~Pasquier, 
\newblock Etiology of IRF models,
\newblock {\em Commun. Math. Phys.} {\bf 118} (1988), 355--364.

\bibitem{RS}
N.~Reshetikhin and F.~Smirnov,
\newblock Hidden quantum group symmetry and integrable 
perturbations of conformal field theory,
\newblock {\em Commun. Math. Phys.} {\bf 131} (1990), 
157--177.

\bibitem{RT} 
N.~Reshetilhin and V.~G.~Turaev, 
\newblock Invariants of 3-manifolds via link polynomials and 
quantum groups, 
\newblock {\em Invent. Math.} {\bf 103} (1991), 547--597. 

\bibitem{Sbk} 
F.~Smirnov, 
\newblock Form factors in completely integrable models in 
quantum field theory, 
\newblock World Scientific, Singapore, 1992.

\bibitem{Abelian} 
F.~Smirnov, 
\newblock On the deformation of Abelian integrals, 
\newblock {\em Lett. Math. Phys.} {\bf 36} (1996), 267--275. 

\bibitem{Sm}
F.~Smirnov, 
\newblock Counting the local fields in SG theory,
\newblock {\em Nucl. Phys. B453} 
{\bf B453} [FS] (1995), 807--824.

\bibitem{SS}
A.~Schilling and M.~Shimozono, 
\newblock Fermionic formulas for level-restricted generalized
Kostka polynomials and coset branching functions, 
\newblock {\em Commun. Math. Phys.} {\bf 220} (2001), 105--164. 

\bibitem{SW}
A.~Schilling and S.~O.~Warnaar,
\newblock Inhomogeneous lattice paths, 
generalized Kostka polynomials 
and $A_{n-1}$ supernomials, 
\newblock math.QA/9802111, 
\newblock {\em Commun. Math. Phys.} {\bf 202} (1999), 359--401.

\bibitem{T}
V.~Tarasov,
\newblock Completeness of the hypergeometric solutions
of the $qKZ$ equations at level $0$,
\newblock {\em Amer. Math. Soc. Translations} {\it Ser.2}
{\bf 201} (2000), 309--321.

\bibitem{TV}
V.~Tarasov and A.~Varchenko,
\newblock Geometry of $q$-hypergeometric functions
as a bridge between Yangians and quantum affine algebras,
\newblock {\em Inventiones Math.} {\bf 128}
(1997), 501--588.

\bibitem{TUY}
A.~Tsuchiya, K.~Ueno and Y.~Yamada, 
\newblock Conformal field theory on universal
family of stable curves with gauge symmetry,
\newblock {\em Adv. Stud. Pure Math.} {\bf 19}
(1989), 459--566.


\bibitem{Zam} 
A.~Zamolodchikov, 
\newblock Integrable field theory from conformal field theory,
\newblock {\em Adv. Stud. Pure Math.} 
{\bf 19} (1989), 641--674.

\end{thebibliography}
\end{document}